\documentclass{aa}  

\usepackage{graphicx}
\usepackage{float} 
\usepackage{longtable}
\usepackage{txfonts}
\usepackage{orcidlink}
\usepackage{xcolor}
\usepackage{colortbl}  

\begin{document}

   \title{Early Light Curve Excess in Type IIb Supernovae Observed by the ATLAS Survey}

   \subtitle{Qualitative constraints on progenitor systems}

        \author{Bastian Ayala\orcidlink{0009-0002-2564-1069}\inst{1,2,3}\thanks{Corresponding author: \email{b.ayalainostroza@gmail.com}}
          \and
          Joseph P. Anderson\orcidlink{0000-0003-0227-3451}\inst{3,2}
          \and
          G. Pignata\orcidlink{0000000-0003-0006-0188}\inst{2,4}
          \and
          Francisco F\"orster\orcidlink{0000-0003-3459-2270}\inst{5,2,6}
          \and
          S.J. Smartt\orcidlink{0000-0002-8229-1731}\inst{7,8}
          \and
          A. Rest\orcidlink{0000-0002-4410-5387}\inst{9,10}
          \and
          Mart\'in Solar\orcidlink{0000-0002-3148-1359}\inst{11}
          \and
          Nicolas Erasmus\orcidlink{0000-0002-9986-3898}\inst{12,13}
          \and
          Raya Dastidar\orcidlink{0000-0001-6191-7160}\inst{1,2}
          \and
          Mauricio Ramirez\orcidlink{0009-0001-3293-7741}\inst{1,2}
          \and
          Jonathan Pineda-García\orcidlink{0000-0003-0737-8463}\inst{1,2}
          }

        \institute{Instituto de Astrof\'isica, Departamento de F\'isica, Facultad de Ciencias Exactas, Universidad Andr\'es Bello, Fern\'andez Concha 700, Las Condes, Santiago RM, Chile\
         \and
             Millennium Institute of Astrophysics, Nuncio Monse\~nor Sotero Sanz 100, Of. 104, Providencia, Santiago, Chile\
         \and
             European Southern Observatory, Alonso de C\'ordova 3107, Vitacura, Santiago, Chile\
         \and
             Instituto de Alta Investigaci\'on, Universidad de Tarapac\'a, Casilla 7D, Arica, Chile\
         \and
             Data and Artificial Intelligence Initiative, University of Chile\
         \and
             Center for Mathematical Modelling, CMM\
         \and
             Department of Physics, University of Oxford, Keble Road, Oxford, OX1 3RH, UK\
         \and
             Astrophysics Research Centre, School of Mathematics and Physics, Queen’s University Belfast, BT7 1NN, UK\
         \and
             Space Telescope Science Institute, 3700 San Martin Drive, Baltimore, MD 21218, USA\
         \and
             Physics and Astronomy Department, Johns Hopkins University, Baltimore, MD 21218, USA\
         \and
             Astronomical Observatory Institute, Faculty of Physics and Astronomy, Adam Mickiewicz University, ul. Słoneczna 36, 60-286, Poznań, Poland\
         \and
             South African Astronomical Observatory, Cape Town, 7925, South Africa\
         \and
             Department of Physics, Stellenbosch University, Stellenbosch, 7602, South Africa
      }

   \date{Received September 15, 1996; accepted March 16, 1997}

 
    \abstract
    {Type IIb supernovae (SNe IIb) often exhibit an early light curve excess (EE) preceding the main peak powered by $^{56}\text{Ni}$ decay. The physical origin of this early emission remains an open question. Among the proposed scenarios, shock cooling emission—resulting from the interaction between the shockwave and extended envelopes—is considered the most plausible mechanism. The frequency of these events remains unconstrained.}
    {This study aims to quantify the frequency of EE in SNe IIb and investigate its physical origin by analyzing optical light curves from the Asteroid Terrestrial-impact Last Alert System (ATLAS) survey.}
    {We selected 74 SNe IIb from 153 spectroscopically classified events in the Transient Name Server (TNS) database, observed by ATLAS, with peak fluxes exceeding 150 \textmu Jy and explosion epoch uncertainties lower than six days. Using light curve model fitting and outlier analysis, we identified SNe IIb exhibiting EE and analyzed their photometric properties.}
    {We found 21 SNe IIb with EE, corresponding to a frequency of approximately 28--40\%, with the higher value obtained under the most stringent data cuts. The EE's duration and color evolution are consistent with shock cooling in extended hydrogen-rich envelopes. We also found that EE SNe IIb have longer rise times and faster post-peak decline rates than non-EE SNe IIb, while both groups share similar peak absolute magnitudes.}
    {Our findings suggest that EE and non-EE SNe IIb likely share similar initial progenitor masses but differ in ejecta mass properties, potentially due to varying degrees of binary interaction. This study constrains the evolutionary pathways of SNe IIb progenitors as compact stars with and without extended hydrogen envelopes.}


   \maketitle
%

\section{Introduction}

Core-collapse supernovae (CCSNe) correspond to the explosion of stars at their final evolutionary stage, with zero-age main sequence (ZAMS) masses \(M_{\text{ZAMS}} > 8M_{\odot}\) \citep[e.g.,][]{2003ApJ...591..288H, 2009ARA&A..47...63S}, leaving a neutron star or a black hole as a remnant. Among CCSNe, stripped-envelope supernovae (SESNe—types Ic, Ib, and IIb) form a subcategory, where the progenitor loses part of its envelope prior to the explosion \citep{1996ApJ...459..547C}. Within the SESNe class, SNe IIb exhibit hydrogen features at early times, which disappear a few weeks after the explosion as Helium lines begin to emerge in the spectrum \citep[e.g.,][]{1993ApJ...415L.103F, 1997ARA&A..35..309F}. SNe~IIb progenitors are thought to retain part of their H envelope, with estimated masses ranging between a few tens and one \(M_{\odot}\) \citep[e.g.,][]{2011MNRAS.412.1522S, 2019ApJ...885..130S, 2022MNRAS.511..691G}. The main mechanism responsible for stripping the material before the explosion remains unclear. However, the likely mechanisms include stellar winds in more massive (\(M_{\text{ZAMS}} \gtrsim 20M_{\odot}\)) isolated stars \citep[e.g.,][]{2006ApJ...645L..45S, 2008A&ARv..16..209P} or mass exchange in close binary systems involving a less massive (\(M_{\text{ZAMS}} \lesssim 17M_{\odot}\)) progenitor star \citep[e.g.,][]{2011A&A...528A.131C, 2013ApJ...762...74B}.

SNe~IIb typically exhibit bell-shaped light curves, reaching the peak approximately 20 days after the explosion \citep[e.g.,][]{2016MNRAS.457..328L, 2011ApJ...741...97D, 2023ApJ...955...71R} powered by the radioactive decay of \({}^{56}\text{Ni} \rightarrow {}^{56}\text{Co} \rightarrow {}^{56}\text{Fe}\) \citep{1994ApJ...429..300W}. However, \citet{2024Natur.628..733R} recently argued that the luminosity of most SESNe (including SNe~IIb) cannot be explained solely by \({}^{56}\text{Ni}\) decay. Regardless of the primary power source (e.g., ${}^{56}\text{Ni}$, central engine or circumstellar material interactions), statistical studies show that SNe~IIb generally exhibits single-peaked light curves \citep[e.g.,][]{2018A&A...609A.136T, 2024Natur.628..733R}. Nevertheless, some SNe~IIb display an additional peak preceding the main peak. This feature has been observed in SNe~IIb such as SN~1993J \citep{1994ApJ...429..300W}, SN~2011fu \citep{2013MNRAS.431..308K}, SN~2011dh \citep{2011ApJ...742L..18A}, SN~2011hs \citep{2014MNRAS.439.1807B}, SN~2013df \citep{2014AJ....147...37V, Morales_Garoffolo_2014}, SN~2016gkg \citep{2018Natur.554..497B, 2017ApJ...837L...2A}, SN~2017jgh \citep{2021MNRAS.507.3125A}, ZTF18aalrxas \citep{2019ApJ...878L...5F}, SN~2020bio \citep{2023ApJ...954...35P}, and SN~2021zby \citep{2023ApJ...943L..15W}.

Various hypotheses have been presented to explain Type IIb SNe with an early flux excess (hereafter EE-SNe), which precedes the $^{56}\text{Ni}$ peak, resulting in double-peaked light curves. For instance, double ${}^{56}\text{Ni}$ distributions could produce an early luminosity peak through jet-like outflows that eject nickel-rich material into low-opacity regions \citep{2022A&A...667A..92O}. Alternatively, Thomson scattering and chemical mixing could influence the strength and shape of the first peak in double-peaked SNe~IIb light curves \citep{Park_2024}.
However, the primary explanation for the ten EE-SNe mentioned above involves the interaction of the shock wave generated during core collapse with an extended hydrogen-rich envelope. This interaction produces a shock-heated envelope that emits radiation as it cools \citep[e.g.,][]{2012ApJ...752...78S, 2014ApJ...788..193N, 2014AJ....147...37V, 2015ApJ...808L..51P, 2017ApJ...838..130S}.

Several studies have investigated the physical conditions required to produce double-peaked light curves, assuming their origin in shock cooling emission, showing that extended envelopes are essential. Using analytical approximations, \citet{2014ApJ...788..193N} demonstrated that a low-mass (\(\sim 0.06 M_{\odot}\)), hydrogen-rich extended envelope can generate a double-peaked light curve. This model was later refined by \citet{2015ApJ...808L..51P}, who argued that when the envelope is sufficiently massive and extended, the shock wave propagates for a longer time, producing a first peak brighter and more pronounced. Similarly, \citet{2017ApJ...838..130S} showed that extended hydrogen-dominated envelopes with masses below \(1 M_{\odot}\), modeled with polytropic density profiles, can also produce early light curve peaks. Supporting these insights, hydrodynamic simulations of SN~2011dh by \citet{2012ApJ...757...31B} suggested that its double-peaked light curve originated from a low-mass extended envelope (\(M_{\text{env}}^{H} \approx 0.1 M_{\odot}\)). Furthermore, \citet{2018A&A...612A..61D} proposed that progenitors with a core-halo structure—where 95\% of the mass is concentrated within 10\% of the stellar radius—could also account for double-peaked light curves due to their tenuous and extended envelopes.

Recently, \citet{2024A&A...685A.169D} simulated the light curves of CCSNe explosions originating from progenitors in binary systems. In their model, they used fixed values of explosion energy and ${}^{56}\text{Ni}$, the primary star had a ZAMS mass of \(12 M_{\odot}\) and a mass ratio 0.9 with respect to the companion, while the initial binary period was varied. These variations led to progenitors with different hydrogen masses and envelope properties, producing a diversity of light curve morphologies, including double-peaked light curves consistent with SNe~IIb. 

The binary progenitor scenario is the most probable explanation for SNe~IIb. Direct and indirect evidence supports this hypothesis. Post-explosion images have revealed companion stars in SN~2011dh \citep{2014ApJ...793L..22F}, SN~2001ig \citep{2018ApJ...856...83R}, and SN~1993J \citep{2004Natur.427..129M}. In addition, pseudo-spectral energy distributions from pre-explosion images of SN~2008ax indicate either a single B-type star or a binary system with a lower-mass progenitor and a main-sequence O9-B0 companion \citep{2015ApJ...811..147F}.

Theoretical studies show that SNe~IIb can originate from both single-star and binary-star progenitors. Single-star models require high ZAMS masses (\(20-26M_{\odot}\)) to strip the hydrogen envelope through strong stellar winds \citep{2012A&A...538L...8G, 2019ApJ...885..130S}. Binary models, on the other hand, demonstrate that mass transfer from lower-mass primaries in close binary systems can reproduce the observed properties of SNe~IIb, such as the He-core mass, under specific parameter configurations highlighting binary interactions as a plausible evolutionary pathway for SNe~IIb \citep{2011A&A...528A.131C, 2017ApJ...840...10Y, 2019ApJ...885..130S, 2020ApJ...903...70S}.

Additional evidence from X-ray and radio observations reveals circumstellar densities higher than those expected from single-star mass loss, suggesting that binary interactions could play an important role in shaping the environments of SNe~IIb \citep[e.g.,][]{2020ApJ...903...70S}. Moreover, more than 70\% of massive stars are estimated to interact in binaries during their evolution \citep{2012Sci...337..444S}. This fraction suggests that binary star systems could be the progenitors of SNe~IIb and that binary interactions could produce the properties of SNe~IIb progenitors. For further details, see \citet{2019ApJ...885..130S, 2020ApJ...903...70S} and references therein.

This paper aims to quantify the fraction of SNe~IIb that exhibit an EE preceding the main peak in the Asteroid Terrestrial-impact Last Alert System (ATLAS) survey \citep{2018PASP..130f4505T}, which provides one of the highest cadence currently available among all-sky surveys. Beyond quantifying this frequency, we characterize the light curves of EE SNe and SNe~IIb without the early excess (hereafter non-EE SNe) using parameters such as post-peak decline (\( \Delta M_{15} \)): the difference in magnitude between the peak and 15 days after the peak), peak absolute magnitude (\( M_{\text{peak}} \)), and rise time (\( t_{\text{rise}} \)). We interpret these light curve parameters in terms of explosion properties, including \(^{56}\text{Ni}\) mass and ejecta mass (\( M_{\text{ej}} \)), based on correlations reported in the literature \citep{2011ApJ...741...97D, 2016MNRAS.457..328L, 2016MNRAS.458.2973P, 2019MNRAS.485.1559P, 2023ApJ...955...71R, 2015MNRAS.450.1295W, 2016MNRAS.458.1618D}. Finally, we attempt to understand and constrain the different progenitor scenarios that can explain the observed frequency of EE and non-EE SNe~IIb.

This paper is structured as follows: Section 2 describes the data sample and photometric cleaning process, including the selection criteria of the light curves. Section 3 introduces the methodology for detecting the EE, including the estimation of explosion epochs and the process of fitting light curves and identifying the EE. Section 4 presents the results, discussing the performance of the model fitting, the frequency of the EE, the characterization of light curve properties, and the verification of the EE detections, frequency estimates, and correlations. This section also explores the distributions and correlations of light curve parameters and the EE's duration and color evolution. Section 5 provides a discussion of the results, including a qualitative physical interpretation of light curve parameter distributions, implications for progenitor systems, conditions for producing double-peaked light curves, scenarios for generating extended envelopes, correlation analysis of light curve parameters, and peculiar objects or potential misclassifications. Finally, Section 6 summarizes the main conclusions of the study.

\section{Data Sample and Data Cleaning}
\label{sec:Data_sample_cleaning}

As of 3rd October 2024, TNS \footnote{\url{https://www.wis-tns.org/}} contains a total of 233 spectroscopically classified IIb SNe, 154 of which had been observed by ATLAS \citep[][]{2020PASP..132h5002S}.

ATLAS is a sky survey funded by the National Aeronautics and Space Administration (NASA) and operated by the University of Hawaii. It consists of four telescopes: two in Hawaii, one in Chile, and another in South Africa. Although its primary mission is to identify near-Earth asteroids, its observing strategy with high cadence and depth makes it an excellent facility to discover and photometrically follow SNe and transient events. With its large field of view (55 deg$^{2}$), ATLAS can scan the entire observable sky every one or two days depending on the weather conditions with four exposures obtained on individual fields on each night\citep{2018PASP..130f4505T}. The survey operates in two bands: cyan (\textit{c}) and orange (\textit{o}), corresponding approximately to the \textit{g+r} and \textit{r+i} filters from Pan-STARRS \citep{2018PASP..130f4505T}, covering wavelength ranges of 420-650 nm and 560-820 nm, respectively. During dark time\footnote{Period around New Moon when the night sky is darkest, optimal for observing faint astronomical objects.}, the ATLAS-o band reaches a limiting flux of $\sim 43.65 \, \mu \text{Jy}$ (corresponding to a magnitude of $19.8$).

We downloaded forced photometry of all SNe IIb from the ATLAS Forced Photometry server \citep{2021TNSAN...7....1S} and used the ATClean tool \citep{2023zndo...7897346R, rest2024atclean} to perform subsequent data cleaning. ATClean employs a data-cleaning methodology that includes several steps. First, it analyzes control light curves (CLCs) defined as forced photometry in a 17'' circular aperture around the SN to measure the sky flux \citep[see figure\,1 in][]{rest2024atclean} and estimates additional noise. Next, chi-square PSF fitting ($\chi^{2}_{\text{PSF}}$) and flux uncertainty cuts are applied. Then, SN photometry is averaged over the four measurements obtained each night. A detailed description of the methodology and parameters used in ATClean are summarized in Appendix~\ref{sec:atclean_appendix}.

Figure~\ref{fig:ejemplo_de_ATclean} (a and b panels) shows an example of the data cleaning results for SN2021zby. In this work, we used only the \textit{o}-band since the \textit{c}-band is significantly less sampled than the \textit{o}-band, as can be seen in Figure~\ref{fig:ejemplo_de_ATclean}.

\begin{figure*}[ht!]
\centering
\includegraphics[width=1\textwidth]{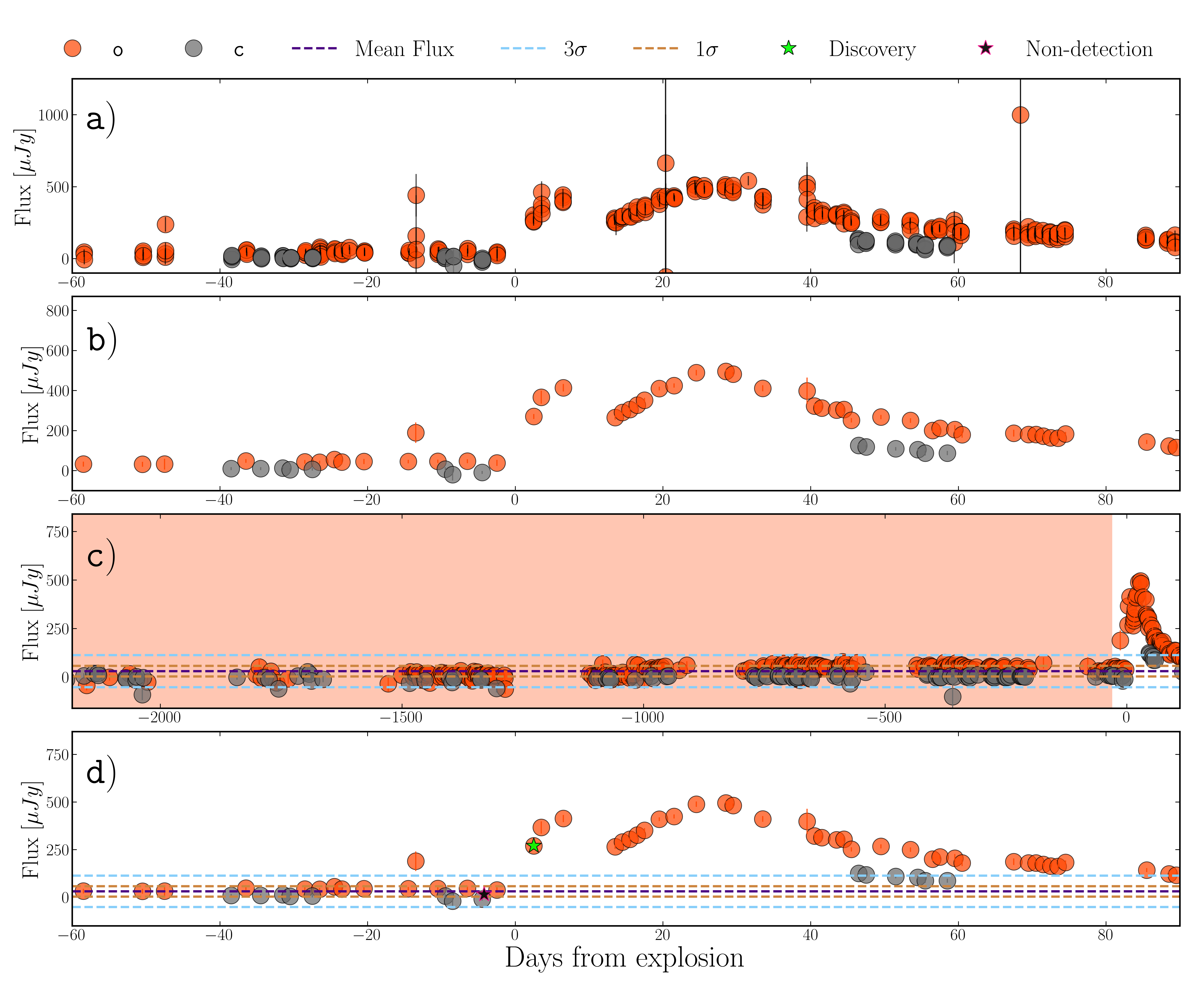} 
\caption{Data cleaning and explosion epoch estimation example for SN2021zby. The panels, labeled as a, b, c, and d from top to bottom, present different stages of the analysis (we detail a and b in Section~\ref{sec:Data_sample_cleaning} and c and d in Section~\ref{sec:met}). Red and grey dots represent photometry in the o and c bands, respectively. Panel a) shows the raw ATLAS forced photometry. Panel b) presents the photometry after applying the ATClean methodology detailed in Section~\ref{sec:Data_sample_cleaning}. Panel c) shows the pre-SN photometry, where the red region highlights the relevant data, and the sky blue and brown dashed lines represent the 3 and 1.5 $\sigma_{\text{pre-SN}}$ values, respectively. In Panel d), green and black star symbols represent the last non-detection and discovery, which we use to estimate the explosion time.
\label{fig:ejemplo_de_ATclean}}
\end{figure*}

\section{Shock Cooling Detection: Methodology}
\label{sec:met}

Figure~\ref{fig:ejemplo_de_ATclean} shows an example SN~IIb light curve of SN~2021zby, which displays the standard broad peak that is understood to be produced by the decay of \(^{56}\text{Ni}\) \citep{1994ApJ...429..300W} and an early excess preceding the \(^{56}\text{Ni}\) peak. 
In this section, we introduce our statistical methodology for detecting or ruling out the presence of excess luminosity in the early light curves. We define early excess detection by identifying positive photometric outliers (i.e., residuals where the flux exceeds the model prediction) before the light curve maximum, compared to a model that does not reproduce double-peaked light curves.

\subsection{Light Curve Properties: Explosion Epoch and Point Density}
\label{sec:light_curve_properties}

Accurate estimation of the explosion epoch and its uncertainty is essential for properly detecting and characterizing early flux excess, given that previously reported cases in the literature have durations ranging from approximately 4 days \citep[e.g., SN~2011dh;][]{2011ApJ...742L..18A} to 12 days \citep[e.g., SN~2021zby;][]{2023ApJ...943L..15W}. We estimate the explosion epoch $t_{\text{exp}}$ and its uncertainty $\delta t_{\text{exp}}$ using pre-explosion SN photometry from ATLAS. Specifically, we determine $t_{\text{exp}}$ as the midpoint between the last non-detection and the discovery epoch, following standard procedures \citep[e.g.,][]{2015A&A...574A..60T}. A detailed description of the discovery time and last non-detection determination is provided in Appendix~\ref{app:explosion_epoch}. Figure~\ref{fig:ejemplo_de_ATclean} (d) illustrates an example of the estimation of the last non-detection and discovery epochs, while panel (c) shows the pre-SN flux.

\begin{figure*}[ht]
\centering
\includegraphics[width=1\textwidth]{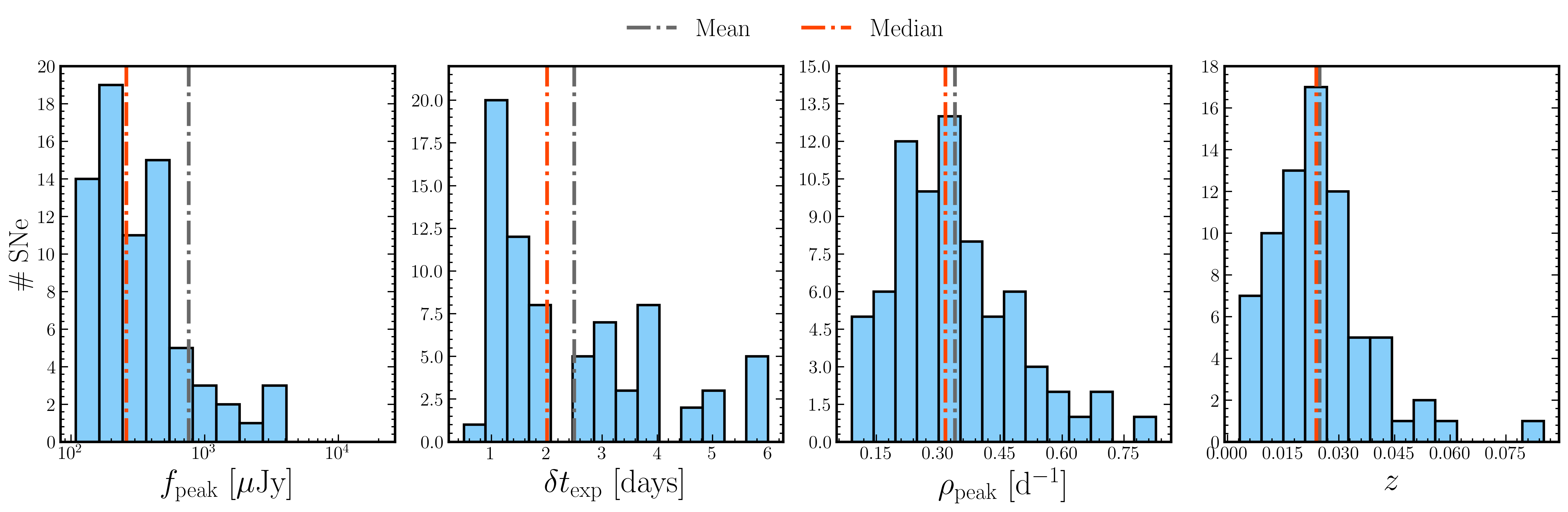} 
\caption{Distributions of $f_{\text{peak}}$, $\delta t_{\text{exp}}$, $\rho_{\text{peak}}$, and $z$ for the SNe sample. Red and grey vertical dashed lines indicate the median and mean, respectively.}
\label{fig:phot_comp}
\end{figure*}

Using our selected sample of 74 SNe with peak flux greater than $\mathrm{f}_{\text{peak}} > 150\,\mu\mathrm{Jy}$, we analyzed the distributions of $\mathrm{f}_{\text{peak}}$, $\delta t_{\text{exp}}$, $\rho_{\text{peak}}$, and redshift ($z$), as shown in Figure~\ref{fig:phot_comp}. Here, $\rho_{\text{peak}}$ represents the numerical density of points during the rise time, defined as the number of photometric data points between the explosion and the peak divided by the time to peak. The sample exhibits a mean and median redshift of 0.02, with a mean and median peak brightness of $\overline{\mathrm{f}_{\text{peak}}} = 560\,\mu\mathrm{Jy}$ and $\tilde{\mathrm{f}}_{\text{peak}} = 339\,\mu\mathrm{Jy}$, respectively. The explosion epoch uncertainty has mean and median values of $\overline{\delta t_{\text{exp}}} = 2.38$ days and $\tilde{\delta t}_{\text{exp}} = 2$ days, respectively. These small uncertainties demonstrate the high precision of ATLAS in determining explosion times, which is crucial for studying early excess flux in SNe~IIb.

\subsection{Light-curve Fitting and Early Excess Identification}
\label{identi}

We detect early flux excess by fitting a model representative of standard SNe~IIb light curves, where we do not expect additional peaks before the maximum. Consequently, any early-time flux excess will result in positive residuals where the observed flux exceeds the model. We define an EE point as a positive outlier that deviates from the model by more than three times the residual standard deviation ($3\sigma_{r}$) between $t_{\text{exp}} - \delta t_{\text{exp}}$ and $t_{\text{peak}}$, where $t_{\text{exp}}$ is the explosion time and $\delta t_{\text{exp}}$ is the uncertainty in the explosion time, and $t_{\text{peak}}$ is the time of the light curve peak. Specifically, a data point at time $t_{i}$ is considered a $3\sigma_{r}$ outlier if its residual, defined as $r_{t_{i}} = |\mathrm{f}_{i}^{\text{observed}} - \mathrm{f}_{i}^{\text{model}}|$, satisfies $r_{t_{i}} > 3\sigma_{r}$. We choose a $3\sigma$ threshold as it represents a highly conservative criterion for outliers.

We fit the Supernova Parametric Model (SPM) introduced by \citet{2019ApJ...884...83V}, which is a six-parameter analytic model that describes SN light curves that take into account factors such as explosion times, initial rise timescales, and post-peak decline rates. We employ the modified version of the SPM as described by \citet{2021AJ....161..141S}, defined by the following equation:

\begin{equation}
\begin{split}
    F &=  \frac{A  \left(1-\beta' \frac{t-t_{0}}{t_{1}-t_{0}}\right)}{1 + \exp\left(- \frac{t-t_{0}}{\tau_{\text{rise}}}\right)} \left[1 - \sigma \left(\frac{t - t_{1}}{3}\right)\right] \\
    & \quad + \frac{A(1-\beta') \exp\left(-\frac{t-t_{1}}{\rho_{\text{fall}}}\right)}{1+\exp\left(-\frac{t-t_{0}}{\rho_{\text{rise}}}\right)} \left[\sigma \left(\frac{t-t_{1}}{3}\right)\right], \\
    & \quad \sigma(t) = \frac{1}{1+\exp(-t)}.
    \label{eq: Villar}
\end{split}
\end{equation}

This modified version includes two key improvements: first, the reparameterization ensures the function remains positive within a valid range; second, incorporating a sigmoid function allows a smooth transition between the two components of the model, enabling more effective optimization and improving parameter estimation. 

We estimate the best fit using the Markov Chain Monte Carlo (MCMC) method, which maximizes the posterior probability \(\mathcal{P}(\theta|D) \propto \mathcal{L}(D|\theta) \pi(\theta)\), where \(\mathcal{L}(D|\theta)\) is the likelihood function and \(\pi(\theta)\) represents the prior probability of the parameters \(\theta\). We employed the Python package \texttt{emcee} \citep{2013PASP..125..306F} for this analysis. We defined the input parameter distributions using Gaussian priors \(\mathcal{N}(\mu, \sigma)\), where \(\mu\) is the mean, \(\sigma\) the standard deviation, and boundary conditions (\(B\)) restrict the priors to zero probability outside the defined range (see Table~\ref{tab:MCMC}). The SPM was fitted to the \(o\)-band light curves at phases between \((t_{\text{exp}} - \delta t_{\text{exp}})\) and 60 days post-explosion, provided the flux exceeded \(60\,\mu\mathrm{Jy}\) to avoid biases near the ATLAS survey's limiting flux of \(\sim43.65\,\mu\mathrm{Jy}\). Best-fit parameter values were derived from the posterior distributions using the median of the samples, with uncertainties estimated as the 16th and 84th percentiles. Figure~\ref{fig:MCMC} illustrates an example of the posterior distributions for SN~2022jpx.

\begin{figure*}[ht!]
\centering
\includegraphics[width=1\textwidth]{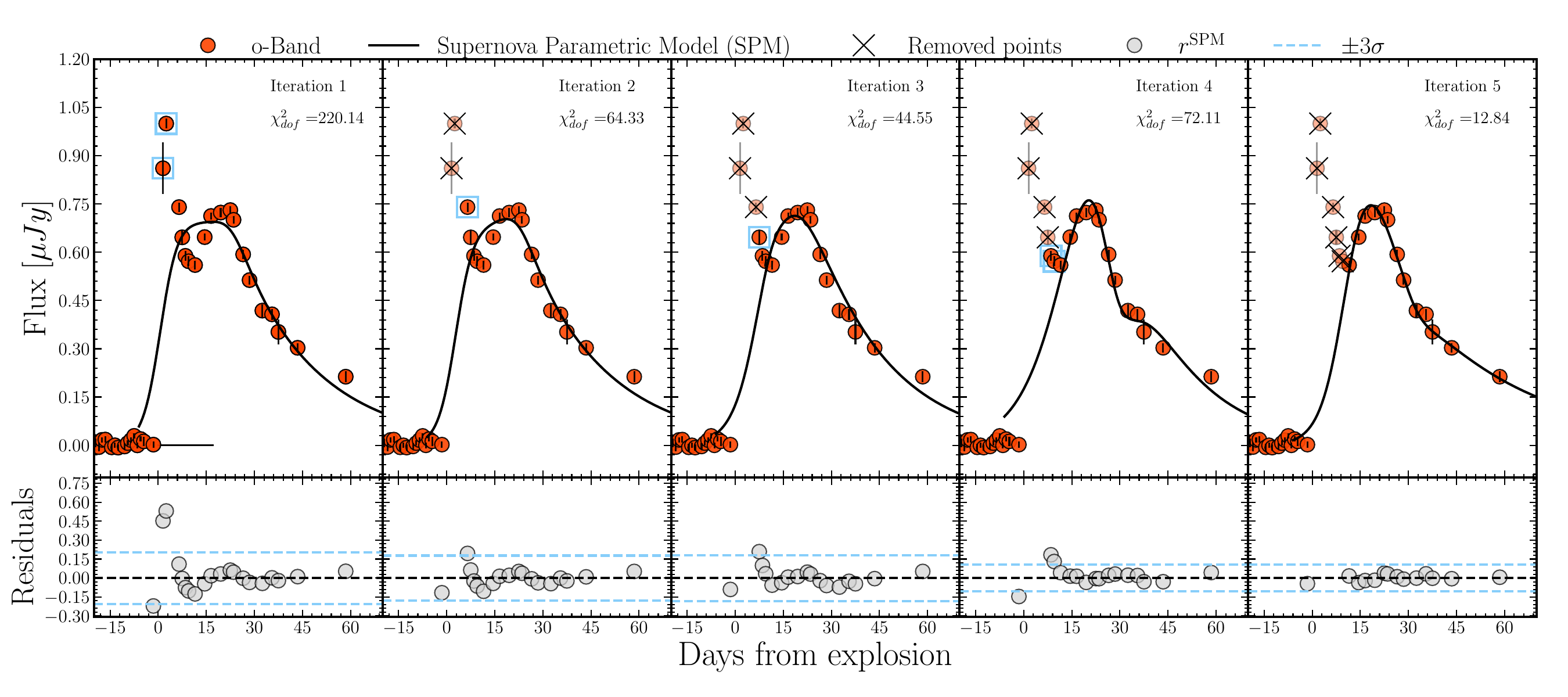}
\caption{Example of the SPM fit and refitting process for SN~2022jpx. The left panel shows the initial SPM fit, with outliers identified and enclosed by open blue squares in the top plot. The bottom plot displays the residuals associated with the SPM fit (\(r^{\text{SPM}}\)), with dashed lines indicating the \(\pm3\sigma_{r}\) thresholds used for outlier identification. From the left panel to the right, we illustrate the iterative SPM fitting process, where outliers are removed in each iteration until no outliers are detected. In this case, no outliers remain after the fifth iteration. The \(\chi^2_{\text{dof}}\) value is reported in each panel for every iteration.
\label{fig:ejemplo_SPM_}
}
\end{figure*}

Given that the early flux excess could affect our light curve fitting (as the SPM does not account for this feature), we mitigate its influence by iteratively refitting the models after removing all positive $3\sigma_{r}$ outliers between $t_{\text{exp}}$ and $t_{\text{peak}}$, as described in Section~\ref{identi}. The refitting process stops at the iteration at which no outliers remain, thus ensuring that our light curve represents the early evolution of normal SNe~IIb (without flux excess), which is fundamental to our methodology for detecting flux excess. Figure~\ref{fig:ejemplo_SPM_}  illustrates an example of the SPM fitting and refitting process after removing outliers for SN~2022jpx.

\begin{table*}
\centering
\begin{tabular}{lllll}
\hline
Parameter & Description      & Prior Distribution                                                             & Bounds ($B$)                & Units     \\ \hline

$\tau_{\text{rise}}$                 & Rise Time        & $\mathcal{N}\left(\max\left(1.0, \frac{t_{\text{peak}} - t_{\text{exp}}}{2.0}\right), \sigma_{B_{\tau_{\text{rise}}}}\right)$    & [1, 70]               & days      \\
$\tau_{\text{fall}}$                 & Decline Time     & $\mathcal{N}(40, \sigma_{B_{\tau_{\text{fall}}}})$                                                       & [1, 100]              & days      \\
$t_{0}$                              & Start Time     & $\mathcal{N}\left(\min\left(30, \max\left(-10, t_{\text{peak}} - t_{\text{exp}}\right)\right), \sigma_{B_{t_{0}}}\right)$ & [-10, 30]            & MJD       \\
$A$                                  & Amplitude        & $\mathcal{N}\left(\max(\text{f}), \sigma_{B_{A}}\right)$                                           & [$\max(\text{Flux})/10$, $2 
 \cdot\max(\text{f})$] & --        \\
$\beta'$                             & Plateau Slope    & $\mathcal{N}(0.5, \sigma_{B_{\beta'}})$                                                                      & [0, 1]               & flux/day  \\
$\gamma \equiv t_{1} - t_{0}$        & Plateau Duration & $\mathcal{N}\left(\min\left(100, \max\left(1, \frac{t_{\text{peak}} - t_{\text{exp}}}{2.0}\right)\right), \sigma_{B_{\gamma}}\right)$ & [1, 100]              & days      \\
\hline
\end{tabular}
\caption{MCMC input parameters for the Supernova Parametric Model (SPM), including parameter descriptions, prior distributions, bounds, and units. Priors are defined with standard deviations ($\sigma_{B}$) based on parameter bounds.}
\label{tab:MCMC}
\end{table*}

\section{Results}
\label{resultados}

\begin{figure*}[ht!]
\centering
\includegraphics[width=1\textwidth]{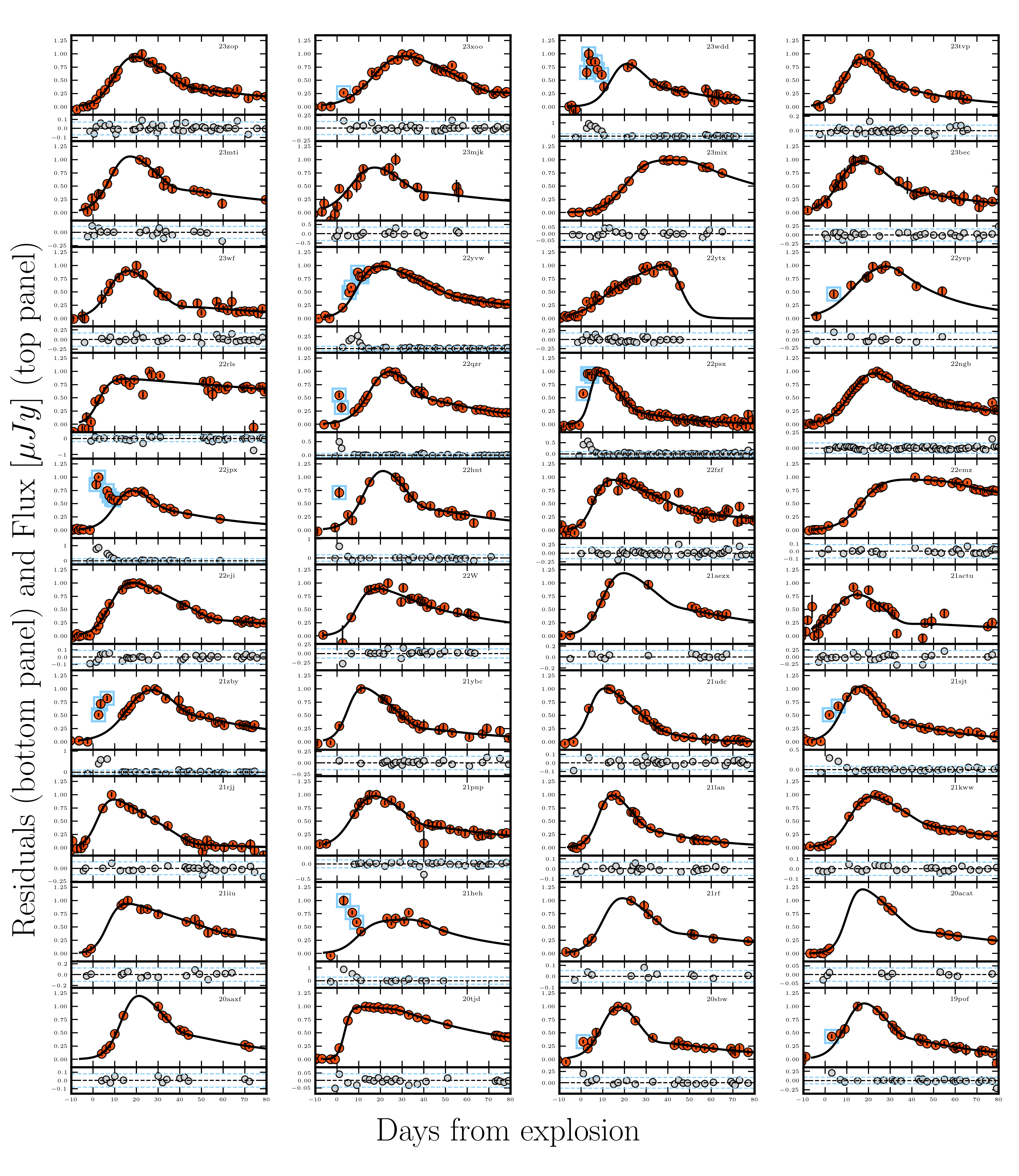} 
\caption{SPM Fits and Residuals for 40 Type IIb Supernovae. The red points show the $o$-band photometry from ATLAS for each of the 40 SNe in our sample. The black continuous line represents each Supernova Parametric Model (SPM) fit. Inside each panel, we indicate the name of the corresponding SN.
Bottom Panels display the residuals of the SPM fit for each supernova. The dashed lines mark the $\pm3\sigma_{r}$ thresholds, which we use to identify outliers. We excluded these outliers from the SPM refitting process. Cyan squares highlight the outliers, which we define as evidence for EE.
\label{fig:LCs_1}}
\end{figure*}

\begin{figure*}[ht!]
\centering
\includegraphics[width=1\textwidth]{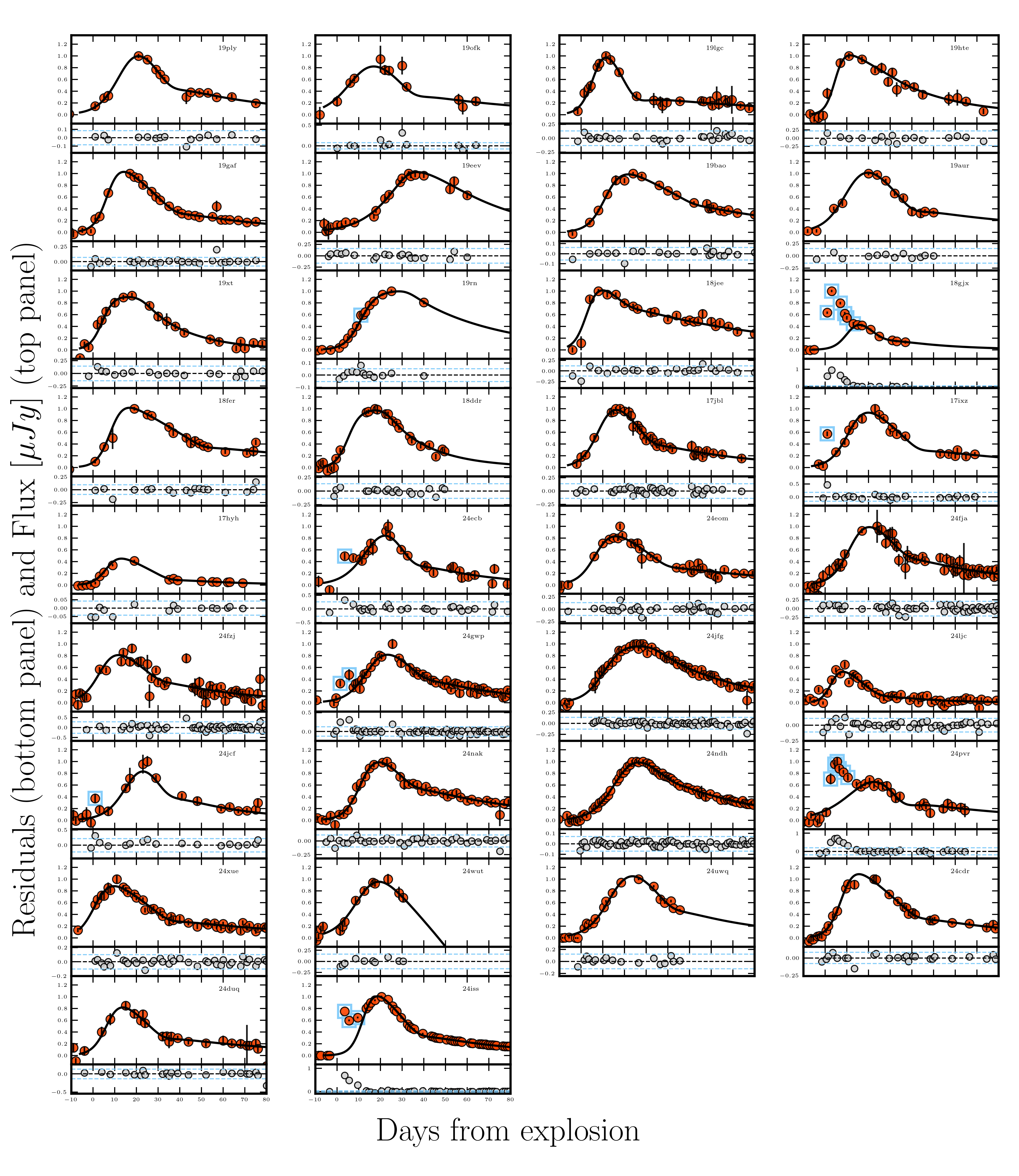} 
\caption{SPM Fits and Residuals for 34 Type IIb Supernovae. \textit{Top Panels:} The red points represent the $o$-band photometry from ATLAS for each of the 34 supernovae in our sample. The black continuous line shows each supernova's Supernova Parametric Model (SPM) fit. The name of each supernova is indicated inside its corresponding panel. 
\textit{Bottom Panels:} Residuals of the SPM fit for each supernova, with the dashed lines indicating the $\pm4\sigma_{r}$ thresholds used for outlier identification. The outliers were excluded from the SPM refitting process. 
\label{fig:LCs_2}}
\end{figure*}

This section presents our results on light curve characterization and the frequency and properties of EE SNe light curves. All SN~IIb light curves in our sample - with and without an EE detection - are displayed in Figures \ref{fig:LCs_1} and \ref{fig:LCs_2}. The fitting process shows relative residuals below 7.2\% for most phases, except between days 10 and 15, where the relative error increases to 13\% due to the rapid flux evolution and limited data points in this phase. The overall performance of the fit and detailed statistical analysis is presented in Appendix~\ref {model_fit_perf}. Next, we quantify the early excess frequency across different photometric densities and explosion epoch uncertainties thresholds (see Section \ref{frecuency}). Subsequently, we characterize our SN~IIb light curves (see Section \ref{distributions}), comparing the distributions of EE and non-EE SNe~IIb and exploring correlations among these parameters. Finally, we study the properties of the EE, specifically its duration and color evolution, and compare these properties between our ATLAS dataset and the literature sample (see Section~\ref{ref_EE_prop}).

\subsection{Early Excess Frequency}
\label{frecuency}

\begin{figure*}[ht!]
\centering
\includegraphics[width=1\textwidth]{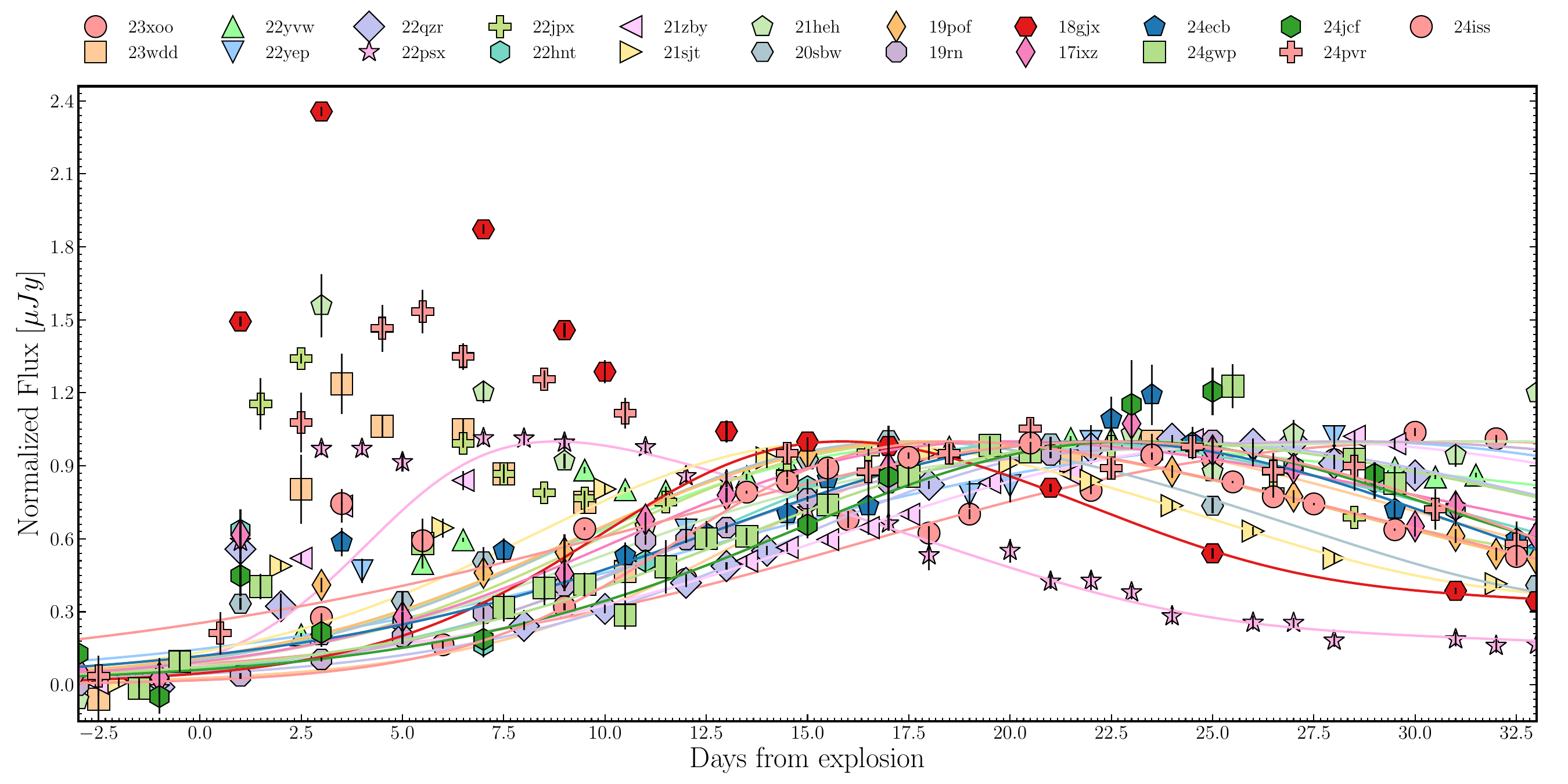} 
\caption{\textbf{Normalized Light Curves of EE Supernovae.} 
Normalized light curves of the 21 supernovae exhibiting early flux excess (hereafter EE). The flux is normalized by the main peak flux in the $o$-band, illustrating the variability and timing of the early excess among the sample.
\label{fig:all_EE_LCs}}
\end{figure*}

Using our defined EE detection methodology, we identify a total of 21 EE SNe from the entire sample of 74 SNe~IIb analyzed by applying the criterion defined in Section~\ref{sec:light_curve_properties}, which selects light curves with a peak flux greater than $f>150\,\mu\mathrm{Jy}$, without considering additional criteria for explosion epoch error ($\delta t_{\text{exp}}$) or observational point density ($\rho_{\text{peak}}$). The light curves of these 21 SNe, normalized by their main peak flux, are shown in Figure~\ref{fig:all_EE_LCs}.

\begin{figure*}[ht!]
\centering
\includegraphics[width=1\textwidth]{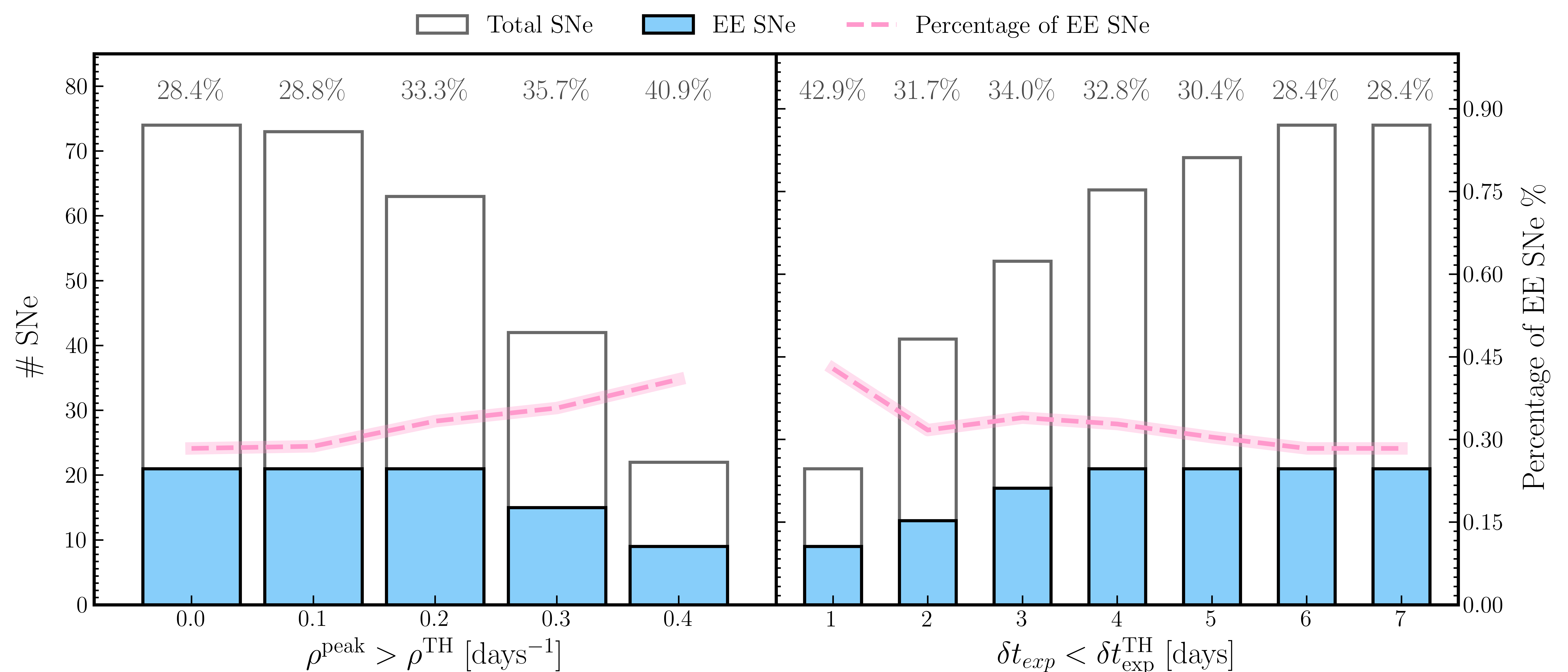} 
\caption{\textbf{Supernovae Counts and EE Percentages Under Specific Criteria.} 
Number of Type IIb supernovae satisfying specific criteria for observational point density ($\rho_{\mathrm{peak}} > \rho_{\mathrm{TH}}$, left panel) and explosion epoch error ($\delta t_{\mathrm{exp}} < \delta t_{\mathrm{TH}}$, right panel) are shown as unfilled bars. Blue-filled bars represent the subset of supernovae exhibiting early flux excess (EE). The dashed pink line indicates the percentage of supernovae with EE for each criterion, with numerical percentages displayed above the bars.
\label{fig:frequency_EE}}
\end{figure*}

We proceed to quantify the frequency of EE SNe under different criteria for $\delta t_{\text{exp}}$ and $\rho_{\text{peak}}$. Specifically, we consider thresholds for the explosion epoch error $\delta t_{\text{TH}} \in [1, 2, \ldots, 7]$ days, and for the observational point density $\rho^{\text{peak}}_{\text{TH}} \in [0.0, 0.1, \ldots, 0.4]$ observations per day, and we apply these thresholds to the sample. Figure~\ref{fig:frequency_EE} illustrates the percentage of SNe~IIb exhibiting early flux excess, which ranges from 28.4\% to 40.9\%, depending on the criteria applied for $\delta t_{\text{exp}}$ and $\rho_{\text{peak}}$. As expected, the frequency of EE SNe increases consistently for smaller explosion epoch errors ($\delta t_{\text{exp}}$) and higher observational point densities ($\rho_{\text{peak}}$).

\subsection{Characterization of Light Curve Properties}

\label{sec:cara_LCss}

After identifying the 21 EE SNe in the ATLAS sample and calculating their frequency, we characterize their photometric properties in order to compare them with the 53 non-EE SNe. 

Our analysis was conducted on the 74 SNe~IIb sample using different criteria for explosion epoch error (\( \delta t_{\text{exp}} \)) and observational point density (\( \rho_{\text{peak}} \)), as defined in Section~\ref{frecuency}. However, we defined a representative subset with \( \rho_{\text{peak}} > 0.3  \text{ days}^{-1}\), which retains 42 SNe and exhibits consistent results across all criteria. This subset is exceptionally reliable due to its well-constrained explosion epoch errors, with a mean of \( 2.12 \, \mathrm{days} \), a median of \( 1.5 \, \mathrm{days} \), and a standard deviation of \( 1.29 \, \mathrm{days} \).

We characterized the light curve properties using the rise time (\(t_{\text{rise}}\)), the decline rate (\(\Delta M_{15}\)), and the peak absolute magnitude (\(M^{\mathrm{\text{peak}}}_{\text{abs}}\)). The absolute magnitudes were estimated using distances derived from recessional redshifts, assuming a local Hubble-Lemaître constant $H_0 = 74.03 \pm 1.42 \, \text{km s}^{-1} \, \text{Mpc}^{-1}$ \citep{2019ApJ...876...85R}, with cosmological parameters $\Omega_m = 0.27$ and $\Omega_\Lambda = 0.73$, and correcting only for Milky Way extinction from galactic dust reddening reported in \cite{2011ApJ...737..103S} using ratio of total to selective extinction (\(R_V=3.1\)). Host galaxy extinction corrections were not applied because the necessary data are unavailable for all SNe in our sample. Even when such data are accessible for methods like the color-color curve \citep{2014AJ....148..107R, 2019MNRAS.483.5459R}, color evolution \citep{2018A&A...609A.135S}, Balmer decrement \citep{2012MNRAS.421..486X}, or sodium absorption lines \citep{2012MNRAS.426.1465P}, the associated uncertainties in estimating \(E(B-V)\) remain significant. As \citet{2018MNRAS.476.4592D} highlighted, these methods are generally unreliable for SNe II and often introduce additional uncertainties. Furthermore, the poorly constrained \(R_V\) exacerbates these issues. Consequently, we did not correct for extinction in the host galaxy to avoid adding further uncertainties to our results.

\subsubsection{Distributions of Light Curve Parameters}
\label{distributions}

In this subsection, we compare the light curve properties of EE and non-EE SNe~IIb, focusing on their rise time (\(t_{\text{rise}}\)), decline rate (\(\Delta M_{15}\)), and peak absolute magnitude (\(M^{\mathrm{peak}}_{\text{abs}}\)).

\begin{figure*}[ht!]
    \centering
    \includegraphics[width=\textwidth]{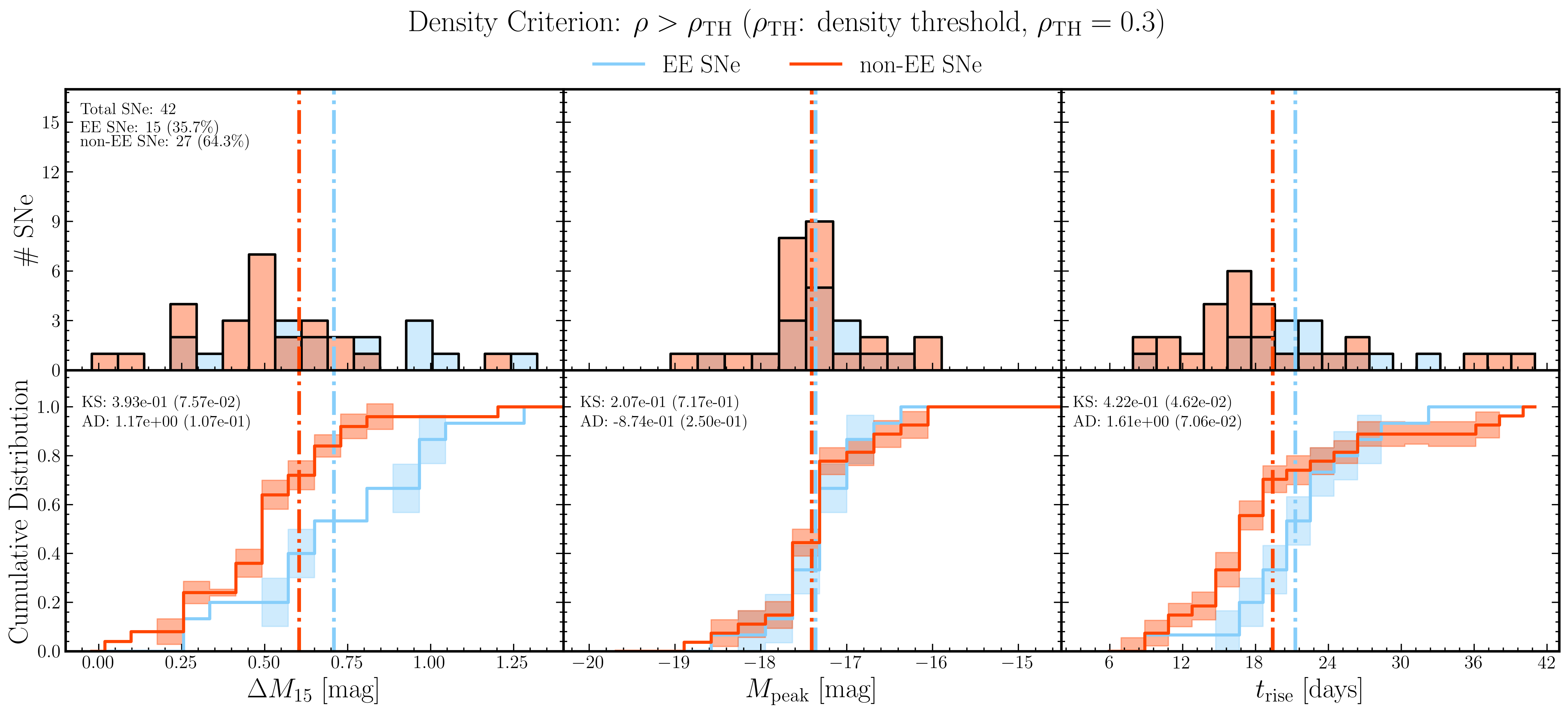}\\[2ex]
    \caption{\textbf{Histograms and cumulative distributions for the EE and non-EE SNe samples, considering a density criterion of $\rho_{\text{peak}} = 0.3 \text{ days}^{-1}$.}
    Red bars represent the non-EE sample, while blue bars represent the EE sample. The total number of SNe, as well as the counts for EE and non-EE SNe, are indicated within each panel. Separate panels display $\Delta M_{15}$, $M^{\text{o}}_{\text{abs}}$, and $t_{\text{rise}}$. The cumulative distributions in the subpanels include the results of the Kolmogorov--Smirnov (KS) and Anderson--Darling (AD) tests for each parameter. 
    }
    \label{fig:corr_dist_rho_0.3}
\end{figure*}

The Kolmogorov--Smirnov (KS) and Anderson--Darling (AD) tests were applied to evaluate differences in the distributions of \(t_{\text{rise}}\), \(\Delta M_{15}\), and \(M^{\mathrm{peak}}_{\text{abs}}\) between the EE and non-EE groups. For the representative subsample with a density threshold of \(\rho_{\text{peak}} = 0.3 \text{ days}^{-1}\), the KS and AD statistics are reported with their corresponding p-values in parentheses. For \(t_{\text{rise}}\), the KS statistic is \(4.22 \times 10^{-1}\) (\(4.62 \times 10^{-2}\)), indicating a significant difference. For \(\Delta M_{15}\), the KS statistic is \(3.93 \times 10^{-1}\) (\(7.57 \times 10^{-2}\)), indicating a marginal difference that does not meet the conventional threshold for statistical significance (\(p < 0.05\)). In addition, for \(M^{\mathrm{peak}}_{\text{abs}}\), the KS statistic is \(2.07 \times 10^{-1}\) (\(7.17 \times 10^{-1}\)), suggesting no significant difference.  
Figure~\ref{fig:corr_dist_rho_0.3} shows the histograms and cumulative distributions of these parameters for our representative density threshold of \(\rho_{\text{peak}} = 0.3 \text{ days}^{-1}\).

We used the KS and AD tests to analyze the distributions of \(t_{\text{rise}}\), \(\Delta M_{15}\), and \(M^{\mathrm{peak}}_{\text{abs}}\) for a grid of density (\(\rho_{\text{peak}}\)) and explosion epoch error thresholds (\(\delta t_{\text{exp}}\)). Table~\ref{tab:tabla_KS_AD} summarizes the full results of these tests. Significant differences in \(t_{\text{rise}}\) between EE and non-EE SNe were found for \(\rho_{\text{peak}} \leq 0.3 \text{ days}^{-1}\) and \(1 \leq \delta t_{\text{exp}} \leq 6\), with p-values below \(0.05\). For \(\Delta M_{15}\), significant differences appeared at \(\rho_{\text{peak}} \leq 0.2\) and most \(1 \leq \delta t_{\text{exp}} \leq 6\) thresholds, except at \(\delta t_{\text{exp}} = 2\). In contrast, \(M^{\mathrm{peak}}_{\text{abs}}\) showed no significant differences, as all p-values were above \(0.05\). Better observational criteria indicate higher qualitative differences, but reduced sample sizes hinder statistical significance.

\subsubsection{Correlations in Light Curve Parameters}
\label{correlations}

\begin{figure*}[ht!]
    \centering
    \includegraphics[width=\textwidth]{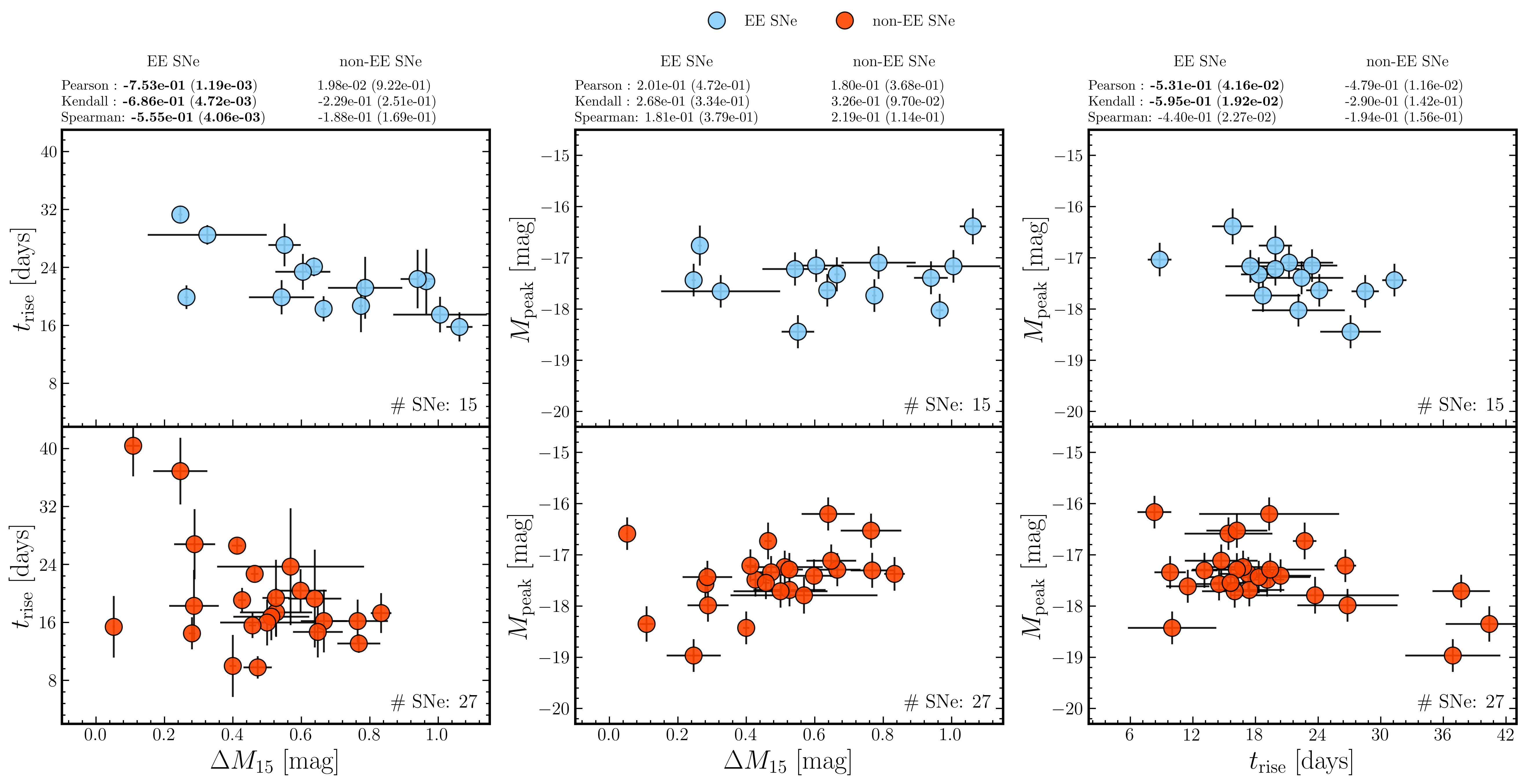}\\[2ex]
    \caption{\textbf{Relation between $\Delta M_{15}$ and $t_{\text{rise}}$, $\Delta M_{15}$ and $M^{\text{o}}_{\text{abs}}$, and $M^{\text{o}}_{\text{abs}}$ and $t_{\text{rise}}$ for the EE and non-EE samples}
    The EE SNe is in blue, and the non-EE SNe is in red. The Pearson, Spearman, and Kendall correlation coefficients for each relationship are reported within the corresponding panels. 
    }
    \label{fig:corr_dist_rho_0.3_}
\end{figure*}

Following the analysis of distributions, we investigated possible correlations within each group of EE and non-EE SNe. For the representative subsample with a density threshold of \(\rho_{\text{peak}} = 0.3 \text{ days}^{-1}\), we observe significant negative correlations in the EE sample for \(\Delta M_{15}\)-\(t_{\text{rise}}\) (\(r = -0.75\), \(p = 1.19 \times 10^{-3}\)) and \(M^{\mathrm{peak}}_{\text{abs}}\)-\(t_{\text{rise}}\) (\(r = -0.53\), \(p = 4.16 \times 10^{-2}\)). In addition, non-correlation between \(\Delta M_{15}\) and \(M^{\mathrm{peak}}_{\text{abs}}\) (\(r = 0.2\), \(p = 4.73 \times 10^{-1}\)) is observed in EE SNe. We find no significant correlations for any parameter pairs in the non-EE sample, as all \(p\)-values exceed \(5 \times 10^{-2}\). Figure~\ref{fig:corr_dist_rho_0.3_} shows the correlations for EE and non-EE samples in our representative subsample.

Expanding to other criteria, significant correlations in the EE sample persist for both \(\Delta M_{15}\)-\(t_{\text{rise}}\) and \(M^{\mathrm{peak}}_{\text{abs}}\)-\(t_{\text{rise}}\) across density thresholds (\(0 \leq \rho_{\text{peak}} \leq 0.3 \text{ days}^{-1}\)) and explosion epoch error thresholds (\(1 \leq \delta t_{\text{exp}} \leq 5\)), with \(p\)-values consistently below \(5 \times 10^{-2}\). For higher density thresholds (\(\rho_{\text{peak}} = 0.4 \text{ days}^{-1}\)), the correlations weaken as \(p\)-values exceed \(5 \times 10^{-2}\), despite some correlation coefficients being more significant than 0.5. This trend can be attributed to the reduced sample size in these cases of better observational characteristics of the light curves, as previously discussed. In contrast, the non-EE sample shows no significant correlations across any density or explosion epoch error thresholds, with all \(p\)-values above \(5 \times 10^{-2}\). Table~\ref{tab:KS_AD_tests} summarizes the full results of these tests

The above results indicate that EE SNe exhibits distinct photometric properties compared to non-EE SNe~IIb. Specifically, brighter SNe at peak tends to reach it at later phases, while SNe with faster post-peak decline rates tends to reach their peak earlier. However, the lack of significant correlation between \(\Delta M_{15}\) and \(M^{\mathrm{peak}}_{\text{abs}}\) in both EE and non-EE samples (\(r = 0.20\), \(p = 4.73 \times 10^{-1}\) for EE; \(r = 0.18\), \(p = 3.68 \times 10^{-1}\) for non-EE) suggests that the decline rate and peak luminosity are not directly linked. Furthermore, the complete absence of significant correlations in the non-EE sample within our ATLAS dataset suggests that early flux excess is associated with unique photometric properties not observed in SNe without this feature. Other hypotheses to explain this absence of correlation include the possibility that the sample is contaminated by peculiar objects or misclassifications, as discussed in Section \ref{pec_miss}.

\subsection{Properties of the Early Excess}
\label{ref_EE_prop}

Motivated by the differences observed in the distributions of light curve parameters between EE and non-EE SNe in Section~\ref{distributions}, as well as the correlations identified for EE SNe in Section~\ref{correlations}, this Section aims to characterize the properties of the EE. To achieve this, we first statistically quantify its duration in Section~\ref{duration_EE_} and subsequently explore differences in color evolution during the phases where the EE is present, comparing it with the color of non-EE SNe at similar phases in Section~\ref{color_evolution_EE}.

\subsubsection{Duration of the Early Excess}
\label{duration_EE_}

To statistically characterize the duration of the EE, we estimated an upper limit for its timescale because the low cadence and the limited number of photometric points during the EE do not allow for an accurate estimation. We define this duration as the time interval between the last photometric data point before the EE and the first photometric data point following it, where the residuals fall below the threshold of three residual standard deviations (\(3\sigma_{r}\)), as defined in Section~\ref{identi} for identifying the EE. If the point preceding the earliest point satisfying this condition occurs before the explosion epoch, we used the explosion epoch as the first epoch before the EE.

The uncertainty in the upper limit duration, \( \delta t_{\mathrm{EE}} \), was estimated using error propagation and incorporated three components: \( \delta t_{\mathrm{prev}} \), the time between the first early excess point and the preceding photometric point; \( \delta t_{\mathrm{post}} \), the time between the last early excess point and the subsequent photometric point; and \( \delta t_{\mathrm{exp}} \), the uncertainty in the explosion epoch. The total uncertainty was calculated as \( \delta t_{\mathrm{EE}} = \sqrt{\delta t_{\mathrm{prev}}^2 + \delta t_{\mathrm{post}}^2 + 2 \delta t_{\mathrm{exp}}^2} \). This approach provides an upper limit for the EE duration, accounting for uncertainties in the explosion epoch and the cadence of the light curve.

For the 21 EE SNe in our sample, the upper limit for the duration of the EE has a mean value of $8.8 \, \mathrm{days}$, a median of $8 \, \mathrm{days}$, and a standard deviation of $3.7 \, \mathrm{days}$. The distribution spans from $3$ to $15 \, \mathrm{days}$. The uncertainty in this upper limit has a mean value of $4.9$, a median of $4.4 \, \mathrm{days}$, and a standard deviation of $2.1 \, \mathrm{days}$, with values ranging between $2.6 \, \mathrm{days}$ and $10.5 \, \mathrm{days}$.

\subsubsection{Color Evolution}
\label{color_evolution_EE}

After statistically estimating the phases where the EE is present, we study the color evolution of EE and non-EE SNe during and after these phases. To do this, we used the Zwicky Transient Facility \citep{2019PASP..131a8002B} (ZTF) Forced Photometry Service \citep{2023arXiv230516279M}, which allowed us to download ZTF-\(g\) and ZTF-\(r\) bands and estimate the \(g - r\) color. We calculated the \(g - r\) color for 59 of the 74 objects in our selected ATLAS sample that the ZTF also observed. The colors were corrected only for extinction in the Milky Way, as described in Section \ref{sec:cara_LCss}.

\begin{figure*}[ht!]
\centering
\includegraphics[width=1\textwidth]{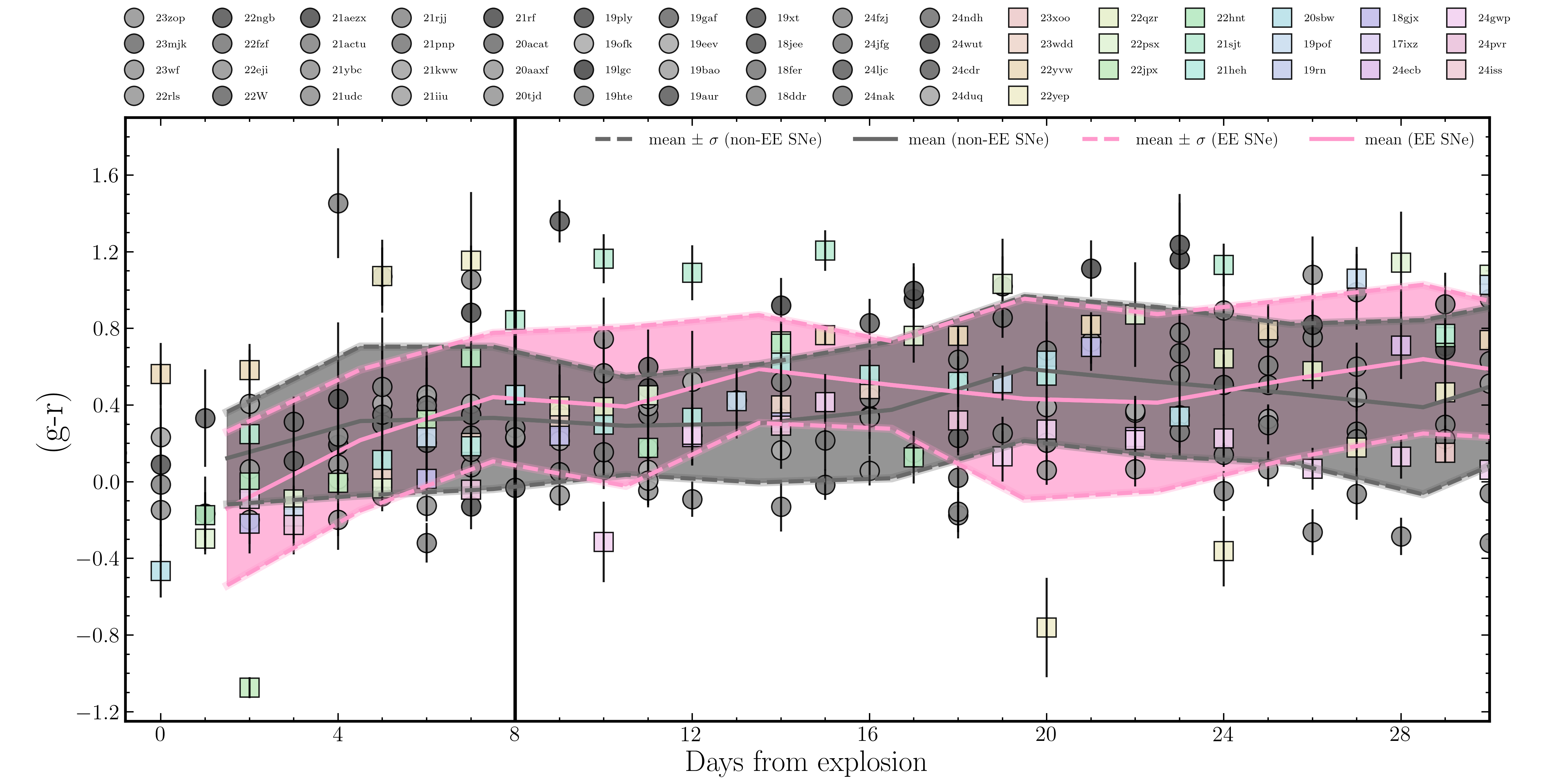} 
\caption{Color $g - r$ evolution for supernovae with and without early excess (EE) up to 30 days post-explosion. Individual data points for supernovae without EE are gray, while those with EE are in pastel-colored markers. The solid lines represent the centered moving averages described in the text: pink for supernovae with EE and gray for those without EE. The dashed lines indicate the standard deviation ($\sigma$) around the mean for each time interval, and the filled regions represent the mean $\pm$ standard deviation ($\sigma$). The vertical black line represents the upper limit average for the duration of the EE. This figure highlights the distinction in color evolution between the two groups, particularly in the early phases.
\label{fig:colores_ZTF}}
\end{figure*}

Of the 59 objects, 40 correspond to non-EE SNe and 19 to EE SNe. We focused on comparing the $g - r$ color of the EE and non-EE SNe. Considering that the median upper limit for the duration of the EE is eight days, we studied the color before and after this epoch. To analyze the average color evolution of the EE and non-EE SNe up to phase 30 days after the explosion, we calculated a centered moving average of the $g - r$ color using consecutive 3-day time windows from 0 to 30 days. Each value was assigned to the window's midpoint, representing the average interval. We computed the mean and standard deviation of the $g - r$ color measurements for each time window. This approach allowed us to compare the average color trends between the two groups, particularly before and after eight days post-explosion  (see Figure~\ref{fig:colores_ZTF}).

EE SNe exhibits a mean \( g - r \) color of \(-0.14\) during the three days post-explosion with a standard deviation of \(0.4\) (based on 11 SNe). In comparison, SNe without EE have a mean \( g - r \) color of \(0.12\) with a standard deviation of \(0.24\) during the three days post-explosion (based on seven SNe). While this difference indicates that EE SNe tends to have bluer colors in this early phase than non-EE SNe, the result is not statistically significant due to the standard deviation. However, as discussed in Section \ref{colors_lit}, we observe a consistent color trend in the literature sample of double-peaked SNe~IIb, which suggests that this color difference is real but limited by the low number of statistics in our sample.

Between three to six days post-explosion, the mean $g - r$ color for EE SNe is $0.22 \pm 0.37$, while for non-EE SNe it's $0.32 \pm 0.39$, indicating less pronounced color differences. From six to 30 days, the mean colors of both groups converge, with overlapping values and similar standard deviations. Between six and nine days, EE SNe has a mean $g - r$ color of $0.44 \pm 0.34$, compared to $0.33 \pm 0.37$ for non-EE SNe. This trend of minimal differences continues in later phases, suggesting a convergence in color evolution.

\subsection{Literature Sample with an Early Excess interpreted as Shock Cooling emission}
\label{lit_comp}

In this section, after identifying 21 EE-SNe and characterizing their light curves and colors, we investigate whether the EE observed in the ATLAS sample can be attributed to shock cooling (SC) emission from the progenitor envelope after shock breakout. To address this, we compare the photometric properties of the ATLAS EE-SNe with those of nine SNe~IIb from the literature that exhibit SC emission, aiming to assess their consistency with this physical origin.

The literature sample is characterized by its higher sampling quality than the ATLAS sample. Specifically, the mean observational point density for these SNe in the V band is $\overline{\rho}_{\text{peak}} = 6.17$ observations per day, significantly higher than the ATLAS sample's $\overline{\rho}_{\text{peak}} = 0.37$ observations per day. Similarly, the mean explosion epoch uncertainty for the literature sample is $\overline{\delta t_{\text{exp}}} = 0.52$ days, substantially lower than the ATLAS sample's $\overline{\delta t_{\text{exp}}} = 2.38$ days. 

The improved sampling properties of the literature sample result from the targeted follow-up campaigns conducted for the literature SNe. In contrast, the ATLAS sample is obtained from a survey with a default sky-scanning cadence, resulting in less frequent observations. The superior sampling of the literature sample enables more precise estimates of light curve parameters than the ATLAS sample.

\subsubsection{Light curves}
\label{lit_comp_LCs}

To characterize the light curves through \(\Delta M_{15}\), \(t_{\text{rise}}\), and \(M_{\text{peak, abs}}\), and to further characterize the SC phase using parameters described below, we employed Automated Loess Regression (ALR). ALR is a non-parametric method that fits local polynomials, with a smoothing parameter that adjusts the fit to local variations in the curve \citep{2019MNRAS.483.5459R}. We chose ALR over SPM because the SPM cannot accurately model the main peak of the light curve due to the high $\rho_{\text{peak}}$ during the SC phase and the inability to fully exclude SC photometric points during the refitting process defined in Section \ref{identi}. In this context, ALR provides a more reliable representation of the SC, transitional phase, and main peak.

\begin{figure*}[ht!]
\centering
\includegraphics[width=1\textwidth]{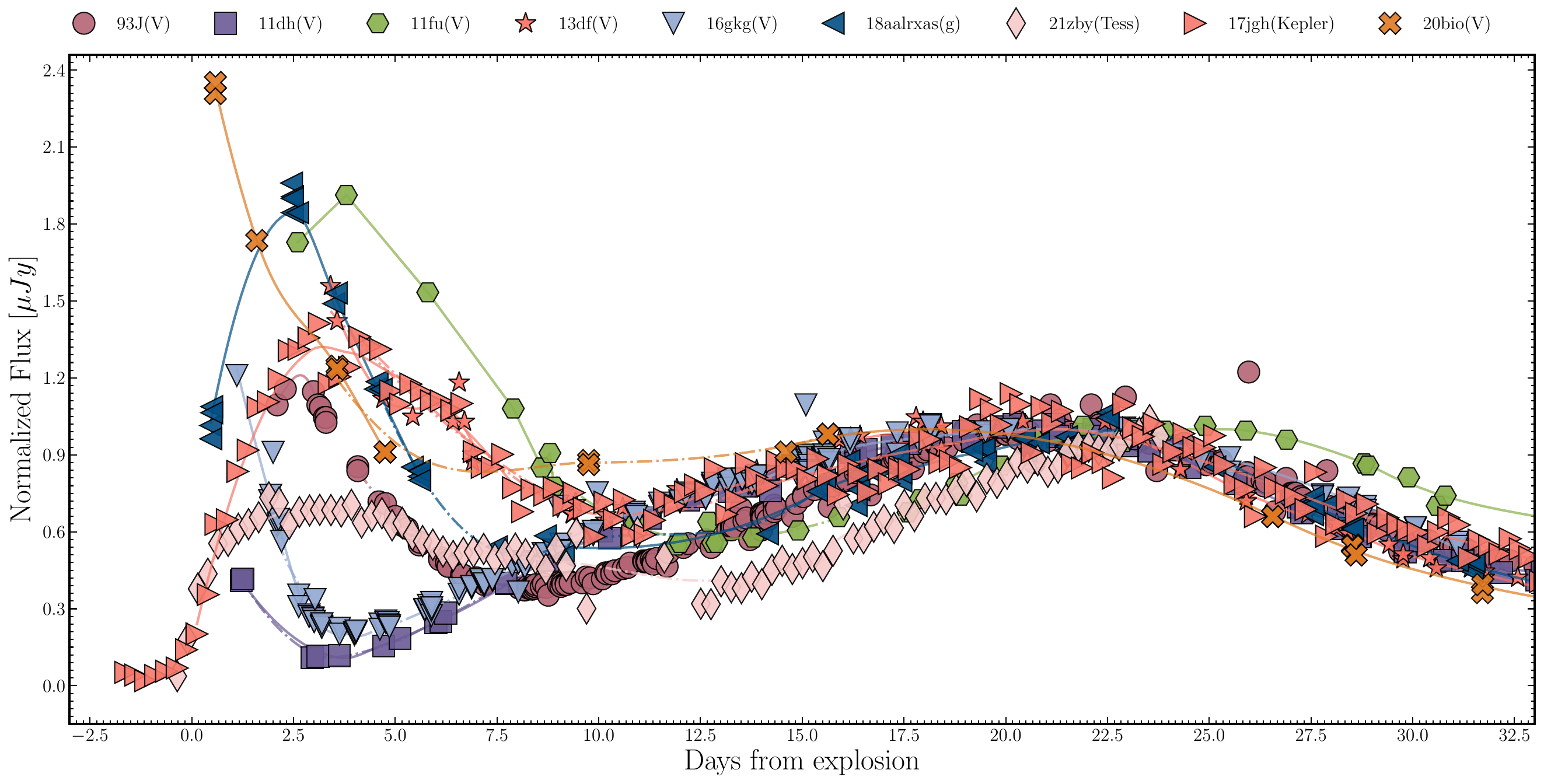} 
\caption{Normalized light curves of nine supernovae with SC emission compiled from the literature, presented in different photometric bands. Analogous to Figure~\ref{fig:all_EE_LCs}, the light curves are normalized by the flux of the second peak, illustrating the diversity of SC properties. Solid lines represent the ALR interpolation of the first and second peaks, while the dash-dotted line represents the transition between the two peaks.
\label{fig:LCs_literature}}
\end{figure*}

Figure~\ref{fig:LCs_literature} illustrates the light curves and their interpolations. We estimate the SC duration \( d_{\text{SC}} \) as the time between the explosion and the minimum flux in the transitional phase ($t_{\text{trans}}$) between the SC and the main peak. The SC duration has a mean of \( \overline{d_{\text{SC}}} = 8.85 \) days, a median of \( \tilde{d}_{\text{SC}} = 9.40 \) days, and a standard deviation of \( \sigma_{d_{\text{SC}}} = 3.37 \) days. 

This methodology to estimate the SC duration is not feasible for the ATLAS EE SNe because their low cadence does not allow for the observation of the transition between the two peaks in most cases. This limitation highlights another reason for investigating the literature sample. Its higher observational quality allows for a more detailed characterization of the SC phase and its properties.

In addition to the SC duration, we also estimate the absolute magnitude during the transition, \( M_{\text{trans}} \), and the peak absolute magnitude in the SC as \( M_{\text{peak, SC}} \). The \( M_{\text{peak, SC}} \) for SNe 2011h, 2016gkg, 2020bio and 2013df was estimated as a lower limit because the SC peak is not sampled. Therefore, we assume the \( M_{\text{peak, SC}} \) value as the earliest SC photometric point. Additionally, we estimate parameters related to the main peak, such as \( t_{\text{rise}} \), \( M^{\mathrm{}}_{\text{abs}} \)\footnote{\( M^{\mathrm{\text{peak}}}_{\text{abs}} \) represents the absolute magnitude corrected for extinction in the Milky Way and the host galaxy, using the values reported in Figure~\ref{tab:sn_observations}.}, and \( \Delta M_{15} \). All the parameters above were calculated for the six SNe\footnote{The \(\Delta M_{15}\) value for SN 2016gkg is not available due to the lack of \(V\)-band photometry 15 days after the second peak.} with photometry in the \(V\)-band, as it provides the largest sample of objects for consistent analysis.

\begin{figure*}[ht!]
\centering
\includegraphics[width=1\textwidth]{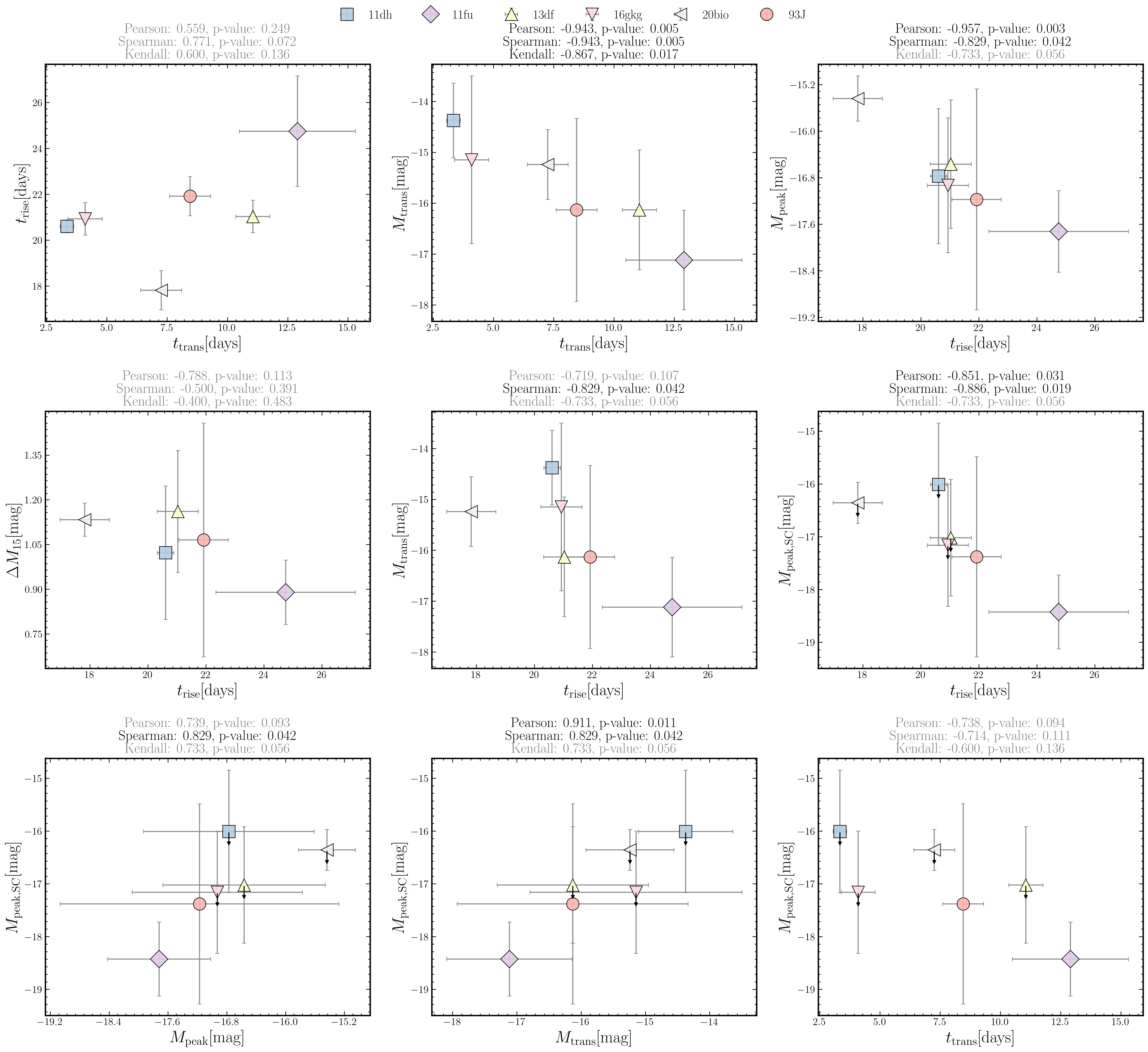} 
\caption{Correlation analysis between parameters measured during the shock cooling (SC) phase and those associated with the main peak of the light curve for the literature sample. Significant correlations are highlighted, including relationships within the SC phase, such as between the transitional absolute magnitude (\(M_{\text{trans}}\)) and the SC rise time (\(t_{\text{rise}}\)), and between the SC peak absolute magnitude (\(M_{\text{peak, SC}}\)) and \(M_{\text{trans}}\). Correlations within the main peak include \(M_{\text{peak}}\) and \(t_{\text{rise}}\). Additionally, correlations between the SC phase and the main peak are shown, including \(t_{\text{trans}}\) with \(t_{\text{rise}}\), \(M_{\text{peak, SC}}\) with \(t_{\text{rise}}\), and \(M_{\text{peak, SC}}\) with \(M_{\text{peak}}\). These relationships provide evidence linking the SC phase to the main peak properties, highlighting the continuity of physical processes across both phases. Non-significant correlations are also shown for completeness.
\label{fig:correlations_literature}}
\end{figure*}

We explored potential correlations between parameters measured during the shock cooling (SC) phase and those associated with the main peak of the light curve. Correlations were considered significant if at least one of the statistical tests (Pearson, Kendall, or Spearman) had a p-value below 0.05. In this case, all significant correlations also presented a correlation coefficient greater than $|0.82|$, providing strong evidence of the relations described. The detailed correlation coefficients and p-values for all parameter pairs across these tests are provided in each top panel in Figure~\ref{fig:correlations_literature}. Below, we describe the significant correlations observed within the SC phase, the main peak, and between these two phases.

Within the SC phase, a strong correlation was observed between the transitional absolute magnitude ($M_{\text{trans}}$) and the transitional time ($t_{\text{trans}}$), indicating that prolonged SC phases correspond to brighter transitions. While a trend between the peak SC absolute magnitude ($M_{\text{peak, SC}}$) and $M_{\text{trans}}$ is observed, it is not statistically significant (Pearson: $-0.74$, $p\text{-value}: 9.4 \times 10^{-2}$). This suggests that brighter SC peaks may tend to produce brighter transitions; however, a larger sample size, along with precise measurements instead of upper limits, is needed to confirm or reject this trend. For the main peak, we identified a correlation between the peak absolute magnitude ($M_{\text{peak}}$) and the rise time ($t_{\text{rise}}$), where brighter SNe tend to exhibit longer rise times.

Correlations between the SC phase and the main peak reveal further links between the early phases and the peak of the light curve. The peak SC absolute magnitude ($M_{\text{peak, SC}}$) was found to correlate with the rise time to the main peak ($t_{\text{rise}}$), showing that brighter SC peaks are associated with longer rise times. Additionally, $M_{\text{peak, SC}}$ correlates with the peak absolute magnitude of the main peak ($M_{\text{peak}}$), demonstrating that brighter SC peaks are linked to brighter main peaks. Lastly, while no significant correlation was found between the duration of the SC phase ($t_{\text{trans}}$) and the rise time to the main peak ($t_{\text{rise}}$), a trend is observed where longer SC phases tend to result in later main peaks.

While these significant correlations provide insight into the statistical relationships between SC parameters and the main peak, several correlations did not reach statistical significance with the current sample size. For instance, the correlation between $\Delta M_{15}$ and $t_{\text{rise}}$ showed moderate trends but lacked sufficient statistical strength to draw definitive conclusions. A larger sample is necessary to confirm or reject these correlations.

\subsubsection{Colors}
\label{colors_lit}

In addition to analyzing the light curve properties of the  SNe with SC from the literature, we studied their optical color evolution. Among the colors, the \( B - V \) color is the most commonly available, with six SC SNe having the necessary photometry to calculate \( B - V \). As previously mentioned in Section \ref{sec:cara_LCss}, the colors were corrected only for extinction in the Milky Way. 

To compare the \( B - V \) color of the six SC SNe, we defined a sample of SNE~IIb without the presence of the SC (non-SC SNe sample). The non-SC sample consists of seven SNe~IIb from the literature with early photometric observations, all at phases earlier than the average SC duration (8.8 days), in which visual inspection suggests the absence of SC in this sample, as no photometric points were consistent with an EE. Of these non-SC SNe, four have data earlier than two days, and three have data earlier than 6.3 days.

\begin{figure*}[ht!]
\centering
\includegraphics[width=1\textwidth]{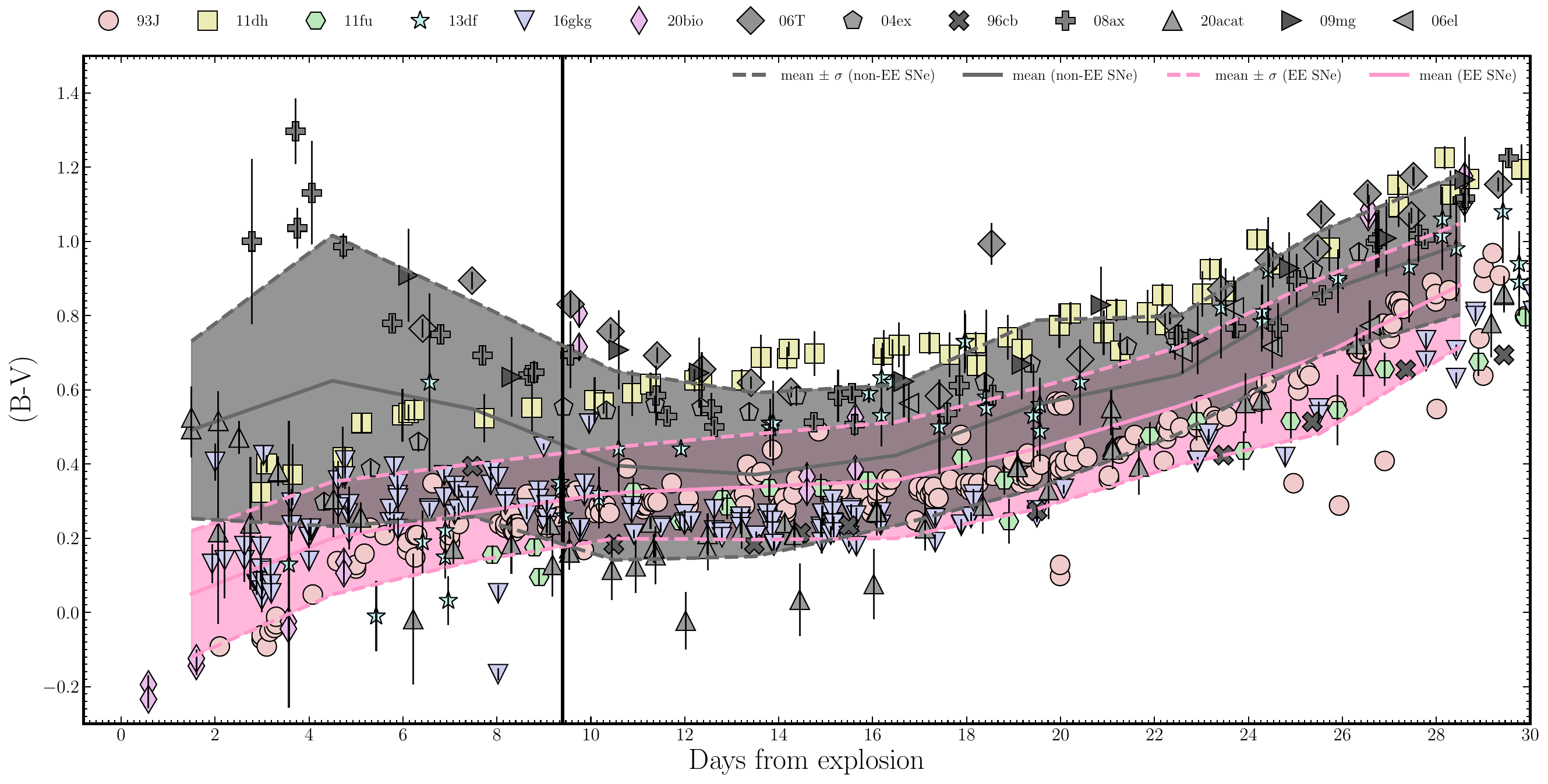} 
\caption{\( B-V \) color evolution for supernovae with (seven) and without (five) SC up to 30 days post-explosion. Individual data points for SNe without SC are gray, while pastel-colored markers represent those with SC. Solid lines show the centered moving averages, with pink representing supernovae with SC and gray for those without SC. Dashed lines indicate the standard deviation (\(\sigma\)) around the mean for each time interval, and the filled regions represent the mean \(\pm \sigma\). The vertical black line marks the average upper limit for the SC duration. Analogous to Figure~\ref{fig:colores_ZTF}, this figure emphasizes the differences in color evolution between the two groups, particularly in the early phases.
\label{fig:colors_literature}}
\end{figure*}

The \( B - V \) color evolution for SNe with and without SC is shown in Figure~\ref{fig:colors_literature}. We estimated the average color using the same methodology described in Section~\ref{color_evolution_EE}.

To analyze the \( B - V \) color evolution during and after the SC phase, we use the average duration of the SC phase (8.8 days) as a reference point. During the first three days after the explosion, SNe with SC exhibit significantly bluer colors (\( B - V = 0.11 \pm 0.14 \)) compared to those without SC (\( B - V = 0.49 \pm 0.24 \)), with the color difference being statistically significant\footnote{Statistically significant indicates that the differences exceed one standard deviation 1$\sigma$.}. After this period, the colors converge, overlapping within the standard deviation range. Beyond the SC phase (after 9 days), the differences between the two groups diminish further. At later phases, such as 18–21 days, the mean \( B - V \) color for SNe with SC is \( 0.44 \pm 0.16 \), compared to \( 0.56 \pm 0.23 \) for those without SC, indicating a convergence in color evolution. However, a larger dataset is necessary to confirm these trends due to the small sample size.
These results show that the \( B - V \) color differences between SNe with and without SC are consistent with the trends observed in the ATLAS EE sample, the implications of these findings, along with their relevance to the physical origin of EE in SNe~IIb, will be discussed in detail in Section~\ref{EE_origin}.

\begin{table*}[t]
\centering
\resizebox{\textwidth}{!}{%
\begin{tabular}{|c|c|c|c|c|c|c|}
\hline
SN          & z         & $t_{\text{exp}}$ & $E(B-V)^{\text{MW}}$ & $E(B-V)^{\text{host}}$ & SC Duration (days) & Distance (Mpc) \\ \hline
1993J       & 4.3$\times 10^{-4}$       & 49073.90 $\pm$ 0.60$^{(6)}$         & 8.0$\times 10^{-2}$ $\pm$ --$^{(3)}$  & 6.0$\times 10^{-2}$ $\pm$ 9.4$\times 10^{-2}$$^{(2)}$ & 8.19 $\pm$ 0.85 & 3.63 $\pm$ 0.34$^{(13)}$ \\ \hline
2011dh      & 1.5$\times 10^{-3}$  & 55712.60 $\pm$ 0.20$^{(1)}$         & 3.1$\times 10^{-2}$ $\pm$ 1.0$\times 10^{-3}$$^{(3)}$  & 7.0$\times 10^{-2}$ $\pm$ 9.4$\times 10^{-2}$$^{(2)}$ & 3.58 $\pm$ 0.28 & 7.10 $\pm$ 1.20$^{(14)}$ \\ \hline
2011fu      & 1.8$\times 10^{-2}$       & 55822.60 $\pm$ 1.70$^{(4)}$         & 6.5$\times 10^{-2}$ $\pm$ 8.0$\times 10^{-4}$$^{(3)}$ & 5.0$\times 10^{-2}$ $\pm$ 1.0$\times 10^{-1}$$^{(2)}$ & 12.35 $\pm$ 2.40 & 77.90 $\pm$ 5.50$^{(15)}$ \\ \hline
2013df      & 2.9$\times 10^{-3}$       & 56450.00 $\pm$ 0.50$^{(7)}$         & 1.7$\times 10^{-2}$ $\pm$ 1.0$\times 10^{-3}$$^{(3)}$   & 8.0$\times 10^{-2}$ $\pm$ 1.6$\times 10^{-2}$$^{(7)}$ & 10.34 $\pm$ 0.71 & 16.60 $\pm$ 0.40$^{(16)}$ \\ \hline
2016gkg     & 4.9$\times 10^{-3}$       & 57651.69 $\pm$ 0.50$^{(8)}$     & 1.7$\times 10^{-2}$ $\pm$ 2.0$\times 10^{-4}$$^{(3)}$ & 1.5$\times 10^{-1}$$^{(8)}$              & 3.89 $\pm$ 0.71 & 26.30 $\pm$ 4.84$^{(17)}$ \\ \hline
2017jgh     & 7.9$\times 10^{-2}$       & 58106.94 $\pm$ 0.13$^{(9)}$    & 2.0$\times 10^{-2}$ $\pm$ 7.0$\times 10^{-4}$$^{(10)}$ & --           & 10.66 $\pm$ 0.18 & 351.52 $\pm$ 0.09$^{(20)}$ \\ \hline
2018aalrxas & 5.8$\times 10^{-2}$  & 58227.37 $\pm$ 0.47$^{(5)}$     & 1.9$\times 10^{-2}$ $\pm$ --$^{(3)}$              & 0.00$^{(5)}$               & 9.88 $\pm$ 0.66 & 263.00 $\pm$ 1.33$^{(18)}$ \\ \hline
2020bio     & 8.5$\times 10^{-3}$       & 58877.27 $\pm$ 0.50$^{(12)}$         & 8.4$\times 10^{-3}$ $\pm$ 1.0$\times 10^{-4}$$^{(3)}$  & 7.0$\times 10^{-2}$ $\pm$ 3.8$\times 10^{-2}$$^{(12)}$           & 7.25 $\pm$ 0.85 & 29.90 $\pm$ 5.10$^{(19)}$ \\ \hline
2021zby     & 2.6$\times 10^{-2}$       & 59474.40 $\pm$ 0.10$^{(11)}$        & 2.1$\times 10^{-1}$ $\pm$ 9.3$\times 10^{-3}$$^{(3)}$  & --           & 12.47 $\pm$ 0.14 & 106.00 $\pm$ 0.09$^{(11)}$ \\ \hline
\end{tabular}%
}
\caption{Summary of SN observations. (1): \citet{2011ApJ...742L..18A}, (2): \citet{2023ApJ...955...71R}, (3): \citet{2011ApJ...737..103S}, (4): Central Bureau Electronic Telegrams (CBET), (5): \citet{2019ApJ...878L...5F}, (6): International Astronomical Union Circulars (IAUC), (7): \citet{2014MNRAS.445.1647M}, (8): \citet{2017ApJ...836L..12T}, (9): \citet{2021MNRAS.507.3125A}, (10): \citet{1998ApJ...500..525S}, (11): \citet{2023ApJ...943L..15W}, (12): \citet{2023ApJ...954...35P}, (13): \citet{1994AJ....107.1022R}, (14): \citet{2006MNRAS.372.1735T}, (15): \citet{2013MNRAS.431..308K}, (16): \citet{2014AJ....147...37V}, (17): \citet{2009AJ....138..323T}, (18): \citet{2019ApJ...878L...5F}, (19): \href{https://ned.ipac.caltech.edu/}{NED}, (20): Derived from recessional redshift.}
\label{tab:sn_observations}
\end{table*}

\subsubsection{Comparison of Early Excess and Shock Cooling Properties}
\label{EE_origin}

At this stage, we have identified the EE, characterized its photometric properties, and compared it with non-EE SNe. We now aim to explore its potential origin by comparing it with SNe exhibiting SC emission, as detailed in Section~\ref{lit_comp}. We focus on their durations, color evolution, and correlations among light curve parameters. 

In Section~\ref{duration_EE_}, we estimated an upper limit for the EE duration using the ATLAS sample, while Section~\ref{lit_comp_LCs} provided the SC emission duration from the literature sample. The methodologies differ mainly due to better explosion epoch determinations and higher observational densities in the literature sample. 

We conducted statistical tests on both samples to compare durations. The KS, AD, and U tests resulted in a KS statistic of 0.27 (p-value = 0.68), an AD statistic of -0.69 (significance = 0.25), and a U statistic of 81 (p-value = 0.9). None showed statistically significant p-values below 0.05, suggesting we cannot reject the null hypothesis that the durations of the EE and SC emissions come from the same distribution. Thus, the upper limit of the EE duration is consistent with the duration of the SC emission.

In addition to comparing EE and SC emission durations, we analyzed their color evolution trends. Both phenomena show similar timescales, with EE lasting a median of 8 days and SC having a mean duration of 8.8 days. During the EE phase, SNe exhibits bluer \( g - r \) colors compared to those without EE\footnote{Although the dispersion in the data prevents statistical significance, we observe a trend where EE SNe show bluer colors at phases earlier than five days compared to non-EE SNe. A larger dataset is necessary to confirm this trend.}. Similarly, during the SC phase, SNe with SC display significantly bluer \( B - V \) colors compared to those without SC. Notably, the \( B - V \) colors of the SC sample are generally bluer than the \( g - r \) colors of the EE sample, likely due to differences in the filter systems used. Despite this, the overall trend of blue color evolution in EE SNe is consistent with SC emission as defined in the literature.

We also examined whether the correlations observed in the ATLAS sample (\(\Delta M_{15}\)-\(t_{\text{rise}}\) and \(M_{\text{peak}}\)-\(t_{\text{rise}}\)) are also present in the SC SNe literature sample.  

In both samples, the general trends are consistent: SNe with slower post-peak declines (\(\Delta M_{15}\)) tend to have longer \(t_{\text{rise}}\), and brighter peak magnitudes (\(M_{\text{peak}}\)) are associated with later phases of \(t_{\text{rise}}\). In comparison, the ATLAS sample shows statistically significant correlations for both pairs under representative criteria (e.g., \(\rho_{\text{peak}} = 0.3\)), the literature sample similarly displays a strong negative correlation for \(t_{\text{rise}}-M_{\text{peak}}\) and a weaker but consistent trend for \(t_{\text{rise}}-\Delta M_{15}\).  

These similarities between the ATLAS sample and the literature suggest that the features we identify as early excess (EE) in ATLAS SNe could indicate the same physical processes driving SC as described in previous studies.

\subsection{Robustness of our Results}

In this section, we analyze the robustness of our results by removing potential false positives. We define a criterion for identifying false positives within the EE sample detected in Section~\ref{frecuency}. After excluding these false positives, we recalculate the observed frequency of SNe with EE in the ATLAS sample and estimate the associated uncertainty in Section~\ref{very_freq}. Subsequently, in Section~\ref{per_dist}, we examine whether the differences in the distributions of light curve parameters between EE and non-EE SNe, observed for the entire sample in Section~\ref{distributions}, persist. Also, Section~\ref{per_dist} assesses whether the correlations observed for SNe with EE remain consistent after removing the three false positives.

\subsubsection{Verification of Early Excess Detections}
\label{verifi_EE}

\begin{figure*}[ht!]
\centering
\includegraphics[width=1\textwidth]{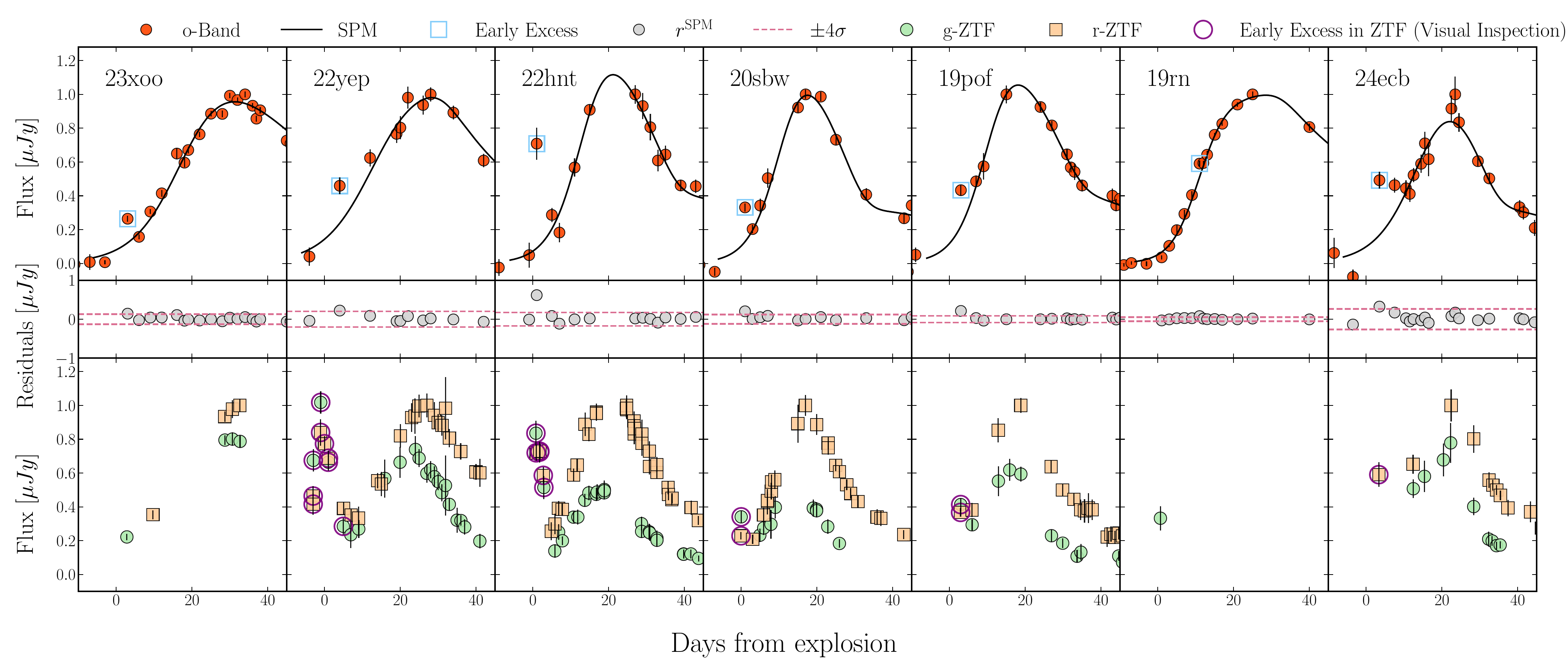} 
\caption{ZTF verification of potential false positives with ZTF photometry. The top panels show the ATLAS \(o\)-band photometry, with potential shock-cooling (SC) points enclosed by blue open squares. The middle panels illustrate the residuals produced by the SPM model. The bottom panels display the ZTF \(g\)- and \(r\)-band photometry, with points consistent with an early excess identified by visual inspection and enclosed in purple open circles.
\label{fig:ZTF_validation}}
\end{figure*}

This section assesses the validity of the 21 SNe identified with EE. We specifically examined cases where our methodology detected only a single photometric point that satisfied the criterion of having a residual, defined as the difference between the observed flux and the SPM, greater than three times the standard deviation of the residuals (\(3\sigma_r\)), considering it as an EE. Nine of the 21 SNe detected meet this condition in the ATLAS light curves, of which seven have additional \(g\)- and \(r\)-band photometry from ZTF. We visually inspected these seven ZTF light curves to verify whether photometric points consistent with the single EE point detected in ATLAS were present. As shown in Figure~\ref{fig:ZTF_validation}, five of these seven objects exhibit flux consistent with the EE detected in ATLAS. However, the remaining two SNe and the two SNe that only have a single point detected in ATLAS and lack ZTF photometry are considered potential false positives.

For these four potential false positives, we redefined the criterion for EE detection by adopting a more conservative threshold of eight times the standard deviation of the residuals (\(8\sigma_r\)) instead of the original \(3\sigma_r\) defined in Section~\ref{identi}. This threshold was chosen for its highly conservative nature, as such extreme values have a negligible probability (\(< 10^{-15}\)) of arising from statistical noise under a normal distribution, significantly reducing the risk of false positives. As a result, only SN 2017ixz satisfies this stricter criterion, while the remaining three objects (SN 2023xoo, SN 2019rn, and SN 2024jcf) were removed from our sample.

\subsubsection{Verification of the Frequency}
\label{very_freq}

\begin{figure*}[ht!]
\centering
\includegraphics[width=1\textwidth]{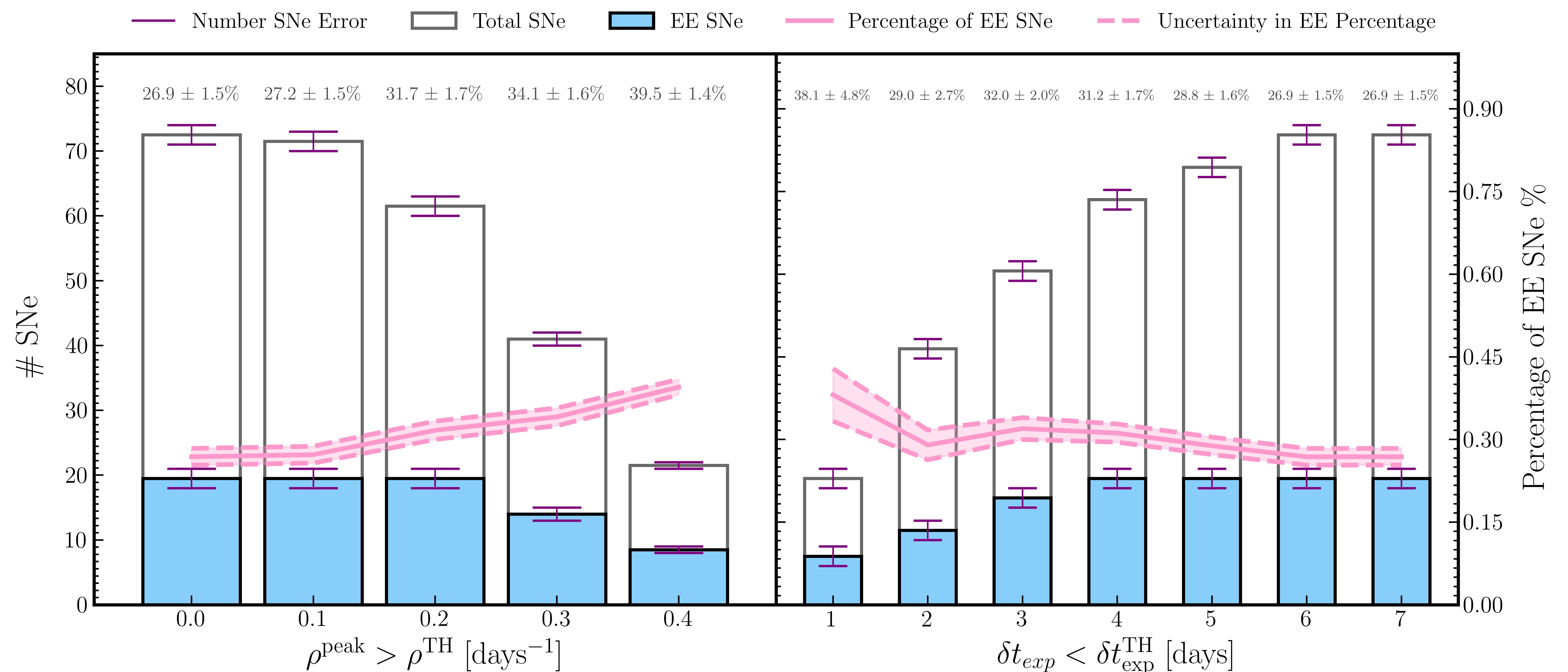} %
\caption{Recalculated SNe counts and EE percentages after rejecting false positives. Analogous to Figure~\ref{fig:frequency_EE}, this analysis recalculates the frequency of SNe with EE after rejecting three potential false positives. Number of Type IIb supernovae satisfying specific criteria for observational point density ($\rho_{\mathrm{peak}} > \rho_{\mathrm{TH}}$, left panel) and explosion epoch error ($\delta t_{\mathrm{exp}} < \delta t_{\mathrm{TH}}$, right panel) are shown as unfilled bars. Blue-filled bars represent EE SNe. The dashed pink line indicates the revised percentage of supernovae with EE for each criterion, estimated as the mean between the initial and recalculated frequencies after removing false positives. Numerical percentages, along with their associated uncertainties, appear above the bars.
\label{fig:frequency_errors}}
\end{figure*}

After rejecting the three potential false positives identified in the previous section, we recalculated the frequency of SNe with EE. This new estimation of the frequency, along with the estimation without removing any objects, as calculated in Section~\ref{frecuency}, was used to reinterpret the frequency of EE SNe and to associate uncertainty. We estimated the number of SNe with and without EE and the percentage of EE SNe, as the average between the estimations without removing any objects (74 SNe) and after removing the three objects that, based on our analysis, are false positives. The error in each of these estimations was calculated as the difference between the average and the estimation without removing any objects. The new frequency and error estimations are illustrated in Figure~\ref{fig:frequency_errors}.

We analyzed the changes in EE frequency across the different density criteria and explosion epoch error thresholds before and after removing the false positives to quantify the impact of removing the three potential false positives.

Focusing on the density criterion of $0.3  \text{ days}^{-1}$, which we selected as the representative case, the frequency of SNe with EE decreases from \(35.71\%\) when considering all objects to \(32.50\%\) after removing the false positives. This reduction highlights the influence of potential false positives on the estimation of EE frequency.

We examined density criteria ranging from \(0 \, \mathrm{ days}^{-1}\) to \(0.4 \, \mathrm{ days}^{-1}\) and found that the frequency of SNe with EE varies from \(25.35\%\) to \(38.10\%\) after removing false positives, compared to \(28.38\%\) to \(40.91\%\) when including all objects. The associated errors in this estimation range from \(1.42\%\) to \(1.67\%\). For explosion epoch error criteria, the frequency decreases from \(28.38\%\) to \(42.86\%\) (including all objects) to \(25.35\%\) to \(33.33\%\) after removing false positives, with errors between \(1.5\%\) and \(4.8\%\). Despite the variations, the results remain stable, demonstrating the robustness of our analysis across both criteria.

\subsubsection{Validation of Distributions and Correlations}
\label{per_dist}

Another essential aspect after removing these three SNe is to determine whether the differences in the distributions of light curve parameters between SNe with and without EE persist. Additionally, we verify if the correlations observed for the entire sample, such as \( \Delta M_{15} \)-\( t_{\text{rise}} \) and \( M_{\text{peak}} \)-\( t_{\text{rise}} \), remain consistent.

We analyzed the differences in light curve parameter distributions for our subsample (\(\rho_{\text{peak}} = 0.3 \text{ days}^{-1}\)). The KS statistic for \(\Delta M_{15}\) is \(0.52\) (p-value \(0.01\)), indicating significant differences, which become more apparent after removing false positives (\(0.52\) vs. \(0.34\)). For \(t_{\text{rise}}\), after removing false positives, the KS statistic is \(0.44\) (p-value \(0.04\)), consistent with the entire sample's findings (\(0.47\), p-value \(0.02\)). Lastly, the distributions of \(M_{\text{peak}}\) show no significant differences between the EE and non-EE groups. The differences persist across various density criteria (\(0.3 \leq \rho_{\text{peak}} \leq 0.4\ \text{ days}^{-1}\)) and explosion epoch error thresholds (\(1 \leq \delta t_{\text{exp}} \leq 4\)), aligning with trends seen in the entire sample.

We examined correlations among light curve parameters, finding significant relations in the EE group: \(\Delta M_{15}\) and \(t_{\text{rise}}\) (\(r = -0.61\), p-value \(0.03\)) and \(M_{\text{peak}}\) and \(t_{\text{rise}}\) (\(r = -0.61\), p-value \(0.03\)). These correlations remained robust across density thresholds of \(\rho_{\text{peak}} = 0.3\) and significant for \(0.3 \leq \rho_{\text{peak}} \leq 0.4\) and \(1 \leq \delta t_{\text{exp}} \leq 4\). In contrast, the non-EE group had weak and statistically insignificant correlations, similar to those before removing false positives.

\section{Discussion}

\subsection{Light Curve Parameter Distributions: Qualitative Physical Interpretation}

In this section, we discuss the cumulative distributions (Section \ref{distributions}) of $\Delta M_{15}$, $t_{\text{rise}}$, and $M_{\text{peak, abs}}$ for EE and non-EE SNe, exploring the physical interpretation of explosion parameters inferred from their correlations with these light curve properties. Given that these parameters exhibit similar trends across individual bands and bolometric light curves \citep{2023ApJ...955...71R}, we base our qualitative interpretation on their measurements in the $o$-band.

Explosion properties such as ejecta mass ($M_{\text{ej}}$), $^{56}\text{Ni}$ mass, and explosion energy ($E$) have been inferred through analytical modeling of bolometric light curves of SESNe, revealing correlations between these physical parameters and observable light curve properties, including rise and decline rates, characteristic timescales, and peak luminosities \citep{2011ApJ...741...97D,2016MNRAS.457..328L,2016MNRAS.458.2973P,2019MNRAS.485.1559P,2023ApJ...955...71R}. A well-established correlation exists between $^{56}\text{Ni}$ mass and peak absolute magnitude for SNe~IIb \citep{2016MNRAS.457..328L, 2016MNRAS.458.2973P}, aligning with expectations from the ``Arnett-rule'' model for SNe~Ia \citep{1982ApJ...253..785A}, where the decay of $^{56}\text{Ni}$ powers the light curve ($M_{\text{Ni}} \propto L_p$). However, this model is inaccurate for SESNe, as their diffusion time ($t_{\text{diff}}$) is comparable to the dynamical timescale ($t_{\text{dyn}}$), leading to adiabatic losses during ejecta expansion. Consequently, the Arnett rule overestimates $^{56}\text{Ni}$ mass by a factor of $\sim2$ compared to values derived from luminosity measurements in the radioactive tail phase \citep[e.g.,][]{2020A&A...641A.177M}. At phases later than $\sim60$ days post-explosion, the ejecta becomes optically thin ($t_{\text{diff}} \ll t_{\text{dyn}}$), and the luminosity directly traces the deposition function of $^{56}\text{Ni}$ decay. Using this method, \citet{2023ApJ...955...71R} also reported a strong correlation between $^{56}\text{Ni}$ mass and peak absolute magnitude for SNe~IIb.

The similarity in the cumulative distributions of peak absolute magnitude for EE and non-EE SNe suggests that their $^{56}\text{Ni}$ mass distributions are comparable. Since $^{56}\text{Ni}$ production is linked to explosion properties and progenitor core structures \citep{2019MNRAS.483.3607S}, this implies that EE and non-EE SNe share similar progenitor characteristics. Moreover, explosion models predict that $^{56}\text{Ni}$ masses increase with ZAMS mass \citep[see figure 14 in][]{2024ApJ...964L..16B}, further reinforcing the idea that both samples originate from progenitors with comparable initial masses.

\citet{2023ApJ...955...71R} found a correlation between $M_{\text{ej}}$, time to peak $t_{\text{peak}}$, and post-peak decline $\Delta M_{15}$, based on an analytical framework derived from Arnett’s model \citep{2015MNRAS.450.1295W}. Their results suggest that SNe with more massive ejecta exhibit slower declines and later peaks, as more massive ejecta result in longer diffusion times. Given the correlation between $M_{\text{ej}}$ and $\Delta M_{15}$ ($r_p = 0.5$, $p$-value $<0.03$) and the statistically higher values of $\Delta M_{15}$ in EE SNe compared to non-EE SNe, one might infer that EE SNe have lower ejecta masses. However, considering the correlation between $M_{\text{ej}}$ and $t_{\text{peak}}$ ($r_p = 0.65$, $p$-value $<0.03$), and the higher cumulative distribution of $t_{\text{peak}}$ for EE SNe within the range $\sim15-21$ days, part of the EE SNe sample may instead have more massive ejecta than non-EE SNe.

The apparent inconsistency in ejecta mass interpretation—where $\Delta M_{15}$ suggests lower masses while $t_{\text{peak}}$ suggests higher masses—stems from the limitations of Arnett’s model in accurately reproducing the light curves of SNe~IIb. Additionally, as discussed in Section~\ref{pec_miss}, this discrepancy is resolved after removing peculiar objects or potential misclassifications.

In summary, the analysis of $M_{\text{peak, abs}}$ distributions suggests that EE and non-EE SNe likely originate from progenitors with similar core structures and initial masses, given the correlation between $M_{\text{peak, abs}}$ and $^{56}\text{Ni}$ production. In contrast, differences in $\Delta M_{15}$ and $t_{\text{rise}}$ distributions indicate variations in ejecta properties, which may be linked to factors such as explosion energy, envelope mass, or progenitor structure. The inconsistencies in ejecta mass interpretation highlight the limitations of Arnett’s model. More detailed modeling is needed to understand the role of extended envelopes and explosion mechanisms in fully shaping the light curves of EE and non-EE SNe.

\subsection{Implications for the progenitor systems}
\label{implications}

Motivated by the similarities in photometric properties between EE and SC discussed in Section \ref{EE_origin}, as well as the differences in ejecta properties between EE and non-EE SNe inferred in the previous Section from the \(\Delta M_{15}\) and \(t_{\text{rise}}\) distributions, we analyze the ejecta conditions suggested by analytical and numerical models necessary to produce SC. These models indicate that an extended, H-rich envelope is required to generate double-peaked SNe~IIb light curves. We further explore the mechanisms that could lead to variations in envelope properties, specifically extended envelopes for EE SNe and non-extended envelopes for non-EE SNe, assuming that both EE and non-EE SNe may originate from similar progenitor types.

\subsubsection{Conditions for Producing Double-Peaked Light Curves: Scenarios for Producing Extended Envelopes}
\label{sce_e_e}

Double-peaked light curves can arise from various mechanisms, including different $^{56}$Ni distributions, interaction with circumstellar material, or central engine activity. However, while this mechanism remains plausible, all ten Type IIb SNe with observed double peaks in the literature have been interpreted as originating from shock cooling emission in extended, H-rich envelopes \citep[e.g.,][]{2015ApJ...808L..51P, 2017ApJ...838..130S, 2012ApJ...757...31B, 2018A&A...612A..61D}. Additionally, both the color evolution and the duration of the early excess in our sample are consistent with the properties observed in the shock cooling (SC) literature sample (Section \ref{EE_origin}), further supporting the interpretation that EE SNe arise from extended, H-rich envelopes. Such findings naturally raise the question: How do the progenitors of EE SNe achieve these properties if they are similar to non-EE SNe?

Considering that the conditions required to produce SC are associated with envelope properties, we hypothesize scenarios that could generate different envelopes for similar progenitors. The most plausible scenario involves progenitors in binary systems. Observational and numerical evidence strongly supports the hypothesis that the probable progenitors of SNe~IIb are binary systems \citep[][]{2011A&A...528A.131C, 2017ApJ...840...10Y, 2019ApJ...885..130S, 2020ApJ...903...70S}. . 

The \cite{2024A&A...685A.169D} model is particularly illustrative of the interpretations of our results because it assumes the same progenitor, with fixed \(^{56}\text{Ni}\) mass and explosion energy, where variations in the initial binary period lead to different envelope properties. However, these models cannot explain why EE SNe exhibits fast post-peak declines while reaching the peak at later phases than non-EE SNe. We calculated \(\Delta M_{15}\), rise time, and peak absolute magnitude for these models, investigating potential trends with increased hydrogen mass but did not identify any significant correlation or dependence on hydrogen mass. 

In summary, we hypothesize that the 21 EE SNe detected in our sample may share progenitors with non-EE SNe but exhibit a core-halo structure with an extended envelope. This hypothesis is interpretated by the observed differences in \(\Delta M_{15}\), \(t_{\text{rise}}\), and \(M_{\text{peak, abs}}\). We propose that binary systems explain these differences in envelope properties.

Alternative explanations appear incompatible, such as circumstellar material (CSM) interactions. Studies incorporating CSM in normal Type II SNe interactions show that it significantly influences early phases, resulting in brighter SNe, faster post-peak declines, and earlier peak times \citep{2017ApJ...838...28M}. Additionally, SN IIb 2013cu, which reached its peak at $\sim$ 12 days after the explosion and exhibited flash-ionized features at early phases ($\sim$ 15 hours after the explosion), indicating a clear CSM interaction displayed a light curve with only a single peak \citep{2014Natur.509..471G}.

\subsection{Correlation Analysis of Light Curve Parameters}

In this section, we qualitatively analyze the correlations observed in our sample, focusing on \(\Delta M_{15}\)-\(t_{\text{rise}}\) and \(M_{\text{peak, abs}}\)-\(t_{\text{rise}}\), and hypothesize potential physical justifications. Additionally, we examine correlations between the SC phase and the main peak based on a literature sample (\ref{lit_comp_LCs}).

\citet{2016MNRAS.458.1618D} identified a correlation between \(\Delta M_{15}\) and \(t_{\text{rise}}\), finding a strong relation in their simulations of SESNe (including SNe IIb) that resulted from the terminal explosions of mass donors in close-binary systems. This correlation arises because more massive ejecta lead to longer diffusion timescales, producing slower post-peak declines and later peak times.
Similarly, \citet{2023ApJ...955...71R} observed this correlation for SNe Ib and Ic, where faster post-peak decline light curves tend to reach the peak at earlier phases. However, they reported no significant correlation for SNe~IIb. Our analysis finds no such correlation for non-EE SNe, consistent with the findings for SNe IIb by \citet{2023ApJ...955...71R}. However, we observe a significant correlation for EE SNe (\(r_p = -0.75, \, p\text{-value} = 0.001\)), where fast post-peak decline light curves also tend to reach the peak at earlier phases. This is consistent with the idea that more massive ejecta lead to longer diffusion timescales, resulting in slower post-peak declines and later peak times.

A non-significant correlation (\(r_k = -0.71\), \(p\)-value = \(1.44 \times 10^{-2}\)) is observed between \(t_{\text{rise}}\) and \(M_{\text{peak, abs}}\) for EE SNe, indicating that brighter objects tend to reach the peak at earlier phases. This correlation is absent for non-EE SNe, suggesting intrinsic differences in the envelope properties between the two populations. In contrast, no correlation is found between \(\Delta M_{15}\) and \(M_{\text{peak, abs}}\) for either EE or non-EE SNe. Previous studies, such as \citet{2011ApJ...741...97D}, \citet{2016MNRAS.458.2973P}, and \citet{2016MNRAS.458.1618D}, also reported a lack of such correlations for SESNe, including SNe~IIb, attributing this to the non-uniform relationship between explosion mechanisms and ejecta properties. \citet{2016MNRAS.458.1618D} further explained this by highlighting the large scatter in \(\Delta M_{15}\) and \(M_{\text{peak}}\) values, as the peak magnitude is primarily controlled by the \(^{56}\text{Ni}\) mass, while the post-maximum decline depends on a combination of \(^{56}\text{Ni}\) mass, ejecta mass, and kinetic energy.

In the literature sample, correlations observed between the SC phase and the main peak suggest a link between the energy source driving the SC and the subsequent light curve evolution. The SC peak absolute magnitude (\(M_{\text{peak, SC}}\)) shows significant correlations with both the rise time to the main peak (\(t_{\text{rise}}\)) and the main peak absolute magnitude (\(M_{\text{peak}}\)), highlighting that brighter SC peaks are associated with brighter and slower-evolving main peaks. These findings suggest a shared underlying energy reservoir or mechanism influencing both phases. In addition, a trend between the SC duration ($t_{\text{trans}}$) and $t_{\text{rise}}$ shows that SNe with longer SC phases tends to take more time to reach their main peak; however, this correlation is not statistically significant ($p$-value $>$ 0.05), a larger sample is required to confirm or reject this potential relationship. Given the small sample six of SNe, further data is essential to validate these correlations robustly.

Assuming the previously described correlations exist, no clear hypothesis currently explains this connection. Analytical models suggest that more massive H-rich envelopes produce longer SC phases, while more extended envelopes result in brighter SC peaks. Furthermore, \citet{Park_2024} demonstrated that the brightness of the first peak can be reduced by more than a factor of three due to Thomson scattering. If an extended H-rich envelope primarily powers SC, the observed correlations between SC timescales, brightness, and the main peak could indicate that H recombination influences the phases of the main peak. Given the limited sample size and the lack of a comprehensive explanation for these correlations, hydrodynamical studies exploring these relationships under specific physical conditions are necessary.

\subsection{Peculiar Objects or Misclassifications}
\label{pec_miss}

Motivated by understanding the presence (EE SNe) and absence (non-EE SNe) of the predicted correlation between \(\Delta M_{15}\)-\(t_{\text{rise}}\) proposed by \citet{2016MNRAS.458.1618D}, which attributes the relation to more massive ejecta creating longer diffusion timescales than less massive ejecta—resulting in slower post-peak declines and longer rise times—we identified objects that deviate from this relation and break the correlation. For these objects, we propose two potential explanations: (1) ``peculiar"\footnote{Objects that exhibit photometric properties deviating from the typical behavior of SNe~IIb.}, or (2) misclassified SNe.

We identify SN~2021iiu, SN~2020tjd, SN~2022rls, and SN~2022fzb, in the non-EE SNe as objects characterized by short rise times (\(t_{\text{rise}} \lesssim 16 \, \text{days}\)) and slow post-peak declines (\(\Delta M_{15} \lesssim 0.18\)). If these objects are indeed SNe IIb, the progenitor or explosion properties responsible for producing slow post-peak declines while reaching the peak in less than 16 days is unclear. However, these objects could be misclassified as SNe IIb because they only have classification spectra and lack additional spectroscopic confirmation, in which their light curves exhibit atypical behavior for SNe IIb, supporting the possibility of misclassification.

Removing these peculiar or misclassified objects (\(t_{\text{rise}} \lesssim 16 \, \text{days}\) and \(\Delta M_{15} \lesssim 0.18\)) from the non-EE sample and recalculating the distributions of \(\Delta M_{15}\), \(t_{\text{rise}}\), and \(M_{\text{peak}}\), we find results that differ from those presented in Section~\ref{distributions}. Specifically, we do not identify any statistically significant differences between EE and non-EE SNe for any of the analyzed light curve parameters (\(\Delta M_{15}\), \(t_{\text{rise}}\), and \(M_{\text{peak}}\)), as the KS tests yield p-values above the \(0.05\) threshold. For \(\Delta M_{15}\), the KS test gives a value of \(3.22 \times 10^{-1}\) (\(2.35 \times 10^{-1}\)), in which p-value are reported in brackets. For \(t_{\text{rise}}\), the KS test results in \(3.45 \times 10^{-1}\) (\(1.72 \times 10^{-1}\)). Finally, for \(M_{\text{peak}}\), the KS test gives \(2.06 \times 10^{-1}\) (\(7.55 \times 10^{-1}\)). These results suggest that the differences in light curve parameters between EE and non-EE SNe are not statistically significant after removing peculiar or misclassified objects. 

We find that the interpretation of the results changes after removing peculiar or misclassified objects. Our analysis now suggests that EE and non-EE SNe likely share similar progenitors and ejecta properties, as we do not observe any statistically significant differences in \(\Delta M_{15}\), \(t_{\text{rise}}\), or \(M_{\text{peak}}\). These findings imply that the ejecta mass for both groups is comparable.

After removing peculiar objects from the non-EE sample, the correlation between \(\Delta M_{15}\) and \(t_{\text{rise}}\) shows the expected trend, with a Kendall correlation coefficient of \(-0.517\) (p-value \(1.15 \times 10^{-2}\)), where slow post-peak decliners reach the peak at later phases than fast post-peak decliners, as predicted by \citet{2016MNRAS.458.1618D}. For the correlations involving \(\Delta M_{15}\)-\(M_{\text{peak}}\) and \(t_{\text{rise}}\)-\(M_{\text{peak}}\), the results remain similar for EE and non-EE SNe. Specifically, we do not find a significant correlation in the first relation. In the second relation, we observe a slight tendency for brighter objects to reach the peak at later phases, with Pearson correlation coefficients of \(-0.531\) (p-value \(4.16 \times 10^{-2}\)) for EE SNe and \(-0.628\) (p-value \(1.34 \times 10^{-3}\)) for non-EE SNe.

Conclusions after removing peculiar objects regarding the change in the interpretation of the results compared to Section~\ref{distributions} are, however, currently not robust. Although these objects exhibit atypical photometric behavior, it is possible that they were indeed SNe~IIb but lacked the necessary spectroscopic confirmation to verify their classification.

\section{Conclusions}

In this work, we have developed a methodology to identify double-peaked SNe~IIb light curves, which we applied to the ATLAS survey, detecting 21 Early Excess (EE) SNe. For the first time, we quantified the frequency of EE SNe under specific sampling conditions of the light curves. Specifically, when the explosion epoch errors are less than one day and the average time between observations is three days, the occurrence rate is approximately \( \sim 39\% \).

The detected 21 EE SNe exhibit similar durations and color trends compared to SNe with Shock Cooling (SC) emission from the literature. Furthermore, the ATLAS EE SNe display distinct light curve properties compared to non-EE SNe, including differences in \(\Delta M_{15}\) and \(t_{\text{rise}}\), but similarities in \(M_{\text{peak, abs}}\). These findings suggest that, in addition to differences in the progenitors' envelopes, variations in the ejecta properties—such as ejecta mass and hydrogen richness—may also play a role. Spectral analyses, combined with hydrodynamical simulations, could provide further insights into these differences.

We propose that these variations arise in binary systems, where differences in the initial binary period for a given ZAMS progenitor can lead to diversity in the envelope's mass and extension. This hypothesis aligns with the observed photometric trends and the core-halo structure required to produce the EE. This structure, characterized by an extended envelope, may explain the presence of the EE in these SNe.

Our results provide constraints on the fraction of massive stars that evolve in such a way—whether through single or binary pathways—to produce the density profiles required for EE. Any population synthesis model aiming to reproduce the observed demographics of massive stars must account for the occurrence rate of EE SNe and the conditions necessary for their progenitor evolution, making EE SNe an important probe for testing and refining models of stellar evolution and binary interactions. In addition, this study highlights the need for further investigation through the hydrodynamical modeling of light curves. Such models could validate the observed correlations and provide precise estimates of progenitor properties and explosion parameters. Additionally, our methodology can be extended to other surveys, such as ZTF, to increase the sample size and enhance statistical significance.

\begin{acknowledgements}
We thank Luc Dessart for their insightful comments and suggestions, which helped to improve the paper.\\
BA acknowledges support from National Agency for Research and Development (ANID) grants ANID-PFCHA/Doctorado Nacional/21221964.\\
MR acknowledges support from National Agency for Research and Development (ANID) grants ANID-PFCHA/Doctorado Nacional/2020-21202606.\\
RD acknowledges funds by ANID grant FONDECYT Postdoctorado N\textdegree 3220449.\\
MS acknowledges that this research was funded in whole or in part by National Science Centre, Poland, through the PRELUDIUM grant 2024/53/N/ST9/00350.\\
FF acknowledges support from Centro de Modelamiento Matem\'atico (CMM) BASAL fund FB210005 for center of excellence from ANID-Chile.\\
SJS acknowledges funding from STFC Grants ST/Y001605/1, ST/X006506/1, ST/T000198/1, a Royal Society Research Professorship, and the Hintze Charitable Foundation.\\
This work was funded by ANID, Millennium Science Initiative, ICN12\_009.
\end{acknowledgements}

%
\bibliographystyle{aa} 
\bibliography{references} 
%

\begin{appendix} 

\appendix
\section{ATClean Methodology and Parameters}
\label{sec:atclean_appendix}

In this appendix, we describe the methodology and parameters used in ATClean for data cleaning.

\begin{enumerate}
    \item \textbf{Compute Eight Control Light Curves (CLCs):} 
    For each SN measurement, ATClean computes eight CLCs, which are forced photometry measurements within a 17" circular aperture around the SN.
    
    \item \textbf{Calculate Median Flux Uncertainty and Typical Error:} 
    For each CLC, ATClean calculates:
    \begin{itemize}
        \item The median flux uncertainty ($\delta\overline{f}$).
        \item The typical error, defined as the flux standard deviation ($\sigma_{f}$) after applying 3-$\sigma$ clipping.
    \end{itemize}
    
    \item \textbf{Analyze Extra Noise Contributions:} 
    Compare the median flux uncertainty with the flux standard deviation:
    \begin{itemize}
        \item If $\sigma_{f} \approx \overline{\delta_{f}}$, no additional noise is considered.
        \item If $\sigma_{f} > \overline{\delta_{f}}$, extra noise is estimated using:
        \[
        \sigma_{\text{extra}}^{2} = \text{median}(\sigma_{f}^{2} - \overline{\delta f ^{2}}).
        \]
        \item If the new flux standard deviation $\sigma_{f}^{\text{new}}$ exceeds the previous value by more than 10\% ($\sigma_{f}^{\text{new}} = (\sigma_{f}^{2} + \sigma_{\text{extra}}^{2})^{1/2}$), the extra noise is incorporated into the SN flux uncertainties:
        \[
        \delta f^{2} |^{\text{new}} = \delta f^{2} + \sigma_{\text{extra}}^{2}.
        \]
    \end{itemize}
    
    \item \textbf{Evaluate the Reliability of SN Photometry:} 
    The expected sky flux ratio for CLCs is $f/\delta f \approx 1$. Significant deviations may indicate observational issues.
    \begin{itemize}
        \item Calculate the $\chi^2_{\text{CLC}}$ based on a constant fit to the CLC flux values.
        \item Reject SN measurements if:
        \begin{itemize}
            \item $\chi^2_{\text{CLC}} > 2.5$,
            \item More than four measurements are clipped,
            \item Fewer than two measurements are averaged.
        \end{itemize}
    \end{itemize}
    
    \item \textbf{Rejection of Measurements Based on Flux Uncertainty:} 
    SN measurements are rejected if:
    \begin{itemize}
        \item $\chi^2_{\text{PSF}} > 5$ or
        \item Flux uncertainty $\delta_{f} > 160\,\mu Jy$ (threshold for bright stars near CCD saturation).
    \end{itemize}
    
    \item \textbf{Apply Corrections for Flux Discontinuities:} 
    ATClean applies corrections for periodic updates to ATLAS’s reference templates in the difference images.
    
    \item \textbf{Average the Four Exposures per Epoch:} 
    After applying corrections, ATClean averages the four exposures for each SN epoch, discarding measurements with:
    \begin{itemize}
        \item High $\chi^2_{\text{PSF}}$,
        \item Significant flux uncertainties, or
        \item Problematic data flagged by clipping.
    \end{itemize}
    
    \item \textbf{Final SN Epoch Quality Check:} 
    Calculate the $\chi^2_{\text{SN}}$ for a constant fit after 3-$\sigma$ clipping. Reject SN epochs if:
    \begin{itemize}
        \item $\chi^2_{\text{SN}} > 4$,
        \item Fewer than two measurements, or
        \item More than one clipped measurement.
    \end{itemize}
\end{enumerate}

\section{Explosion Epoch Determination}
\label{app:explosion_epoch}

In this appendix, we detail the methodology used to determine the explosion epoch $t_{\text{exp}}$, including the estimation of the last non-detection and discovery epochs.

\begin{enumerate}
    \item \textbf{Pre-SN Flux Determination:} 
    \begin{itemize}
        \item The pre-SN flux $\mathrm{f}_{\text{pre-SN}}$ is defined as the median flux of all photometric points obtained before $t_{\text{exp}}^{\text{TNS}} - 10$ days.
        \item The median $\overline{\mathrm{f}}_{\text{pre-SN}}$ and standard deviation $\sigma_{f_\text{pre-SN}}$ are computed using the median absolute deviation (MAD), which is robust against outliers:
        \[
        \sigma_{f_\text{pre-SN}} \approx 1.487 \cdot \text{MAD}.
        \]
        \item Figure~\ref{fig:ejemplo_de_ATclean} (c) illustrates an example of the pre-SN flux determination.
    \end{itemize}
    
    \item \textbf{Discovery Epoch Estimation:} 
    \begin{itemize}
        \item The discovery time $t_{\text{d}}$ is defined as the first photometric point satisfying:
        \[
        \mathrm{f}_{i} - \delta \mathrm{f}_{i} - \overline{\mathrm{f}}_{\text{pre-SN}} > 3 \times \sigma_{f_\text{pre-SN}},
        \]
        where $\delta \mathrm{f}_{i}$ is the flux uncertainty.
        \item The following two consecutive epochs must also satisfy the above condition to confirm detection.
    \end{itemize}

    \item \textbf{Last Non-Detection Epoch Estimation:} 
    \begin{itemize}
        \item The last non-detection epoch $t_{\text{lnd}}$ is the latest photometric point before $t_{\text{d}}$ that satisfies:
        \[
        \left|\mathrm{f}_{i} - \delta\mathrm{f}_{i} - \overline{\mathrm{f}}_{\text{pre-SN}}\right| < 1.5 \times \sigma_{f_\text{pre-SN}}.
        \]
        \item The choice of $1.5\sigma$ ensures that flux variations remain consistent with the pre-SN flux within reasonable statistical limits.
    \end{itemize}

    \item \textbf{Explosion Epoch and Uncertainty Calculation:} 
    \begin{itemize}
        \item The explosion epoch is estimated as the midpoint between the last non-detection and discovery:
        \[
        t_{\text{exp}} = \frac{t_{\text{lnd}} + t_{\text{d}}}{2}.
        \]
        \item The uncertainty in the explosion epoch is given by:
        \[
        \delta t_{\text{exp}} = \frac{t_{\text{d}} - t_{\text{lnd}}}{2}.
        \]
        \item This methodology follows \citet{2015A&A...574A..60T}.
    \end{itemize}
    
    \item \textbf{Example of Explosion Epoch Estimation:} 
    \begin{itemize}
        \item Figure~\ref{fig:ejemplo_de_ATclean} (d) illustrates an example of the determination of $t_{\text{lnd}}$ and $t_{\text{d}}$ in a real dataset.
    \end{itemize}
\end{enumerate}

\section{Posterior Distributions from the MCMC Fit for SN~2022jpx}

\begin{figure*}[ht]
\centering
\includegraphics[width=0.88\textwidth]{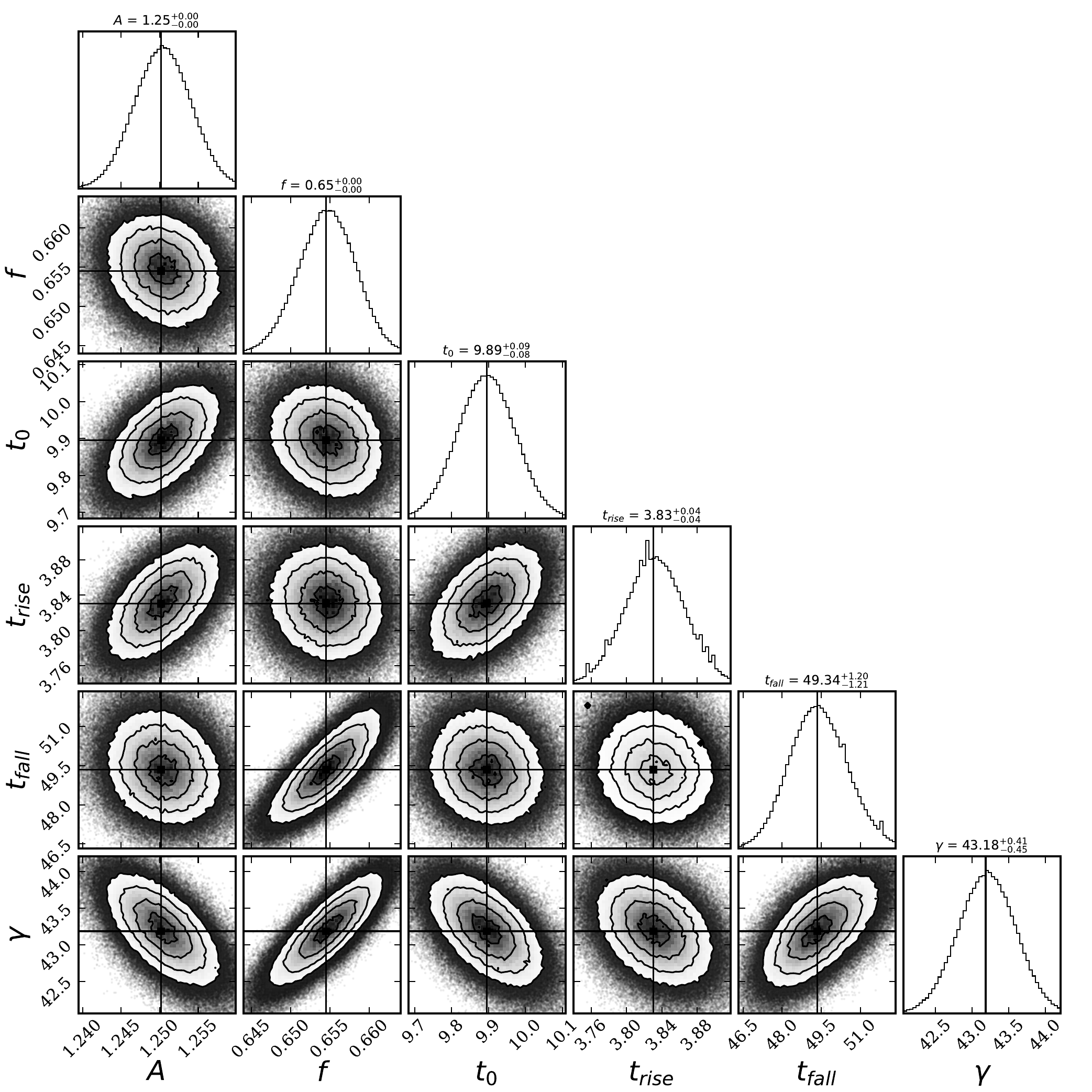} 
\caption{Posterior distributions of the six SPM parameters (\(A\), \(f\), \(t_0\), \(t_{\text{rise}}\), \(t_{\text{fall}}\), \(\gamma\)) for SN~2022jpx. The distributions are based on MCMC sampling, with the median (50th percentile) indicated by vertical dashed lines, and shaded regions representing the 16th and 84th percentiles. The best-fit values and their associated errors are reported at the top of each panel.
\label{fig:MCMC}}
\end{figure*}

Figure~\ref{fig:MCMC} presents the posterior distributions of the six SPM parameters (\(A\), \(f\), \(t_0\), \(t_{\text{rise}}\), \(t_{\text{fall}}\), \(\gamma\)) for SN~2022jpx. These distributions were obtained using the MCMC method, with the median (50th percentile) indicated by vertical dashed lines and shaded regions representing the 16th and 84th percentiles. The best-fit values and associated errors are provided at the top of each panel.

\section{Performance of the Supernova Parametric Model}
\label{model_fit_perf}

\begin{figure*}[ht]
\centering
\includegraphics[width=0.95\textwidth]{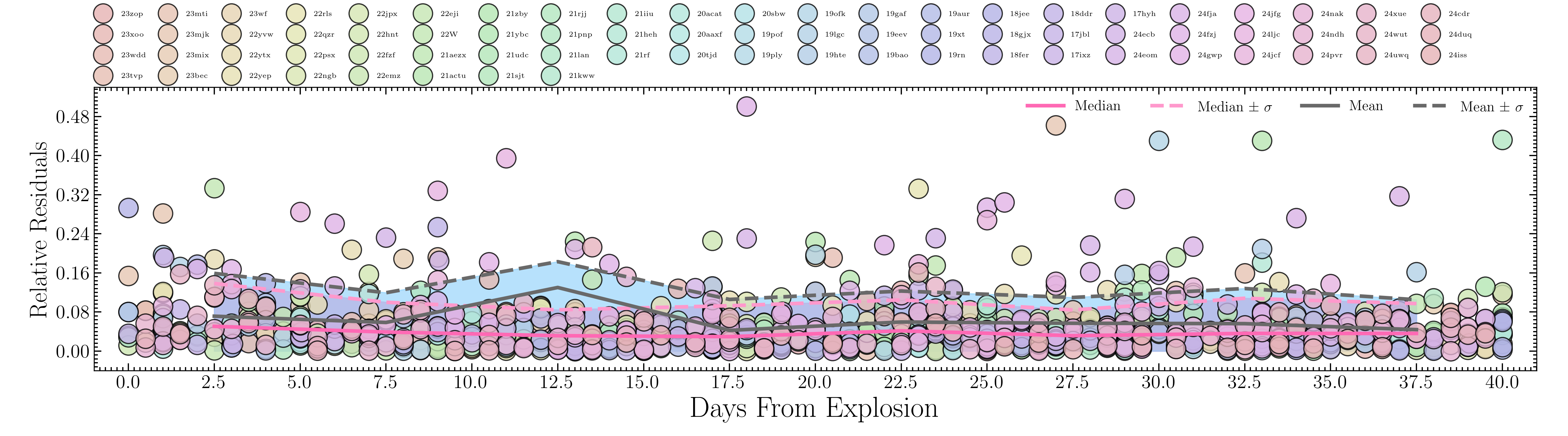} 
\caption{Relative residuals calculated concerning the average flux between the explosion and 40 days for the 74 SNe in the ATLAS sample. The pink and gray solid lines represent the mean and median estimates computed within a comoving 5-day window, respectively, while the dashed lines represent the standard deviations.  
\label{fig:stat_residuals}}
\end{figure*}

To evaluate the performance of the SPM, we analyzed the residuals and the $\chi^2_{\text{dof}}$. To ensure consistent comparisons for all objects, we calculated residuals relative to the average flux between the explosion and 40 days later, removing biases associated with the brightness of specific SNe. Thus, we analyzed the residuals by computing the mean, median, and standard deviation within a comoving window of five days. Figure~\ref{fig:stat_residuals} shows the relative residuals and their statistics. Generally, the relative errors are below 7.2\% across most phases, except between days 10 and 15, where the relative error increases to 13\%. This higher error is attributed to the rapid flux evolution during the rise time and the limited data points available in this phase, with an average of seven observations.

The $\chi^2$, which accounts for both residuals and the associated flux uncertainties, tends to be overestimated for brighter objects ($\text{f}_{\text{peak}} > 1000 \, \mu\mathrm{Jy}$). The reason is that brighter objects exhibit minor relative uncertainties than fainter ones, consistent with the Poissonian nature of flux distributions, where the flux uncertainty scales as $\sqrt{\text{f}}$, leading to higher $\chi^2$ values for brighter objects.

\begin{figure}[ht]
\centering
\includegraphics[width=0.48\textwidth]{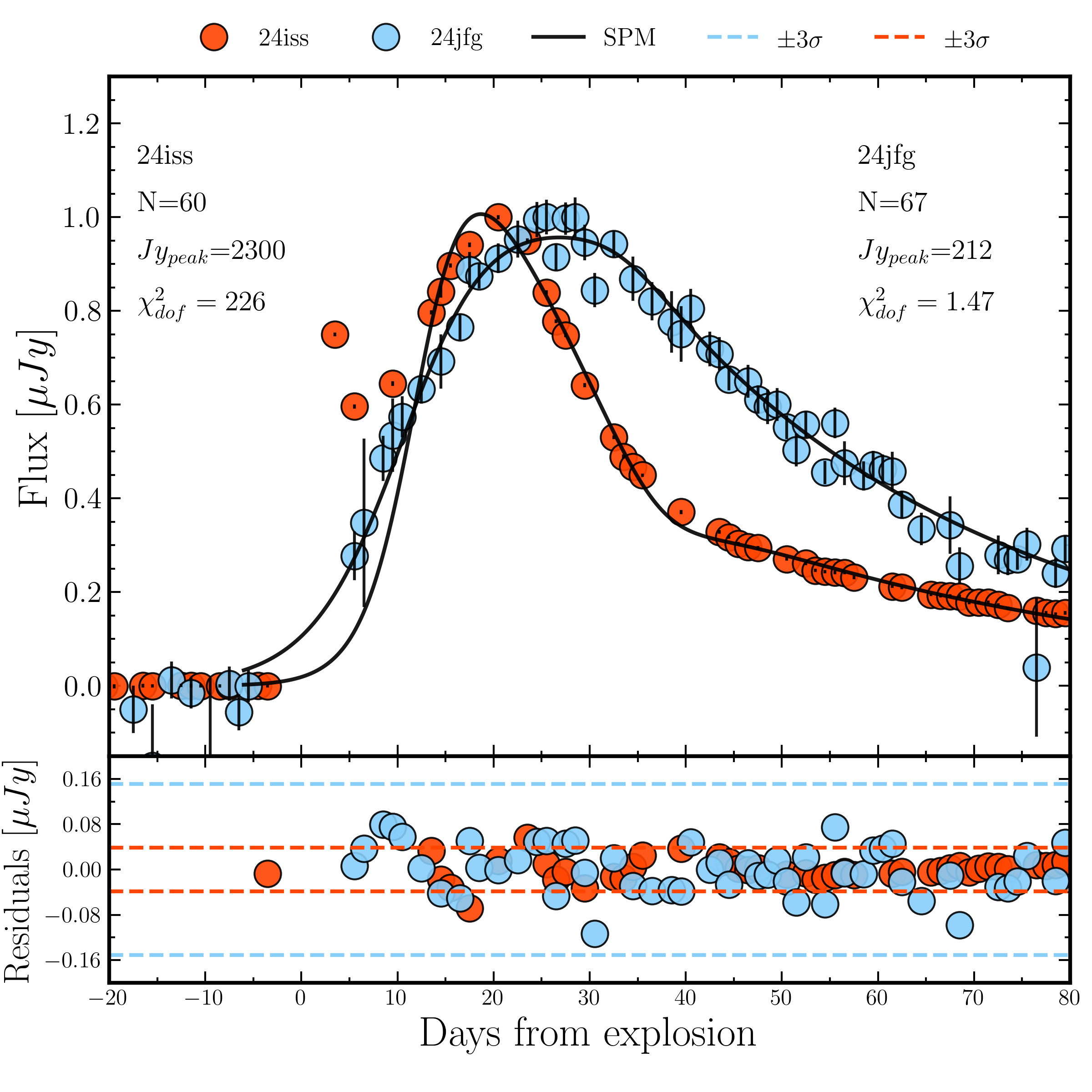} 
\caption{An illustrative example of $\chi^2$ overestimation for bright objects. The upper panel displays the light curves, and SPM fits for SN 2024iss (red dots) and SN 2024jfg (blue dots), while the bottom panel shows their residuals. Within the upper panel, we report the number of observations, peak flux, and $\chi^2_{\text{dof}}$. The dashed lines in the bottom panel represent the values for three residual standard deviations. This example demonstrates how brighter objects with smaller residuals and lower relative errors can lead to an overestimation of $\chi^2$.} 
\label{fig:ejmplo_chi}
\end{figure}

For the 63 SNe with a peak flux below $1000\mu\mathrm{Jy}$, the mean, median, and standard deviation of the $\chi^2_{\text{dof}}$ are 3.5, 2.6, 3.1 respectively. In contrast, for the complete sample of 74 SNe, these values increase to a mean of 10.7, a median of 3, and a standard deviation of 35. The increase in $\chi^2_{\text{dof}}$ when including brighter objects highlights the effect above. This trend is further illustrated in Figure~\ref{fig:ejmplo_chi}, where a fainter object with larger mean residuals but higher relative errors exhibits a lower $\chi^2_{\text{dof}}$ compared to a brighter object with smaller mean residuals but lower relative errors.

Analyzing the implications of these results for EE detection and light curve characterization, we define EE detection in our methodology based on residual analysis, showing a maximum error of 13\%, an acceptable uncertainty level. Additionally, during the post-peak phases, we measure all the parameters we use to characterize the light curves. The relative residuals are below 6\% (see Figure~\ref{fig:stat_residuals}), ensuring robust analysis for subsequent studies.

\definecolor{lightpink}{rgb}{1.0, 0.9, 0.9}
\definecolor{pastelpurple}{rgb}{0.8, 0.8, 1.0}
\definecolor{lightskyblue}{rgb}{0.88, 0.94, 1.0}
\definecolor{pastelred}{rgb}{1.0, 0.8, 0.8}

\section{Summary of Distribution and Correlation Statistics}

This appendix presents the full statistical analysis of light curve parameters for EE and non-EE SNe~IIb discussed in Sections~\ref{distributions} and~\ref{correlations}. Table~\ref{tab:tabla_KS_AD} provides the results of the Kolmogorov--Smirnov (KS) and Anderson--Darling (AD) tests applied to the distributions of \(t_{\text{rise}}\), \(\Delta M_{15}\), and \(M^{\mathrm{peak}}_{\text{abs}}\) across a grid of explosion epoch error thresholds \((\delta t_{\text{exp}} \in [1, 2, \dots, 7] \text{ days})\) and observational point density thresholds \((\rho_{\text{peak}} \in [0.0, 0.1, \dots, 0.4] \text{ days}^{-1})\). Table~\ref{tab:KS_AD_tests} summarizes the Pearson correlation coefficients and p-values for the relationships between these parameters within the EE and non-EE samples, also evaluated across the same grid of \(\delta t_{\text{exp}}\) and \(\rho_{\text{peak}}\) thresholds.

\onecolumn
\begin{longtable}{|c|c|c|}
\caption{KS and AD Test Results for Parameter Distributions of Supernovae} \\
\hline
\multicolumn{3}{|c|}{\textbf{Density Threshold: 0, Total SNe: 74 (28.38\% EE SNe, 71.62\% non-EE SNe)}} \\
\hline
\textbf{Parameter} & \textbf{KS Statistic (p-value)} & \textbf{AD Statistic (significance)} \\
\hline
$\Delta M_{15}$ & 4.02$\times 10^{-1}$ \cellcolor{lightgray} (1.07$\times 10^{-2}$) & \cellcolor{lightgray}4.14 (7.16$\times 10^{-3}$) \\
\hline
$M_{\mathrm{peak}}$ & 2.31$\times 10^{-1}$ (3.40$\times 10^{-1}$) & 2.63$\times 10^{-1}$ (2.50$\times 10^{-1}$) \\
\hline
$t_{\mathrm{rise}}$ & 4.42$\times 10^{-1}$ \cellcolor{lightgray} (3.34$\times 10^{-3}$) & \cellcolor{lightgray}5.48 (2.33$\times 10^{-3}$) \\
\hline
\multicolumn{3}{|c|}{\textbf{Density Threshold: 0.1, Total SNe: 73 (28.77\% EE SNe, 71.23\% non-EE SNe)}} \\
\hline
$\Delta M_{15}$ & 3.97$\times 10^{-1}$ \cellcolor{lightgray} (1.20$\times 10^{-2}$) & \cellcolor{lightgray}4.09 (7.44$\times 10^{-3}$) \\
\hline
$M_{\mathrm{peak}}$ & 2.26$\times 10^{-1}$ (3.60$\times 10^{-1}$) & 9.43$\times 10^{-2}$ (2.50$\times 10^{-1}$) \\
\hline
$t_{\mathrm{rise}}$ & 4.34$\times 10^{-1}$ \cellcolor{lightgray} (4.77$\times 10^{-3}$) & \cellcolor{lightgray}5.21 (2.92$\times 10^{-3}$) \\
\hline
\multicolumn{3}{|c|}{\textbf{Density Threshold: 0.2, Total SNe: 63 (33.33\% EE SNe, 66.67\% non-EE SNe)}} \\
\hline
$\Delta M_{15}$ & 4.05$\times 10^{-1}$ \cellcolor{lightgray} (1.76$\times 10^{-2}$) & \cellcolor{lightgray}3.83 (9.36$\times 10^{-3}$) \\
\hline
$M_{\mathrm{peak}}$ & 2.14$\times 10^{-1}$ (5.25$\times 10^{-1}$) & 1.78$\times 10^{-1}$ (2.50$\times 10^{-1}$) \\
\hline
$t_{\mathrm{rise}}$ & 4.05$\times 10^{-1}$ \cellcolor{lightgray} (1.76$\times 10^{-2}$) & \cellcolor{lightgray}3.78 (9.75$\times 10^{-3}$) \\
\hline
\multicolumn{3}{|c|}{\textbf{Density Threshold: 0.3, Total SNe: 42 (35.71\% EE SNe, 64.29\% non-EE SNe)}} \\
\hline
$\Delta M_{15}$ & 3.93$\times 10^{-1}$ (7.57$\times 10^{-2}$) & 1.17 (1.07$\times 10^{-1}$) \\
\hline
$M_{\mathrm{peak}}$ & 2.07$\times 10^{-1}$ (7.17$\times 10^{-1}$) & -8.74$\times 10^{-1}$ (2.50$\times 10^{-1}$) \\
\hline
$t_{\mathrm{rise}}$ & 4.22$\times 10^{-1}$ \cellcolor{lightgray} (4.62$\times 10^{-2}$) & 1.61 (7.06$\times 10^{-2}$) \\
\hline
\multicolumn{3}{|c|}{\textbf{Density Threshold: 0.4, Total SNe: 22 (40.91\% EE SNe, 59.09\% non-EE SNe)}} \\
\hline
$\Delta M_{15}$ & 4.70$\times 10^{-1}$ (1.39$\times 10^{-1}$) & -2.47$\times 10^{-2}$ (2.50$\times 10^{-1}$) \\
\hline
$M_{\mathrm{peak}}$ & 2.91$\times 10^{-1}$ (6.40$\times 10^{-1}$) & -8.01$\times 10^{-1}$ (2.50$\times 10^{-1}$) \\
\hline
$t_{\mathrm{rise}}$ & 3.59$\times 10^{-1}$ (4.00$\times 10^{-1}$) & -4.22$\times 10^{-1}$ (2.50$\times 10^{-1}$) \\
\hline
\multicolumn{3}{|c|}{\textbf{Explosion Time Error Threshold: 1, Total SNe: 21 (42.86\% EE SNe, 57.14\% non-EE SNe)}} \\
\hline
$\Delta M_{15}$ & 6.11$\times 10^{-1}$ \cellcolor{lightgray} (2.60$\times 10^{-2}$) & \cellcolor{lightgray}3.41 (1.35$\times 10^{-2}$) \\
\hline
$M_{\mathrm{peak}}$ & 2.22$\times 10^{-1}$ (9.32$\times 10^{-1}$) & -1.06 (2.50$\times 10^{-1}$) \\
\hline
$t_{\mathrm{rise}}$ & 6.11$\times 10^{-1}$ \cellcolor{lightgray} (2.60$\times 10^{-2}$) & \cellcolor{lightgray}2.50 (3.06$\times 10^{-2}$) \\
\hline
\multicolumn{3}{|c|}{\textbf{Explosion Time Error Threshold: 2, Total SNe: 41 (31.71\% EE SNe, 68.29\% non-EE SNe)}} \\
\hline
$\Delta M_{15}$ & 4.78$\times 10^{-1}$ \cellcolor{lightgray} (2.33$\times 10^{-2}$) & 1.66 (6.73$\times 10^{-2}$) \\
\hline
$M_{\mathrm{peak}}$ & 3.30$\times 10^{-1}$ (2.25$\times 10^{-1}$) & 2.49$\times 10^{-1}$ (2.50$\times 10^{-1}$) \\
\hline
$t_{\mathrm{rise}}$ & 4.48$\times 10^{-1}$ \cellcolor{lightgray} (3.93$\times 10^{-2}$) & \cellcolor{lightgray}2.25 (3.85$\times 10^{-2}$) \\
\hline
\multicolumn{3}{|c|}{\textbf{Explosion Time Error Threshold: 3, Total SNe: 53 (33.96\% EE SNe, 66.04\% non-EE SNe)}} \\
\hline
$\Delta M_{15}$ & 4.38$\times 10^{-1}$ \cellcolor{lightgray} (1.30$\times 10^{-2}$) & \cellcolor{lightgray}2.46 (3.19$\times 10^{-2}$) \\
\hline
$M_{\mathrm{peak}}$ & 2.98$\times 10^{-1}$ (1.84$\times 10^{-1}$) & 6.43$\times 10^{-1}$ (1.79$\times 10^{-1}$) \\
\hline
$t_{\mathrm{rise}}$ & 5.19$\times 10^{-1}$ \cellcolor{lightgray} (2.00$\times 10^{-3}$) & \cellcolor{lightgray}5.43 (2.43$\times 10^{-3}$) \\
\hline
\multicolumn{3}{|c|}{\textbf{Explosion Time Error Threshold: 4, Total SNe: 64 (32.81\% EE SNe, 67.19\% non-EE SNe)}} \\ 
\hline
$\Delta M_{15}$ & 4.34$\times 10^{-1}$ \cellcolor{lightgray} (6.93$\times 10^{-3}$) & \cellcolor{lightgray}3.64 (1.10$\times 10^{-2}$) \\
\hline
$M_{\mathrm{peak}}$ & 1.97$\times 10^{-1}$ (5.68$\times 10^{-1}$) & -1.36$\times 10^{-1}$ (2.50$\times 10^{-1}$) \\
\hline
$t_{\mathrm{rise}}$ & 5.08$\times 10^{-1}$ \cellcolor{lightgray} (6.55$\times 10^{-4}$) & \cellcolor{lightgray}6.02 (1.53$\times 10^{-3}$) \\
\hline
\multicolumn{3}{|c|}{\textbf{Explosion Time Error Threshold: 5, Total SNe: 69 (30.43\% EE SNe, 69.57\% non-EE SNe)}} \\
\hline
$\Delta M_{15}$ & 4.38$\times 10^{-1}$ \cellcolor{lightgray} (4.78$\times 10^{-3}$) & \cellcolor{lightgray}4.02 (7.97$\times 10^{-3}$) \\
\hline
$M_{\mathrm{peak}}$ & 2.05$\times 10^{-1}$ (4.96$\times 10^{-1}$) & 2.67$\times 10^{-1}$ (2.50$\times 10^{-1}$) \\
\hline
$t_{\mathrm{rise}}$ & 4.61$\times 10^{-1}$ \cellcolor{lightgray} (2.41$\times 10^{-3}$) & \cellcolor{lightgray}5.36 (2.58$\times 10^{-3}$) \\
\hline
\multicolumn{3}{|c|}{\textbf{Explosion Time Error Threshold: 6, Total SNe: 74 (28.38\% EE SNe, 71.62\% non-EE SNe)}} \\
\hline
$\Delta M_{15}$ & 4.02$\times 10^{-1}$ \cellcolor{lightgray} (1.07$\times 10^{-2}$) & \cellcolor{lightgray}4.14 (7.16$\times 10^{-3}$) \\
\hline
$M_{\mathrm{peak}}$ & 2.12$\times 10^{-1}$ (4.41$\times 10^{-1}$) & 2.68$\times 10^{-1}$ (2.50$\times 10^{-1}$) \\
\hline
$t_{\mathrm{rise}}$ & 4.42$\times 10^{-1}$ \cellcolor{lightgray} (3.34$\times 10^{-3}$) & \cellcolor{lightgray}5.48 (2.33$\times 10^{-3}$) 
\label{tab:tabla_KS_AD}
\end{longtable}
\twocolumn

\onecolumn
\begin{longtable}{|l|c|c|c|}
\caption{Correlation coefficients under different pair-parameters.} \\
\hline
\textbf{Pair-Parameter} & \textbf{Pearson (p-value)} & \textbf{Spearman (p-value)} & \textbf{Kendall (p-value)} \\
\hline
\endfirsthead
\caption[]{(Continuation)} \\
\hline
\textbf{Pair-Parameter} & \textbf{Pearson (p-value)} & \textbf{Spearman (p-value)} & \textbf{Kendall (p-value)} \\
\hline
\endhead
\hline
\multicolumn{4}{|c|}{Continued on the next page...} \\
\hline
\endfoot
\hline
\endlastfoot
\rowcolor{pastelpurple}
\multicolumn{4}{|c|}{\textbf{Total Sample}} \\
\hline
\rowcolor{pastelpurple}
\multicolumn{4}{|c|}{\textbf{Density Threshold: 0, Total Supernovae: 74}} \\
\hline
\textbf{$\Delta M_{15}$ - $M_{\mathrm{peak}}$} & \cellcolor{lightgray}0.23 ($4.75 \times 10^{-2}$) & \cellcolor{lightgray}0.30 ($1.05 \times 10^{-2}$) & \cellcolor{lightgray}0.21 ($8.84 \times 10^{-3}$) \\
\textbf{$\Delta M_{15}$ - $t_{\mathrm{rise}}$} & -0.12 ($3.20 \times 10^{-1}$) & -0.18 ($1.31 \times 10^{-1}$) & -0.12 ($1.21 \times 10^{-1}$) \\
\textbf{$M_{\mathrm{peak}}$ - $t_{\mathrm{rise}}$} & -0.20 ($8.64 \times 10^{-2}$) & -0.16 ($1.67 \times 10^{-1}$) & -0.11 ($1.59 \times 10^{-1}$) \\
\hline
\rowcolor{pastelpurple}
\multicolumn{4}{|c|}{\textbf{Density Threshold: 0.1, Total Supernovae: 73}} \\
\hline
\textbf{$\Delta M_{15}$ - $M_{\mathrm{peak}}$} & \cellcolor{lightgray}0.23 ($4.57 \times 10^{-2}$) & \cellcolor{lightgray}0.31 ($8.45 \times 10^{-3}$) & \cellcolor{lightgray}0.21 ($7.23 \times 10^{-3}$) \\
\textbf{$\Delta M_{15}$ - $t_{\mathrm{rise}}$} & -0.12 ($3.21 \times 10^{-1}$) & -0.17 ($1.42 \times 10^{-1}$) & -0.12 ($1.31 \times 10^{-1}$) \\
\textbf{$M_{\mathrm{peak}}$ - $t_{\mathrm{rise}}$} & -0.23 ($5.35 \times 10^{-2}$) & -0.20 ($8.97 \times 10^{-2}$) & -0.14 ($8.73 \times 10^{-2}$) \\
\hline
\rowcolor{pastelpurple}
\multicolumn{4}{|c|}{\textbf{Density Threshold: 0.2, Total Supernovae: 63}} \\
\hline
\textbf{$\Delta M_{15}$ - $M_{\mathrm{peak}}$} & 0.21 ($9.55 \times 10^{-2}$) & \cellcolor{lightgray}0.27 ($3 \times 10^{-2}$) & \cellcolor{lightgray}0.19 ($2.86 \times 10^{-2}$) \\
\textbf{$\Delta M_{15}$ - $t_{\mathrm{rise}}$} & -0.13 ($2.92 \times 10^{-1}$) & -0.22 ($8.39 \times 10^{-2}$) & -0.16 ($6.25 \times 10^{-2}$) \\
\textbf{$M_{\mathrm{peak}}$ - $t_{\mathrm{rise}}$} & \cellcolor{lightgray}-0.25 ($4.92 \times 10^{-2}$) & \cellcolor{lightgray}-0.26 ($3.84 \times 10^{-2}$) & \cellcolor{lightgray}-0.18 ($3.27 \times 10^{-2}$) \\
\hline
\rowcolor{pastelpurple}
\multicolumn{4}{|c|}{\textbf{Density Threshold: 0.3, Total Supernovae: 42}} \\
\hline
\textbf{$\Delta M_{15}$ - $M_{\mathrm{peak}}$} & 0.19 ($2.33 \times 10^{-1}$) & 0.26 ($9.82 \times 10^{-2}$) & 0.18 ($8.89 \times 10^{-2}$) \\
\textbf{$\Delta M_{15}$ - $t_{\mathrm{rise}}$} & -0.12 ($4.64 \times 10^{-1}$) & -0.27 ($8.69 \times 10^{-2}$) & -0.20 ($6.23 \times 10^{-2}$) \\
\textbf{$M_{\mathrm{peak}}$ - $t_{\mathrm{rise}}$} & \cellcolor{lightgray}-0.48 ($1.29 \times 10^{-3}$) & \cellcolor{lightgray}-0.38 ($1.36 \times 10^{-2}$) & \cellcolor{lightgray}-0.26 ($1.61 \times 10^{-2}$) \\
\hline
\rowcolor{pastelpurple}
\multicolumn{4}{|c|}{\textbf{Density Threshold: 0.4, Total Supernovae: 22}} \\
\hline
\textbf{$\Delta M_{15}$ - $M_{\mathrm{peak}}$} & 0.11 ($6.23 \times 10^{-1}$) & 0.10 ($6.51 \times 10^{-1}$) & 0.06 ($6.96 \times 10^{-1}$) \\
\textbf{$\Delta M_{15}$ - $t_{\mathrm{rise}}$} & 0.03 ($8.81 \times 10^{-1}$) & -0.30 ($1.73 \times 10^{-1}$) & -0.23 ($1.28 \times 10^{-1}$) \\
\textbf{$M_{\mathrm{peak}}$ - $t_{\mathrm{rise}}$} & \cellcolor{lightgray}-0.50 ($1.74 \times 10^{-2}$) & \cellcolor{lightgray}-0.49 ($2.10 \times 10^{-2}$) & \cellcolor{lightgray}-0.34 ($2.78 \times 10^{-2}$) \\
\hline
\rowcolor{pastelpurple}
\multicolumn{4}{|c|}{\textbf{Explosion Time Error Threshold: 1, Total Supernovae: 21}} \\
\hline
\textbf{$\Delta M_{15}$ - $M_{\mathrm{peak}}$} & 0.18 ($4.28 \times 10^{-1}$) & 0.19 ($3.97 \times 10^{-1}$) & 0.11 ($4.92 \times 10^{-1}$) \\
\textbf{$\Delta M_{15}$ - $t_{\mathrm{rise}}$} & -0.19 ($3.98 \times 10^{-1}$) & -0.09 ($7.01 \times 10^{-1}$) & -0.08 ($6.08 \times 10^{-1}$) \\
\textbf{$M_{\mathrm{peak}}$ - $t_{\mathrm{rise}}$} & -0.03 ($9.09 \times 10^{-1}$) & -0.14 ($5.42 \times 10^{-1}$) & -0.09 ($5.66 \times 10^{-1}$) \\
\hline
\rowcolor{pastelpurple}
\multicolumn{4}{|c|}{\textbf{Explosion Time Error Threshold: 2, Total Supernovae: 41}} \\
\hline
\textbf{$\Delta M_{15}$ - $M_{\mathrm{peak}}$} & 0.17 ($2.94 \times 10^{-1}$) & 0.18 ($2.64 \times 10^{-1}$) & 0.12 ($2.71 \times 10^{-1}$) \\
\textbf{$\Delta M_{15}$ - $t_{\mathrm{rise}}$} & -0.00 ($9.81 \times 10^{-1}$) & -0.16 ($3.20 \times 10^{-1}$) & -0.11 ($3.12 \times 10^{-1}$) \\
\textbf{$M_{\mathrm{peak}}$ - $t_{\mathrm{rise}}$} & -0.21 ($1.97 \times 10^{-1}$) & -0.22 ($1.75 \times 10^{-1}$) & -0.15 ($1.64 \times 10^{-1}$) \\
\hline
\rowcolor{pastelpurple}
\multicolumn{4}{|c|}{\textbf{Explosion Time Error Threshold: 3, Total Supernovae: 53}} \\
\hline
\textbf{$\Delta M_{15}$ - $M_{\mathrm{peak}}$} & 0.21 ($1.23 \times 10^{-1}$) & \cellcolor{lightgray}0.28 ($4.02 \times 10^{-2}$) & \cellcolor{lightgray}0.19 ($3.98 \times 10^{-2}$) \\
\textbf{$\Delta M_{15}$ - $t_{\mathrm{rise}}$} & -0.02 ($8.90 \times 10^{-1}$) & -0.14 ($3.32 \times 10^{-1}$) & -0.09 ($3.30 \times 10^{-1}$) \\
\textbf{$M_{\mathrm{peak}}$ - $t_{\mathrm{rise}}$} & -0.19 ($1.70 \times 10^{-1}$) & -0.21 ($1.33 \times 10^{-1}$) & -0.14 ($1.31 \times 10^{-1}$) \\
\hline
\rowcolor{pastelpurple}
\multicolumn{4}{|c|}{\textbf{Explosion Time Error Threshold: 4, Total Supernovae: 64}} \\
\hline
\textbf{$\Delta M_{15}$ - $M_{\mathrm{peak}}$} & 0.23 ($6.97 \times 10^{-2}$) & \cellcolor{lightgray}0.27 ($3.13 \times 10^{-2}$) & \cellcolor{lightgray}0.18 ($3.30 \times 10^{-2}$) \\
\textbf{$\Delta M_{15}$ - $t_{\mathrm{rise}}$} & -0.12 ($3.32 \times 10^{-1}$) & -0.20 ($1.15 \times 10^{-1}$) & -0.14 ($9.98 \times 10^{-2}$) \\
\textbf{$M_{\mathrm{peak}}$ - $t_{\mathrm{rise}}$} & -0.24 ($5.84 \times 10^{-2}$) & -0.19 ($1.27 \times 10^{-1}$) & -0.14 ($1.05 \times 10^{-1}$) \\
\hline
\rowcolor{pastelpurple}
\multicolumn{4}{|c|}{\textbf{Explosion Time Error Threshold: 5, Total Supernovae: 69}} \\
\hline
\textbf{$\Delta M_{15}$ - $M_{\mathrm{peak}}$} & \cellcolor{lightgray}0.25 ($3.94 \times 10^{-2}$) & \cellcolor{lightgray}0.31 ($8.45 \times 10^{-3}$) & \cellcolor{lightgray}0.21 ($9.04 \times 10^{-3}$) \\
\textbf{$\Delta M_{15}$ - $t_{\mathrm{rise}}$} & -0.12 ($3.17 \times 10^{-1}$) & -0.19 ($1.15 \times 10^{-1}$) & -0.13 ($1.05 \times 10^{-1}$) \\
\textbf{$M_{\mathrm{peak}}$ - $t_{\mathrm{rise}}$} & -0.20 ($9.62 \times 10^{-2}$) & -0.16 ($1.77 \times 10^{-1}$) & -0.12 ($1.54 \times 10^{-1}$) \\
\hline
\rowcolor{pastelpurple}
\multicolumn{4}{|c|}{\textbf{Explosion Time Error Threshold: 6, Total Supernovae: 74}} \\
\hline
\textbf{$\Delta M_{15}$ - $M_{\mathrm{peak}}$} & \cellcolor{lightgray}0.24 ($3.90 \times 10^{-2}$) & \cellcolor{lightgray}0.30 ($9.90 \times 10^{-3}$) & \cellcolor{lightgray}0.20 ($1.07 \times 10^{-2}$) \\
\textbf{$\Delta M_{15}$ - $t_{\mathrm{rise}}$} & -0.12 ($3.20 \times 10^{-1}$) & -0.18 ($1.31 \times 10^{-1}$) & -0.12 ($1.21 \times 10^{-1}$) \\
\textbf{$M_{\mathrm{peak}}$ - $t_{\mathrm{rise}}$} & -0.20 ($8.77 \times 10^{-2}$) & -0.17 ($1.59 \times 10^{-1}$) & -0.12 ($1.45 \times 10^{-1}$) \\
\hline
\rowcolor{lightskyblue}
\multicolumn{4}{|c|}{\textbf{EE Sample}} \\
\hline
\rowcolor{lightskyblue}
\multicolumn{4}{|c|}{\textbf{Density Threshold: 0, Total EE: 21 (28.38\%)}} \\
\hline
\textbf{$\Delta M_{15}$ - $M_{\mathrm{peak}}$} & 0.27 ($2.41 \times 10^{-1}$) & 0.26 ($2.51 \times 10^{-1}$) & 0.17 ($2.94 \times 10^{-1}$) \\
\textbf{$\Delta M_{15}$ - $t_{\mathrm{rise}}$} & \cellcolor{lightgray}-0.77 ($3.91 \times 10^{-5}$) & \cellcolor{lightgray}-0.69 ($4.76 \times 10^{-4}$) & \cellcolor{lightgray}-0.55 ($5.70 \times 10^{-4}$) \\
\textbf{$M_{\mathrm{peak}}$ - $t_{\mathrm{rise}}$} & \cellcolor{lightgray}-0.52 ($1.58 \times 10^{-2}$) & \cellcolor{lightgray}-0.65 ($1.35 \times 10^{-3}$) & \cellcolor{lightgray}-0.46 ($3.71 \times 10^{-3}$) \\
\hline
\rowcolor{lightskyblue}
\multicolumn{4}{|c|}{\textbf{Density Threshold: 0.1, Total EE: 21 (28.77\%)}} \\
\hline
\textbf{$\Delta M_{15}$ - $M_{\mathrm{peak}}$} & 0.27 ($2.41 \times 10^{-1}$) & 0.26 ($2.51 \times 10^{-1}$) & 0.17 ($2.94 \times 10^{-1}$) \\
\textbf{$\Delta M_{15}$ - $t_{\mathrm{rise}}$} & \cellcolor{lightgray}-0.77 ($3.91 \times 10^{-5}$) & \cellcolor{lightgray}-0.69 ($4.76 \times 10^{-4}$) & \cellcolor{lightgray}-0.55 ($5.70 \times 10^{-4}$) \\
\textbf{$M_{\mathrm{peak}}$ - $t_{\mathrm{rise}}$} & \cellcolor{lightgray}-0.52 ($1.58 \times 10^{-2}$) & \cellcolor{lightgray}-0.65 ($1.35 \times 10^{-3}$) & \cellcolor{lightgray}-0.46 ($3.71 \times 10^{-3}$) \\
\hline
\rowcolor{lightskyblue}
\multicolumn{4}{|c|}{\textbf{Density Threshold: 0.2, Total EE: 21 (33.33\%)}} \\
\hline
\textbf{$\Delta M_{15}$ - $M_{\mathrm{peak}}$} & 0.27 ($2.41 \times 10^{-1}$) & 0.26 ($2.51 \times 10^{-1}$) & 0.17 ($2.94 \times 10^{-1}$) \\
\textbf{$\Delta M_{15}$ - $t_{\mathrm{rise}}$} & \cellcolor{lightgray}-0.77 ($3.91 \times 10^{-5}$) & \cellcolor{lightgray}-0.69 ($4.76 \times 10^{-4}$) & \cellcolor{lightgray}-0.55 ($5.70 \times 10^{-4}$) \\
\textbf{$M_{\mathrm{peak}}$ - $t_{\mathrm{rise}}$} & \cellcolor{lightgray}-0.52 ($1.58 \times 10^{-2}$) & \cellcolor{lightgray}-0.65 ($1.35 \times 10^{-3}$) & \cellcolor{lightgray}-0.46 ($3.71 \times 10^{-3}$) \\
\hline
\rowcolor{lightskyblue}
\multicolumn{4}{|c|}{\textbf{Density Threshold: 0.3, Total EE: 15 (35.71\%)}} \\
\hline
\textbf{$\Delta M_{15}$ - $M_{\mathrm{peak}}$} & 0.20 ($4.73 \times 10^{-1}$) & 0.27 ($3.34 \times 10^{-1}$) & 0.18 ($3.79 \times 10^{-1}$) \\
\textbf{$\Delta M_{15}$ - $t_{\mathrm{rise}}$} & \cellcolor{lightgray}-0.75 ($1.19 \times 10^{-3}$) & \cellcolor{lightgray}-0.69 ($4.72 \times 10^{-3}$) & \cellcolor{lightgray}-0.56 ($4.06 \times 10^{-3}$) \\
\textbf{$M_{\mathrm{peak}}$ - $t_{\mathrm{rise}}$} & \cellcolor{lightgray}-0.53 ($4.16 \times 10^{-2}$) & \cellcolor{lightgray}-0.60 ($1.92 \times 10^{-2}$) & \cellcolor{lightgray}-0.44 ($2.27 \times 10^{-2}$) \\
\hline
\rowcolor{lightskyblue}
\multicolumn{4}{|c|}{\textbf{Density Threshold: 0.4, Total EE: 9 (40.91\%)}} \\
\hline
\textbf{$\Delta M_{15}$ - $M_{\mathrm{peak}}$} & 0.34 ($3.65 \times 10^{-1}$) & 0.37 ($3.32 \times 10^{-1}$) & 0.28 ($3.58 \times 10^{-1}$) \\
\textbf{$\Delta M_{15}$ - $t_{\mathrm{rise}}$} & \cellcolor{lightgray}-0.75 ($2.06 \times 10^{-2}$) & \cellcolor{lightgray}-0.70 ($3.58 \times 10^{-2}$) & \cellcolor{lightgray}-0.61 ($2.47 \times 10^{-2}$) \\
\textbf{$M_{\mathrm{peak}}$ - $t_{\mathrm{rise}}$} & \cellcolor{lightgray}-0.66 ($5.15 \times 10^{-2}$) & \cellcolor{lightgray}-0.83 ($5.27 \times 10^{-3}$) & \cellcolor{lightgray}-0.67 ($1.27 \times 10^{-2}$) \\
\hline
\rowcolor{lightskyblue}
\multicolumn{4}{|c|}{\textbf{Explosion Time Error Threshold: 1, Total EE: 9 (42.86\%)}} \\
\hline
\textbf{$\Delta M_{15}$ - $M_{\mathrm{peak}}$} & 0.53 ($1.41 \times 10^{-1}$) & 0.52 ($1.54 \times 10^{-1}$) & 0.33 ($2.60 \times 10^{-1}$) \\
\textbf{$\Delta M_{15}$ - $t_{\mathrm{rise}}$} & \cellcolor{lightgray}-0.93 ($3.04 \times 10^{-4}$) & \cellcolor{lightgray}-0.88 ($1.59 \times 10^{-3}$) & \cellcolor{lightgray}-0.78 ($2.43 \times 10^{-3}$) \\
\textbf{$M_{\mathrm{peak}}$ - $t_{\mathrm{rise}}$} & -0.61 ($7.87 \times 10^{-2}$) & -0.70 ($3.58 \times 10^{-2}$) & -0.44 ($1.19 \times 10^{-1}$) \\
\hline
\rowcolor{lightskyblue}
\multicolumn{4}{|c|}{\textbf{Explosion Time Error Threshold: 2, Total EE: 13 (31.71\%)}} \\
\hline
\textbf{$\Delta M_{15}$ - $M_{\mathrm{peak}}$} & 0.36 ($2.30 \times 10^{-1}$) & 0.40 ($1.81 \times 10^{-1}$) & 0.26 ($2.52 \times 10^{-1}$) \\
\textbf{$\Delta M_{15}$ - $t_{\mathrm{rise}}$} & \cellcolor{lightgray}-0.81 ($7.52 \times 10^{-4}$) & \cellcolor{lightgray}-0.78 ($1.61 \times 10^{-3}$) & \cellcolor{lightgray}-0.63 ($2.74 \times 10^{-3}$) \\
\textbf{$M_{\mathrm{peak}}$ - $t_{\mathrm{rise}}$} & -0.50 ($8.34 \times 10^{-2}$) & \cellcolor{lightgray}-0.71 ($6.88 \times 10^{-3}$) & \cellcolor{lightgray}-0.48 ($2.37 \times 10^{-2}$) \\
\hline
\rowcolor{lightskyblue}
\multicolumn{4}{|c|}{\textbf{Explosion Time Error Threshold: 3, Total EE: 18 (33.96\%)}} \\
\hline
\textbf{$\Delta M_{15}$ - $M_{\mathrm{peak}}$} & 0.36 ($1.37 \times 10^{-1}$) & 0.42 ($8.10 \times 10^{-2}$) & 0.28 ($1.12 \times 10^{-1}$) \\
\textbf{$\Delta M_{15}$ - $t_{\mathrm{rise}}$} & \cellcolor{lightgray}-0.79 ($1.11 \times 10^{-4}$) & \cellcolor{lightgray}-0.72 ($7.53 \times 10^{-4}$) & \cellcolor{lightgray}-0.57 ($9.67 \times 10^{-4}$) \\
\textbf{$M_{\mathrm{peak}}$ - $t_{\mathrm{rise}}$} & \cellcolor{lightgray}-0.57 ($1.42 \times 10^{-2}$) & \cellcolor{lightgray}-0.78 ($1.35 \times 10^{-4}$) & \cellcolor{lightgray}-0.56 ($1.26 \times 10^{-3}$) \\
\hline
\rowcolor{lightskyblue}
\multicolumn{4}{|c|}{\textbf{Explosion Time Error Threshold: 4, Total EE: 21 (32.81\%)}} \\
\hline
\textbf{$\Delta M_{15}$ - $M_{\mathrm{peak}}$} & 0.28 ($2.14 \times 10^{-1}$) & 0.27 ($2.39 \times 10^{-1}$) & 0.17 ($2.94 \times 10^{-1}$) \\
\textbf{$\Delta M_{15}$ - $t_{\mathrm{rise}}$} & \cellcolor{lightgray}-0.77 ($3.91 \times 10^{-5}$) & \cellcolor{lightgray}-0.69 ($4.76 \times 10^{-4}$) & \cellcolor{lightgray}-0.55 ($5.70 \times 10^{-4}$) \\
\textbf{$M_{\mathrm{peak}}$ - $t_{\mathrm{rise}}$} & \cellcolor{lightgray}-0.52 ($1.52 \times 10^{-2}$) & \cellcolor{lightgray}-0.66 ($1.06 \times 10^{-3}$) & \cellcolor{lightgray}-0.46 ($3.71 \times 10^{-3}$) \\
\hline
\rowcolor{lightskyblue}
\multicolumn{4}{|c|}{\textbf{Explosion Time Error Threshold: 5, Total EE: 21 (30.43\%)}} \\
\hline
\textbf{$\Delta M_{15}$ - $M_{\mathrm{peak}}$} & 0.28 ($2.14 \times 10^{-1}$) & 0.27 ($2.39 \times 10^{-1}$) & 0.17 ($2.94 \times 10^{-1}$) \\
\textbf{$\Delta M_{15}$ - $t_{\mathrm{rise}}$} & \cellcolor{lightgray}-0.77 ($3.91 \times 10^{-5}$) & \cellcolor{lightgray}-0.69 ($4.76 \times 10^{-4}$) & \cellcolor{lightgray}-0.55 ($5.70 \times 10^{-4}$) \\
\textbf{$M_{\mathrm{peak}}$ - $t_{\mathrm{rise}}$} & \cellcolor{lightgray}-0.52 ($1.52 \times 10^{-2}$) & \cellcolor{lightgray}-0.66 ($1.06 \times 10^{-3}$) & \cellcolor{lightgray}-0.46 ($3.71 \times 10^{-3}$) \\
\hline
\rowcolor{lightskyblue}
\multicolumn{4}{|c|}{\textbf{Explosion Time Error Threshold: 6, Total EE: 21 (28.38\%)}} \\
\hline
\textbf{$\Delta M_{15}$ - $M_{\mathrm{peak}}$} & 0.28 ($2.14 \times 10^{-1}$) & 0.27 ($2.39 \times 10^{-1}$) & 0.17 ($2.94 \times 10^{-1}$) \\
\textbf{$\Delta M_{15}$ - $t_{\mathrm{rise}}$} & \cellcolor{lightgray}-0.77 ($3.91 \times 10^{-5}$) & \cellcolor{lightgray}-0.69 ($4.76 \times 10^{-4}$) & \cellcolor{lightgray}-0.55 ($5.70 \times 10^{-4}$) \\
\textbf{$M_{\mathrm{peak}}$ - $t_{\mathrm{rise}}$} & \cellcolor{lightgray}-0.52 ($1.52 \times 10^{-2}$) & \cellcolor{lightgray}-0.66 ($1.06 \times 10^{-3}$) & \cellcolor{lightgray}-0.46 ($3.71 \times 10^{-3}$) \\
\hline
\rowcolor{pastelred}
\multicolumn{4}{|c|}{\textbf{Non-EE Sample}} \\
\hline
\rowcolor{pastelred}
\multicolumn{4}{|c|}{\textbf{Density Threshold: 0, Total Non-EE: 53 (71.62\%)}} \\
\hline
\textbf{$\Delta M_{15}$ - $M_{\mathrm{peak}}$} & 0.18 ($1.93 \times 10^{-1}$) & \cellcolor{lightgray}0.30 ($3.09 \times 10^{-2}$) & \cellcolor{lightgray}0.21 ($2.94 \times 10^{-2}$) \\
\textbf{$\Delta M_{15}$ - $t_{\mathrm{rise}}$} & -0.03 ($8.22 \times 10^{-1}$) & -0.20 ($1.53 \times 10^{-1}$) & -0.14 ($1.33 \times 10^{-1}$) \\
\textbf{$M_{\mathrm{peak}}$ - $t_{\mathrm{rise}}$} & -0.18 ($1.96 \times 10^{-1}$) & -0.06 ($6.82 \times 10^{-1}$) & -0.04 ($6.90 \times 10^{-1}$) \\
\hline
\rowcolor{pastelred}
\multicolumn{4}{|c|}{\textbf{Density Threshold: 0.1, Total Non-EE: 52 (71.23\%)}} \\
\hline
\textbf{$\Delta M_{15}$ - $M_{\mathrm{peak}}$} & 0.19 ($1.82 \times 10^{-1}$) & \cellcolor{lightgray}0.32 ($1.97 \times 10^{-2}$) & \cellcolor{lightgray}0.22 ($1.95 \times 10^{-2}$) \\
\textbf{$\Delta M_{15}$ - $t_{\mathrm{rise}}$} & -0.03 ($8.41 \times 10^{-1}$) & -0.18 ($1.94 \times 10^{-1}$) & -0.13 ($1.65 \times 10^{-1}$) \\
\textbf{$M_{\mathrm{peak}}$ - $t_{\mathrm{rise}}$} & -0.21 ($1.38 \times 10^{-1}$) & -0.10 ($4.60 \times 10^{-1}$) & -0.07 ($4.68 \times 10^{-1}$) \\
\hline
\rowcolor{pastelred}
\multicolumn{4}{|c|}{\textbf{Density Threshold: 0.2, Total Non-EE: 42 (66.67\%)}} \\
\hline
\textbf{$\Delta M_{15}$ - $M_{\mathrm{peak}}$} & 0.15 ($3.30 \times 10^{-1}$) & 0.28 ($7.50 \times 10^{-2}$) & 0.18 ($8.49 \times 10^{-2}$) \\
\textbf{$\Delta M_{15}$ - $t_{\mathrm{rise}}$} & -0.03 ($8.40 \times 10^{-1}$) & -0.23 ($1.42 \times 10^{-1}$) & -0.18 ($1.02 \times 10^{-1}$) \\
\textbf{$M_{\mathrm{peak}}$ - $t_{\mathrm{rise}}$} & -0.23 ($1.50 \times 10^{-1}$) & -0.17 ($2.87 \times 10^{-1}$) & -0.12 ($2.64 \times 10^{-1}$) \\
\hline
\rowcolor{pastelred}
\multicolumn{4}{|c|}{\textbf{Density Threshold: 0.3, Total Non-EE: 27 (64.29\%)}} \\
\hline
\textbf{$\Delta M_{15}$ - $M_{\mathrm{peak}}$} & 0.18 ($3.68 \times 10^{-1}$) & 0.33 ($9.70 \times 10^{-2}$) & 0.22 ($1.14 \times 10^{-1}$) \\
\textbf{$\Delta M_{15}$ - $t_{\mathrm{rise}}$} & 0.02 ($9.22 \times 10^{-1}$) & -0.23 ($2.51 \times 10^{-1}$) & -0.19 ($1.69 \times 10^{-1}$) \\
\textbf{$M_{\mathrm{peak}}$ - $t_{\mathrm{rise}}$} & \cellcolor{lightgray}-0.48 ($1.16 \times 10^{-2}$) & -0.29 ($1.42 \times 10^{-1}$) & -0.19 ($1.56 \times 10^{-1}$) \\
\hline
\rowcolor{pastelred}
\multicolumn{4}{|c|}{\textbf{Density Threshold: 0.4, Total Non-EE: 13 (59.09\%)}} \\
\hline
\textbf{$\Delta M_{15}$ - $M_{\mathrm{peak}}$} & 0.01 ($9.83 \times 10^{-1}$) & 0.04 ($8.87 \times 10^{-1}$) & 0.00 ($1.00 \times 10^{0}$) \\
\textbf{$\Delta M_{15}$ - $t_{\mathrm{rise}}$} & 0.27 ($3.68 \times 10^{-1}$) & -0.27 ($3.64 \times 10^{-1}$) & -0.23 ($3.06 \times 10^{-1}$) \\
\textbf{$M_{\mathrm{peak}}$ - $t_{\mathrm{rise}}$} & -0.43 ($1.42 \times 10^{-1}$) & -0.27 ($3.74 \times 10^{-1}$) & -0.15 ($5.10 \times 10^{-1}$) \\
\hline
\rowcolor{pastelred}
\multicolumn{4}{|c|}{\textbf{Explosion Time Error Threshold: 1, Total Non-EE: 12 (57.14\%)}} \\
\hline
\textbf{$\Delta M_{15}$ - $M_{\mathrm{peak}}$} & 0.07 ($8.39 \times 10^{-1}$) & 0.00 ($1.00 \times 10^{0}$) & 0.00 ($1.00 \times 10^{0}$) \\
\textbf{$\Delta M_{15}$ - $t_{\mathrm{rise}}$} & 0.04 ($8.91 \times 10^{-1}$) & -0.01 ($9.74 \times 10^{-1}$) & -0.05 ($8.37 \times 10^{-1}$) \\
\textbf{$M_{\mathrm{peak}}$ - $t_{\mathrm{rise}}$} & 0.19 ($5.47 \times 10^{-1}$) & 0.15 ($6.40 \times 10^{-1}$) & 0.05 ($8.37 \times 10^{-1}$) \\
\hline
\rowcolor{pastelred}
\multicolumn{4}{|c|}{\textbf{Explosion Time Error Threshold: 2, Total Non-EE: 28 (68.29\%)}} \\
\hline
\textbf{$\Delta M_{15}$ - $M_{\mathrm{peak}}$} & 0.08 ($6.79 \times 10^{-1}$) & 0.07 ($7.15 \times 10^{-1}$) & 0.04 ($7.69 \times 10^{-1}$) \\
\textbf{$\Delta M_{15}$ - $t_{\mathrm{rise}}$} & 0.17 ($3.92 \times 10^{-1}$) & -0.13 ($5.24 \times 10^{-1}$) & -0.08 ($5.40 \times 10^{-1}$) \\
\textbf{$M_{\mathrm{peak}}$ - $t_{\mathrm{rise}}$} & -0.17 ($3.95 \times 10^{-1}$) & -0.16 ($4.24 \times 10^{-1}$) & -0.10 ($4.41 \times 10^{-1}$) \\
\hline
\rowcolor{pastelred}
\multicolumn{4}{|c|}{\textbf{Explosion Time Error Threshold: 3, Total Non-EE: 35 (66.04\%)}} \\
\hline
\textbf{$\Delta M_{15}$ - $M_{\mathrm{peak}}$} & 0.13 ($4.56 \times 10^{-1}$) & 0.20 ($2.59 \times 10^{-1}$) & 0.13 ($2.74 \times 10^{-1}$) \\
\textbf{$\Delta M_{15}$ - $t_{\mathrm{rise}}$} & 0.16 ($3.74 \times 10^{-1}$) & -0.14 ($4.20 \times 10^{-1}$) & -0.09 ($4.43 \times 10^{-1}$) \\
\textbf{$M_{\mathrm{peak}}$ - $t_{\mathrm{rise}}$} & -0.14 ($4.11 \times 10^{-1}$) & -0.15 ($3.83 \times 10^{-1}$) & -0.11 ($3.63 \times 10^{-1}$) \\
\hline
\rowcolor{pastelred}
\multicolumn{4}{|c|}{\textbf{Explosion Time Error Threshold: 4, Total Non-EE: 43 (67.19\%)}} \\
\hline
\textbf{$\Delta M_{15}$ - $M_{\mathrm{peak}}$} & 0.18 ($2.57 \times 10^{-1}$) & 0.26 ($8.85 \times 10^{-2}$) & 0.17 ($1.00 \times 10^{-1}$) \\
\textbf{$\Delta M_{15}$ - $t_{\mathrm{rise}}$} & -0.03 ($8.54 \times 10^{-1}$) & -0.25 ($1.00 \times 10^{-1}$) & -0.19 ($8.04 \times 10^{-2}$) \\
\textbf{$M_{\mathrm{peak}}$ - $t_{\mathrm{rise}}$} & -0.22 ($1.61 \times 10^{-1}$) & -0.08 ($6.24 \times 10^{-1}$) & -0.05 ($6.08 \times 10^{-1}$) \\
\hline
\rowcolor{pastelred}
\multicolumn{4}{|c|}{\textbf{Explosion Time Error Threshold: 5, Total Non-EE: 48 (69.57\%)}} \\
\hline
\textbf{$\Delta M_{15}$ - $M_{\mathrm{peak}}$} & 0.20 ($1.77 \times 10^{-1}$) & \cellcolor{lightgray}0.32 ($2.45 \times 10^{-2}$) & \cellcolor{lightgray}0.22 ($3.01 \times 10^{-2}$) \\
\textbf{$\Delta M_{15}$ - $t_{\mathrm{rise}}$} & -0.03 ($8.16 \times 10^{-1}$) & -0.23 ($1.16 \times 10^{-1}$) & -0.17 ($9.29 \times 10^{-2}$) \\
\textbf{$M_{\mathrm{peak}}$ - $t_{\mathrm{rise}}$} & -0.18 ($2.22 \times 10^{-1}$) & -0.05 ($7.27 \times 10^{-1}$) & -0.04 ($7.15 \times 10^{-1}$) \\
\hline
\rowcolor{pastelred}
\multicolumn{4}{|c|}{\textbf{Explosion Time Error Threshold: 6, Total Non-EE: 53 (71.62\%)}} \\
\hline
\textbf{$\Delta M_{15}$ - $M_{\mathrm{peak}}$} & 0.19 ($1.73 \times 10^{-1}$) & \cellcolor{lightgray}0.30 ($3.06 \times 10^{-2}$) & \cellcolor{lightgray}0.20 ($3.56 \times 10^{-2}$) \\
\textbf{$\Delta M_{15}$ - $t_{\mathrm{rise}}$} & -0.03 ($8.22 \times 10^{-1}$) & -0.20 ($1.53 \times 10^{-1}$) & -0.14 ($1.33 \times 10^{-1}$) \\
\textbf{$M_{\mathrm{peak}}$ - $t_{\mathrm{rise}}$} & -0.18 ($2.06 \times 10^{-1}$) & -0.06 ($6.54 \times 10^{-1}$) & -0.04 ($6.56 \times 10^{-1}$)
\\
\label{tab:KS_AD_tests}
\end{longtable}
\twocolumn

\end{appendix}

\end{document}